\documentclass[10pt,final,3p,fleqn]{elsarticle}

\usepackage{caption}
\usepackage{subcaption}

\usepackage{gensymb}

\usepackage[bitstream-charter]{mathdesign}
\usepackage[T1]{fontenc}
\usepackage[utf8]{inputenc} 
\usepackage[normalem]{ulem}

\newcommand{\floor}[1]{\left\lfloor #1 \right\rfloor}

\usepackage{hyperref}
\usepackage{url}

\hypersetup{
    colorlinks=true,
    linkcolor=blue,
    filecolor=magenta,
    urlcolor=cyan,
}
\usepackage{booktabs}
\usepackage{xcolor}
\usepackage{color}

\usepackage{xcolor}

\definecolor{bostonuniversityred}{rgb}{0.8, 0.0, 0.0}
\definecolor{blue(pigment)}{rgb}{0.2, 0.2, 0.6}
\definecolor{blue(ncs)}{rgb}{0.0, 0.53, 0.74}
\definecolor{antiquefuchsia}{rgb}{0.57, 0.36, 0.51}

\definecolor{mygreen}{rgb}{0,0.6,0}
\definecolor{mygray}{rgb}{0.5,0.5,0.5}
\definecolor{mymauve}{rgb}{0.58,0,0.82}

\definecolor{l1}{RGB}{  0,   0, 255}
\definecolor{l2}{RGB}{  0, 255, 253}
\definecolor{l3}{RGB}{109, 238,   0}
\definecolor{l4}{RGB}{  0,  96,  19}
\definecolor{l5}{RGB}{177, 153,  37}
\definecolor{l6}{RGB}{255, 122,   0}
\definecolor{l7}{RGB}{255,   0,   0}

\def \cand {\mathrm{c}}
\def \neigh {\mathrm{s}}

\usepackage{nicefrac}

\usepackage{array}
\usepackage{mathtools}
\usepackage{wasysym}
\usepackage{verbatim}
\usepackage{relsize}
\usepackage{multirow}

\usepackage{amsmath}

\DeclareMathOperator{\atantwo}{atan_2}

\usepackage{enumerate}
\usepackage{enumitem}
\setlist{nosep}

\newcommand{\lstin}[1]{{\ttfamily #1}}
\usepackage{listings}
\newcommand{\rev}[1]{\textcolor{magenta}{#1}}
\newcommand{\revA}[1]{\textcolor{red}{#1}}

\newcommand*{\tran}{^{\mkern-1.5mu\mathsf{T}}}

\DeclareMathAlphabet{\pazocal}{OMS}{zplm}{m}{n}

\usepackage{lscape}		
\usepackage{fancyvrb}

\usepackage[mathscr]{euscript}

\newcommand{\DD} { \ensuremath{\pazocal{D}}}  

\newcommand{\vor}{Vorono{\"\i}}

\newcommand{\xDP}{{\ensuremath{ \pmb{x}^\star}}}
\newcommand{\xDPi[1]}{{\ensuremath{ \pmb{x}^\star_{#1}}}}
\newcommand{\balpha}{{\ensuremath{\pmb{\alpha}}}}

\usepackage{textcomp}

\def \p  {\pmb{p}} 
\def \X   {\ensuremath{{\pmb{X}}}} 
\def \x   {\ensuremath{{\pmb{x}}}} 
\def \gx  {\ensuremath{g( \pmb{x})}} 

\def \sobol  {{Sobol\textquotesingle}}

\usepackage{bigints}

\usepackage[switch]{lineno}

\newcounter{MVremark}

\journal{Computer Methods in Applied Mechanics and Engineering}

\usepackage{dsfont}

\def \IS {\textsf{IS}}

\newcommand{\dd}[1]{\mathrm{\,d}\hspace{0.05em}#1}

\newcommand{\Ns}{{\ensuremath{N_{\mathrm{sim}}}}}
\newcommand{\Nv}{{\ensuremath{N_{\mathrm{var}}}}}

\newcommand{\nT}{{\ensuremath{n_{\pazocal{T}}  } }}
\newcommand{\NT}{{\ensuremath{N_{\pazocal{T}}  } }}
\newcommand{\nF}{{\ensuremath{n_\pazocal{F}}}}

\newcommand{\PF}[1] {\ensuremath{{p_{\pazocal{F}}^{\left(#1\right)}}}}

\newcommand{\PT}[1]  {\ensuremath{{p_{\pazocal{T}}^{\left(#1\right)}}}}  
\newcommand{\PO}[1] {\ensuremath{{p_{\pazocal{O}}^{\left(#1\right)}}}}

\def \pFv {{\ensuremath{p_{\pazocal{F},v}}}} %
\def \pF  {{\ensuremath{p_{\pazocal{F}}}}} 
\def \pT  {{\ensuremath{p_{\pazocal{T}}}}} 

\def \IF  {{\ensuremath{\pmb{1}_{\pazocal{F}}\left(\x \right)}}} 
\def \IFN  {{\ensuremath{\pmb{1}_{\pazocal{F}}^{\Ns}\left(\x \right)}}} 
\def \ITX  {{\ensuremath{\pmb{1}_{\pazocal{T}}\left(\X \right)}}} 
\def \ITx  {{\ensuremath{\pmb{1}_{\pazocal{T}}\left(\x \right)}}} 
\def \ITN  {{\ensuremath{\pmb{1}_{\pazocal{T}}^{(\Ns)}\left(\x \right)}}} 

\def \ITNi  {{\ensuremath{\pmb{1}_{\pazocal{T}}^{(\Ns)}\left(\x_i \right)}}} 

\def \Fexact  {{\ensuremath{\pazocal{F}}}} 
\def \FS      {{\ensuremath{\partial\Fexact}}} 

\def \RSp  {{ \ensuremath{\mathcal{R}}}} 
\def \GSp  {{ \ensuremath{\mathcal{G}}}} 

\usepackage{textcomp}

\begin{document}
\begin{frontmatter}



\title{
    Reliability analysis of discrete-state performance functions
    via adaptive sequential sampling with detection of failure surfaces
}


\author{Miroslav Vo\v{r}echovsk\'{y}}
\address{Corresponding author. \\
Institute of~Structural Mechanics, Brno University of~Technology, \\ Veve\v{r}\'{i} 331/95, 602 00 Brno, Czech Republic,\\
e-mail: vorechovsky.m@vut.cz}


\begin{abstract}
The paper presents a~new efficient and robust method for rare event probability estimation for computational models of an engineering product or a~process returning categorical information only, for example, either success or \emph{failure}. For such models, most of the methods designed for the estimation of failure probability, which use the numerical value of the outcome to compute gradients or to estimate the proximity to the failure surface, cannot be applied. Even if the performance function provides more than just binary output, the state of the system may be a~non-smooth or even a~discontinuous function defined in the domain of continuous input variables. This often happens because the mathematical model features non-smooth components or discontinuities (e.g., in the constitutive laws), bifurcations, or even domains in which \emph{no reasonable model response} is obtained. In these cases, the classical gradient-based methods usually fail.
We propose a~simple yet efficient algorithm, which performs a~sequential adaptive selection of points from the input domain of random variables to extend and refine a~simple distance-based surrogate model.
    Two different tasks can be accomplished at any stage of sequential sampling:
    (i) estimation of the failure probability, and
    (ii) selection of the best possible candidate for the subsequent model evaluation if further improvement is  necessary.
The proposed criterion for selecting the next point for model evaluation maximizes the expected probability \revA{classified} by using the candidate. Therefore, the perfect balance between \emph{global exploration} and \emph{local exploitation} is maintained automatically.
If there are more rare events such as failure modes, the method can be generalized to estimate the probabilities of all these event types. Moreover, when the numerical value of model evaluation can be used to build a~smooth surrogate, the algorithm can accommodate this information to increase the accuracy of the estimated probabilities.
Lastly, we define a~new simple yet general geometrical measure of the global sensitivity of the rare-event probability to individual variables,
which is obtained as a~by-product of the proposed refinement algorithm.

\end{abstract}

\begin{highlights}

    \item
    accurate estimation of failure probability with error prediction, resistance to multiple design points, and strong nonlinearities

    \item
    the method needs at least a~categorical limit state function (including no response output) for continuously distributed inputs

    \item
    adaptive sequential extension of experimental design using the $\psi$ criterion, minimizing the number of limit state evaluations

    \item the novel criterion automatically balances between exploration and exploitation in the input space

    \item
    efficient importance sampling in important rings, excluding the safe ball and unimportant exterior

    \item
    global sensitivity indices quantifying the importance of individual variables to rare event probability

    \item
    adaptive expansion and refinement near the failure surface avoiding clusters of points (no redundant information)

\end{highlights}

\begin{keyword}
        $\psi$ criterion \sep
        failure probability \sep
        failure surface \sep
        Gaussian space \sep
        extension of experimental design\sep
        gradient-free optimization \sep
        categorical limit state function \sep
        binary surrogate model \sep
        importance sampling
\end{keyword}
\end{frontmatter}



\section{Introduction}

As summarized in a~2001 review by \citet{Rackwitz2001}, the issue of the determination of failure probability attracted considerable attention in the second half of the last century. Currently, the interest of industry and  researchers shows that the importance of the topic is even increasing.
Freudenthal, whose landmark paper \cite{Freudenthal1947} dates back to 1947, is often considered the father of modern structural reliability.
The standard problem setting is a~limit state function \gx\ (also called a~performance function or computational model) with continuously distributed random inputs
\rev{
\X\ forming the random vector with dimension $\Nv$. The probability density function $f_{\X}(\x)$ is assumed as known. The spectrum of existing methods for reliability estimation is very rich and they can be classified based on the way how and how much they use and also make targeted modifications to the three groups of input input information
    (i) the information about the limit state function $\gx$,
    (ii) the density of the input vector $\X$, and
    (iii) the geometry and topology of the input space.
Crude Monte Carlo sampling estimation of probabilistic integrals does not modify or make special use of any of these inputs which makes it very robust method, but also quite inefficient. Other methods utilize more or less one or more pieces of information and also make various kinds of assumptions about these inputs to improve their efficiency.
}

The development of the mathematical theory of the first order reliability method (FORM)
\cite{Hasofer1974ExactAI,Rackwitz1978}
and the second-order reliability theory (SORM)  \cite{Fiessler1979,Hohenbichler1987,BreitungHohen1989,Tvedt1990} with the related FORM/SORM asymptotic approximations for multinormal integrals using Taylor series expansions of the first/second order
\cite{Breitung1984,Breitung1994}
was completed in the eighties \cite{DitlMadsen:StructRelMeth:96,MadsenKrenkLind:MethStructSaf:86,MelchBeck:2017}. These foundational methods are still perhaps the most important results on which modern design codes for engineers are built.
In 1983, two papers
\cite{Harbitz,Shinozuka1983} brought importance sampling (\IS) concepts to the attention of the reliability community. The great advantage of having error estimation alongside probability estimation made \IS\ an excellent tool for reliability  estimation updating, and it also improved FORM/SORM estimations \cite{Hohenbichler1988,Bucher1988,Maes1993}.
Importance sampling around the mean value and the method of asymptotic sampling \cite{Bucher2009} both exploit the fact that in most engineering reliability problems the failure regions are located further from the central region of the joint density, and the sampling density increases the spread compared to the joint density. Another version of importance sampling used a~sampling density centered on the design point, which expresses the premise that the most central failure point is surrounded by a~small neighborhood, which contains most of \revA{the failure probability} \pF\ (a~result derived by  \citet{BreitungHohen1989}; see also  \citet{Breitung1994}). Indeed, a~design point in the space of independent standard normal variables is a~failure point that maximizes the joint normal density, or equivalently, minimizes the distance to the origin. Moving away from this point into the failure domain as well as moving along the boundary between failure and safe regions decreases the probability density. These points are searched based on the premise that the value of the performance function describing the state of the analyzed product or process decreases towards them in the safe region and becomes negative in the failure region. FORM usually operates in the standard normal space where the ``design points''  have to be found.

As pointed out by Rackwitz \cite{Rackwitz2001}, in practical engineering the \emph{reliability analysis} of an engineering product (be it a~structure, a~mechanical system, or a~process) is itself perhaps less interesting than the \emph{optimization} \cite{BeySen:CMAME:07} of those products or processes with reliability constraints. However, reliability optimization methods encompass reliability analysis and call it repeatedly \cite{Valdebenito2010,Rajan2020}.

With ever-growing computational power, numerical tools are now, and more than ever, being widely applied to the representation and solution of complex problems in engineering and the sciences. Unfortunately, in many cases, the problem of evaluating reliability via the estimation of failure probability remains intractable. This is often because of the large computational expenses associated with the model evaluations that are needed to run a~selected sampling method for failure probability estimation, such as importance sampling \cite{Melchers1989,Au1999,AU2003139,Papaio:Papa:Straub:SeqIS:SS:16},
line sampling \cite{Schuller2004,deAngelis2015},
directional simulation \cite{Bjerager1988,Nie2000},
asymptotic sampling \cite{Bucher2009} or subset simulation \cite{Au2001,au_engineering_2014,PapBet-15}. High-fidelity computational models describing the performance of engineering systems are computationally intensive, and the application of  advanced (and often adaptive) sampling methods in combination with a~high-fidelity model to obtain sufficiently accurate estimations of failure probability is not feasible.



In order to build an approximate representation of a~true performance function, which is inexpensive to analyze via sampling strategies, a~variety of surrogate models have been developed. Smart strategies exploit the information from already analyzed points from the design domain, and they can adaptively refine the surrogate model; see, e.g. \cite{Sundar2016, TeixOconn:AdaptiveMetamodel:Review:SS2021}.
Many \emph{adaptive} reliability analysis methods have been developed, some using radial basis functions \cite{Li2018,Shi2019}, support vector regression surrogates \cite{Li2006,BOURINET2011343:fourbranch,Bourinet2016,Pan2017,Roy2022},
artificial neural networks \cite{deSantanaGomes2019,Gomes2020,SaraygordAfshari2022},
sparse polynomial chaos \cite{Marelli2018,Zhou2020} or Kriging \cite{Echard2011,Wang2022}.
Another alternative for probabilistic calibration as new information emerges is to use Bayesian updating with Kriging \cite{SONG2022114578}.
Authors have invented various combinations of methods, such as a~combination of Importance Sampling and Kriging \cite{Echard2013}. Unfortunately, most of these methods are developed with the assumption that the surrogate  approximates a~well-behaved function defined over the whole design space. Moreover, methods based on Kriging (a.k.a. Gaussian process regression) are often criticized for the strong dependence of the results on the selection of the kernel, which is a~user choice, often subjective, and hardly supported by sufficient data. Recently, the sample space partitioning strategy has been proposed along with an adaptive Kriging model \cite{Song2021}, which relaxes this assumption. However, most adaptive strategies often use various kinds of learning functions in which the numerical value of the performance function plays an important role.

In most of the existing methods, it is assumed that the performance of a~system is a~smooth function $\gx$ defined over the whole input domain \DD. Traditionally, the performance function \gx\ is defined such that its negative values signal a~\emph{failure event} and positive values success. The boundary between the safe and failure sets is called the \emph{failure surface} and will be denoted as \FS. We assume it is an $(\Nv-1)$-dimensional object, and it is often illustrated as the zero-valued contour of $\gx$. When the failure surface \FS\ is a~differentiable function and the function is almost linear in the standard normal space, the failure probability has a~simple relationship to the shortest distance $\beta$ from the origin to the failure surface: $\pF \approx \Phi(-\beta)$. The rotational symmetry of the standard normal density makes this FORM approximation simple because the failure domain is approximated by a~half-space fitted
to the true failure domain at the distinct \emph{nearest failure point}.
Many optimization algorithms have been developed \cite{Liu1991} to find the nearest failure point for smooth failure surfaces. The most important is the method by \citet{Hasofer1974ExactAI} for second-moment reliability analysis, which \citet{Rackwitz1978} later extended to include distribution information (the famous ``HL-RF'' iterative scheme).
We remark that while this point marks the failure event with the maximum Gaussian density, it may not be the most likely failure point in terms of the original non-Gaussian density due to the nonlinearity of the probability transformation. The main task then becomes the location of the most central failure point (often called the design point or the $\beta$ point).

On the other hand, methods guided by the numerical values of the performance function might be misled into heading in an incorrect search direction. Not only the ``design point search'' needed for FORM and SORM relies on the assumption that the performance function decreases towards the most central failure point. Subset Simulation (SuS) \cite{Au2001} or some sequential adaptive variants of Importance Sampling methods \cite{Bucher1988,Papaio:Papa:Straub:SeqIS:SS:16} are also based on this assumption. In this way, SuS, for instance, can be viewed as a~stochastic version of the gradient (downhill) optimization method and, as exemplified by  \citet{Breitung:19:RESS} using numerous examples, the method can fail for functions with complicated landscapes.
Similar to the design point search for FORM, such downhill optimization methods may not be successful in global optimization because the search may be caught in local minima. Repeated SuS runs may not help if the search is always initiated from the origin and the information about the geometry of the performance function is not stored.  The problem of multiple design points has not been satisfactorily solved even in FORM. Moreover, various different formulations of \gx\ that lead to the same failure set may alter the result of \pF\ estimation because the formulations change the evolution of the estimation process, which is obviously wrong.
\rev{
The idea of extrapolation with a~sequence of modified problems, which was focused on modifications of the sampling density in Asymptotic Sampling \cite{Bucher2009}, can be seen in an analogy to making modification to the limit state function \cite{Naess2009,Luo2022}, thus making assumptions about the role of supposedly smooth landscape of \gx. A~different group of reliability methods which can be termed ``moment methods'' also use the numerical values of \gx: the basic idea is to fit a~proper probability distribution to the output variable of the limit state function based on the knowledge of its estimated moments of certain type (see, e.g., integer moment based methods \cite{ZhaoLu:ReliabMoments:21,ZHOUpeng:AdaptBayesQuadMomEstStructRel:20}, fractional moment based methods \cite{ZhaPandey:FractMoment:13,XuKong:unscented:18,DangXu:MixtureFractMom:20}, moment-generating function (or Laplace transform) based methods \cite{XuDang:fractMomMaxEntReliab:19,DangXu:Laplace:20,DangWeiBeer:Seismic:21}).
}

Moreover, it sometimes happens that complex computational models are not able to provide any answer for some input values (e.g., a~nonlinear finite element solver of a~structure is not able to converge for a~combination of input parameters). Or, the performance of a~system is not a~smooth function or contains discontinuities which pose a~problem for most of the gradient-based algorithms embedded in the reliability analyses.
The gradient-based algorithms are used to locate the most central failure point (or more of these points for different parts of the design domain), and the shape of the supposedly continuous performance function is used to estimate the distance to the boundary between failure and safe regions (the safety margin). However, some performance functions of a~system may be just a~discrete number of states or just a~binary function returning either ``success'' or ``failure''.
In these cases, the majority of the above-mentioned algorithms and methods fail entirely, and thus the motivation behind the present work was to develop a~robust technique that can solve all these problems related to the performance function.

There is thus a~need for a~method that is resistant to noise or a~non-smooth shape in the performance function, its jumps or even discrete values, and yet is able to provide a~reasonable failure probability estimation with a~small number of function evaluations.
The method should balance between the global \emph{exploration} of the design domain in search for new input space  territories leading to failure
and the local \emph{exploitation} of the previously discovered boundary between safe and failure sets (failure surface) in order to refine its description. We see it as the problem of ``how to divide a~territory?": to detect and geometrically describe the boundaries between the safe and failure sets. Optimally, the method should sequentially extend the experimental design (ED) in single steps to maximize the gain from the already obtained information at each stage.
The method should keep adapting its representation of the true performance via some form of a~surrogate model, which is inexpensive to evaluate and which uses a~tailored sampling strategy for quick on-the-fly reliability estimation.
The method should not make unnecessary calls of the expensive response function
in regions surrounded by ``safe'' samples (no matter ``how'' safe they are) as these are almost sure to be also a~part of the safe set. Analogously, we want to avoid unnecessary new samples in the (almost sure) failure regions. It is evident that the most precious information is the refinement of the boundary separating safe and failure regions in proportion to the local density. It is there that the expensive information needs to be obtained by calling the true performance function, i.e., the limit state function.


\section{Problem statement \label{sec:problemStatement}}

Let us assume a~$\Nv$-dimensional vector $\X$ of continuously distributed basic random variables with known joint probability density function $f_{\X}(\x)$. Vector $\X$ is the input to the performance function of a~system/process whose reliability is to be evaluated. Assume that for any realization (a~point from the \emph{design domain}, $\DD$), $\x \in \DD$, we can obtain, albeit at high expense, information as to whether or not the system fails, or generally about the system performance.
We define the failure set (domain) $\Fexact \subset \DD$ as a~union of all regions within the design domain in which failure occurs.
The probability of failure \pF\ is then defined as
\begin{align}
    \label{eq:pf:definition1}
    \pF \equiv
    \idotsint_{\Fexact} f_{\X}(\x) \; \dd \x
    =
    \idotsint_\Fexact \dd F_{\X}(\x) ,
\end{align}
where $F_{\X}(\x)$ is the cumulative density functions of the random vector \X.
The integral over the whole \DD\ equals one (exhaustion of all possible events).

Let us assume that the performance of a~system is evaluated via a~performance function, sometimes referred to as the \emph{limit state function} \gx. From here on, we will consider the crudest case in which \gx\ returns either 1 (failure) or zero for a~safe state (cases when the performance function does not provide any answer in some input points will be discussed separately). Limiting the response to a~\emph{binary function} does not affect the definition of \pF\  in Eq.~\eqref{eq:pf:definition1} as it never included any other information.
When working with systems for which we can obtain responses other than a~binary response via \gx, we use it to define the \emph{indicator function} $\IF$, which returns one for $\x$ falling to the failure set (typically $\gx < 0$) and zero otherwise
\begin{align}
    \IF
    &=
     \left\{\begin{array}{ll}
        1,          & \x \in \Fexact  \text{ (failure event)},\\
        0           & \text{otherwise} .
        \end{array}\right.
\end{align}
Note that the failure set may be composed of more types of failure or  indications of malfunctioning, and the corresponding failure probabilities may also be desired. We address this eventuality by generalizing the dichotomous nature to handle more than two event types. A~rare event will be generally denoted as $\pazocal{T}$. The method described in this paper focuses on the approximation of the boundaries between various sets.
However, as illustrated below, we assume that the rare-event sets are not scattered over too many disjoint sets, which are closed and have individually almost zero measure.

By including this indicator function in the integrand in Eq.~\eqref{eq:pf:definition1},
the integral over the whole design space gets narrowed down to the failure domain
\begin{align}
    \label{eq:pf:definition2}
    \pF =
     \idotsint_{\DD} \IF  \, f_{\X}(\x) \, \dd \x.
\end{align}

Without loss of generality, we assume that the joint probability distribution function (PDF) of \X\ is the $\Nv$-dimensional standard normal density with independent components.
If the original problem features non-normal marginals and dependencies, we assume that a~probability preserving transformation exists, such as the Nataf model \cite{Nataf,DerKiureghian1986,Vorech:08:CrossCorr,Lebrun2009} (sometimes referred to as the Gaussian copula \cite{Nelsen:Copulas:06}) or the Rosenblatt-transformation, via a~product of conditional distributions \cite{Rosenblatt1952,MelchBeck:2017}. The Gaussian space is particularly suitable for the proposed method as it is rationally invariant and open (unbounded).
We remark that in many practical problems, the information about the joint PDF is often not complete. If the information is limited to univariate marginals and the correlation matrix only, there exist many possible forms of the joint PDF fulfilling this input, and it may happen that none of them can be represented via the Nataf transformation. New ways to transform the original joint PDF to a~Gaussian space which are more flexible
\cite{Lebrun2009} exist, and are still appearing. They may employ polynomial chaos representation \cite{Sakamoto2002} or other distributions 
\cite{Zhao2021}.
We deliberately refrain from discussing these transformations further as they are not the main focus of this paper.

This paper proceeds with a~quick presentation of tools useful in rotationally symmetrical standard Gaussian space (Sec.~\ref{sec:stdGauss}), and these are immediately used to present the proposed sequential adaptive extension of sample size in Sec.~\ref{sec:AdaptSeq:ED}. The desired failure probability can be estimated by a~quick and simple evaluation of the information obtained at the current ED via binary surrogate, which we propose should be constructed on-the-fly; see Sec.~\ref{Sec:surrogate:pf}. The byproduct of this estimation is the information about novel global sensitivity indices which we propose in Sec.~\ref{sec:sens}.

\section{Useful tools in the standard Gaussian space
\label{sec:stdGauss}
}

We assume that the probability density in the standard Gaussian space is jointly Gaussian, i.e., the individual standard normal random variables are independent.
Therefore, the probability density function of any point $\x = \{ x_1,\ldots,x_{\Nv} \}$ becomes the product of univariate Gaussian densities of the individual marginals
\begin{equation}
    \label{eq:stdGauss:point:x}
    f_{\X}(\x)
    =
    \prod_{v=1}^{\Nv}  \varphi(x_i)
    =
    \frac{1}{(2\pi)^{\Nv/2}}\exp{ \left(-\frac{1}{2} \x \tran \x \right) } ,
\end{equation}
where $\varphi(x) = {(2\pi)^{-1/2}}\exp{ \left(-{x^2}/{2} \right) } $ is the standard univariate Gaussian density.

Due to the rotational symmetry with respect to the origin, the space can be indexed using (i) $\Nv-1$ independent directions and (ii) one univariate Euclidean distance from the origin
\begin{equation}
    \label{eq:rhodist}
    \rho =  \lVert \x \rVert  =  \sqrt{ \x \tran \x } = \sqrt{\sum_{v=1}^{\Nv} x_v^2},
    \quad
    v=1,\ldots,\Nv.
\end{equation}
Therefore, the standard Gaussian density $f_{\X}(\x)$ at a~point $\x$ simplifies to univariate Gaussian density depending on the radial distance $\rho$ only
\begin{equation}
 \label{eq:stdGauss:point:rho}
    f_{\X}(\x)
    = \varphi(\rho)
    =
    \frac{1}{(2\pi)^{\Nv/2}}\exp{ \left(-\frac{\rho^2}{2} \right) }
\end{equation}

\begin{figure*}[!t]
    \centering
    \includegraphics[width=19cm]{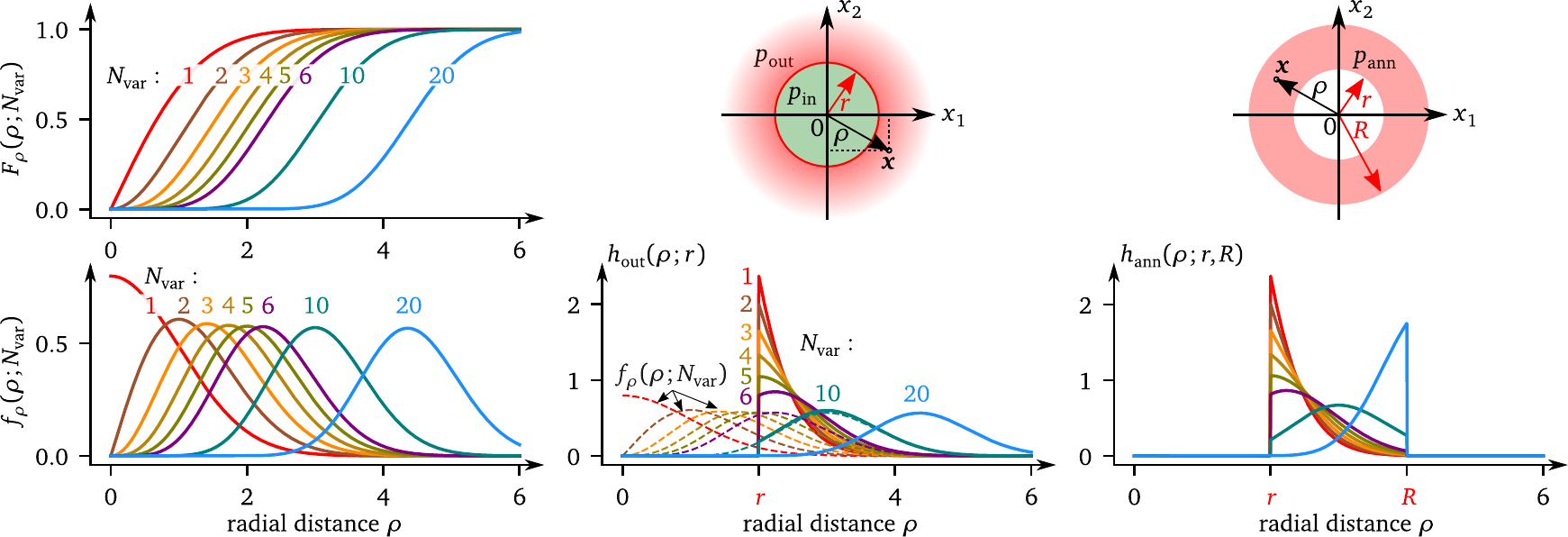}
    \caption{
    Left: probability density and distribution function for a~random radial distance in standard Gaussian space.
    Middle: radial distance density of a~point selected outside an \Nv-ball of radius $r$.
    Right: radial distance density of a~point selected in between two \Nv-balls of radii $r$ and $R$.
    }
 \label{fig:chidist}
\end{figure*}

An important ingredient in the proposed algorithm is the distribution function for a~random distance $\rho$ in the Gaussian space. Assume a~randomly selected point, $\x$. Its Euclidean distance from the origin has $\chi$ (chi) distribution with $\Nv$ degrees of freedom.
The probability density function of a~distance $\rho > 0$ reads (see Fig.~\ref{fig:chidist} bottom left)
\begin{equation}
\label{eq:chi-pdf}
    f_{\rho} (\rho; \Nv)
    =   \frac{2^{1-\Nv/2} }
           { \Gamma\left(\frac{\Nv}{2} \right)  }
        \rho^{\Nv-1}
        \exp{ \left(-\frac{ \rho^2}{2} \right) } ,
\end{equation}
where $\Gamma(\cdot)$ is the standard (complete) gamma function.
The corresponding cumulative distribution function,
defined as
$ F_{\rho} (\rho; \Nv)  = \int_{0}^{\rho}  f_{\rho} (t, \Nv) \dd t$,
reads  (see Fig.~\ref{fig:chidist} top left)
\begin{align}
\label{eq:chi-cdf}
    F_{\rho} (\rho; \Nv)
    =
    \frac{2^{1-\Nv/2} }
    { \Gamma\left(\frac{\Nv}{2} \right)  }
    \int_{0}^{\rho}
    t^{\Nv-1}
    \exp{ \left(-\frac{ t^2}{2} \right) }
    \dd t
    = P\left( \frac{\Nv}{2} , \frac{\rho^2}{2}  \right),
\end{align}
where $P ( s, x) = \gamma(s,x) / \Gamma(s)$ is the \emph{regularized lower incomplete gamma function}.
It follows from the lower incomplete gamma function
$\gamma(s,x) = \int_0^x t^{s-1} \exp(-t) \dd t$  by its regularization via the complete gamma function $\Gamma(s)$.  $P ( s, x)$ is one of the ``special functions'' and it is a~standard part of various mathematical libraries, such as \lstin{scipy} \cite{2020SciPy-NMeth}, which is available in \lstin{Python} (function \lstin{gammainc($\cdot,\cdot$)}).

Therefore, when a~point is selected randomly from $\Nv$-dimensional standard Gaussian space, its distance from origin has a~distribution as if it were selected from a~univariate $\chi$ distribution with \Nv\ degrees of freedom. Its mean Euclidean distance $\mu_{\rho}$, the approximate median (50 \% percentile $\mathrm{me}_{\rho}$), and the modus (the most probable distance $\mathrm{mo}_{\rho}$) read
\begin{align}
   \label{eq:mu:rho}
    \mu_{\rho} &= \sqrt{2} \frac{ \Gamma \left( (\Nv+1)/2 \right) }{\Gamma(\Nv/2)}
     \xrightarrow{\Nv \to \infty}
     \sqrt{\Nv-1/2},
    \quad \mathrm{me}_{\rho} \approx \sqrt{\Nv }  \left(  1- \frac{2}{9\Nv}\right)^{3/2}    \\
    \label{eq:mo:rho}
    \mathrm{mo}_{\rho} &= \sqrt{\Nv-1}
\end{align}
The variance of a~random distance $\rho$ is simply $\sigma^2_{\rho} = \Nv -  \mu_{\rho}^2$, which asymptotically equals $1/2$, independently of $\Nv$.

\subsection{Interior of an $\Nv$-ball in standard Gaussian space}

An important geometrical entity in the proposed method is an \Nv-ball with radius $r$, which is centered at the origin of the coordinate system; see the green domain in Fig.~\ref{fig:chidist} top middle.
The volume and the surface of the $\Nv$-ball, denoted as $B_{r}  \in \mathbb{R}^{\Nv}$, read
\begin{align}
\label{eq:vol}
    \mathrm{Vol} \left[ B_{r} \right]
    &=
    \frac{\pi^{\Nv/2}}{\Gamma\left(  \frac{\Nv}{2}+1 \right)}
    r^{\Nv},
    \\
\label{eq:sur}
    \mathrm{Sur}\left[  B_{r} \right]
    &=
    \frac{2 \pi^{\Nv/2}}{\Gamma\left(  \frac{\Nv}{2}\right)}
    r^{\Nv-1}.
\end{align}
We remark that the $\chi$ density presented in Eq.~\eqref{eq:chi-pdf}
 can be obtained simply by multiplying the point Gaussian density
from Eq.~\eqref{eq:stdGauss:point:rho} by the ball
surface from Eq.~\eqref{eq:sur}, for which we take the radius $r$ equal to the random radial distance $\rho$, i.e.,
$f_{\rho} = \varphi(\rho)  \cdot \mathrm{Sur}\left[  B_{\rho} \right] $.

Given this probability density function and the corresponding cumulative distribution function in Eq.~\eqref{eq:chi-cdf}, one can easily evaluate the probability content of the $\Nv$-ball interior. The probability that a~random Gaussian point falls inside the ball $B_r $ reads:
\begin{align}
    p_{\mathrm{in}} (r,\Nv)
    \equiv
    F_{\rho} (r; \Nv)
    = P\left( \Nv/2 , r^2/2  \right).
\end{align}
The inverse of the regularized lower incomplete gamma function, $P^{-1}(\cdot)$, implemented as
\lstin{gammaincinv($\cdot,\cdot$)}) in  \lstin{scipy},  can be used to compute the  \emph{radius} of an \Nv-ball $B_{r}$ that contains a~point with prescribed probability $p_{\mathrm{in}}$:
\begin{equation}
\label{eq:radius}
    r (p_{\mathrm{in}}; \Nv)
    \equiv
    F_{\rho}^{-1} (p_{\mathrm{in}}; \Nv)
    = \sqrt{  2 \, \left[ P^{-1}(\Nv/2, p_{\mathrm{in}} )\right]    }.
\end{equation}

\subsection{Exterior of an $\Nv$-ball in standard Gaussian space
\label{sec:exterior}}

Analogously, one can also define the exterior of $B_{\rho} $ schematically represented by the red domain in Fig.~\ref{fig:chidist} top middle. Therefore, the probability corresponding to the region outside the ball, i.e., all points with the radial distance $\rho \in \left(r, \infty \right)$, must be complementary to the function $ F_{\rho} (\rho; \Nv)  $
\begin{align}
\label{eq:pout}
 \nonumber
    p_{\mathrm{out}} (r;\Nv)
    & \equiv
    1- p_{\mathrm{in}} (r;\Nv)
    =
    1- F_{\rho} (r; \Nv)
    \\
    &= 1- P\left( \Nv/2 , r^2/2  \right)
    \\
    \nonumber
    &=  Q\left( \Nv/2 , r^2/2  \right),
\end{align}
where $Q\left( \cdot , \cdot  \right)$ is known as the  \emph{regularized upper incomplete gamma function}. This ``special function'' is again a~standard part of various mathematical libraries such as \lstin{scipy} available in \lstin{Python} (function \lstin{gammaincc($\cdot,\cdot$)}).
Analogously to Eq.~\eqref{eq:radius}, the radius of the ball \emph{not} containing a~point with given probability $ p_{\mathrm{out}}$ can be obtained by the inversion of Eq.~\eqref{eq:pout}
\begin{equation}
\label{eq:radius2}
    r (p_{\mathrm{out}}; \Nv) = \sqrt{  2 \, \left[ Q^{-1}(\Nv/2, p_{\mathrm{out}} )\right]    }.
\end{equation}
Of course, the radii $r$ from Eqs.~\eqref{eq:radius} and \eqref{eq:radius2} match when
$  p_{\mathrm{in}}  +   p_{\mathrm{out}}  = 1$.

The exterior of an \Nv-ball is an important geometrical entity for reliability analysis. Consider a~situation in which one samples from the exterior via importance sampling. The \emph{sampling density} is nonzero only outside the \Nv-ball and is obtained by re-scaling the standard Gaussian density limited to $ \lVert  \x \rVert \in( {\color{red} r} , \infty)$  so that the probability content outside the ball is still unit
\begin{align}
    h_{\mathrm{out}} \left( \x; r \right)
    &=
        \displaystyle
        \frac{
            \Pi_{v=1}^{\Nv}  \varphi (\x_v)
        }
        {
        p_{\mathrm{out}} (r,\Nv)
        } .
\end{align}
In practice, it is more convenient to obtain this sampling density for point $\x$ in terms of its radial distance  $\rho$ from the origin. Such left-bounded $\xi$ density is just the re-scaled density from Eq.~\eqref{eq:chi-pdf}
\begin{align}
    h_{\mathrm{out}} \left( \rho; r \right)
    &=
        \displaystyle
        \frac{
            f_{\rho} (\rho; \Nv)
        }
        {
        p_{\mathrm{out}} (r,\Nv)
        }
       =
        \displaystyle
        \frac{
            f_{\rho} (\rho; \Nv)
        }
        {
        1 - F_{\rho} (r; \Nv)
        }.
\end{align}
Indeed, in order to randomly sample a~point from the exterior of an $\Nv$ ball $B_r$, we exploit the rotational symmetry of the density in standard Gaussian space and decompose the task into two sub-tasks:
(i)  sampling a~random direction with the uniform distribution (which is an $(\Nv-1)$ dimensional problem) and
(ii) sampling a~random Euclidean distance from the origin, which is simply a~one-dimensional problem.
A~practical method of performing the first step, i.e., sampling a~random unit direction in \Nv\ dimensions, is covered in \ref{sec:rand:dir}, and we propose a~method to improve the spread of multiple random directions in \ref{sec:unif:dir}.
What then remains is to choose a~univariate random Euclidean distance from the origin, which is a~one-dimensional problem.
Component-wise multiplication of the previously obtained $\Nv$ unit direction vectors $\pmb{s}$ by random distances that obey a~left-bounded $\chi$ variable will move the points along their radii-vectors to the desired distance from the origin $d \in \left( {\color{red} r}, \infty \right)$.  This random distance corresponding to a~sampling probability $p \in (0,1)$ can be obtained via inverse transformation as
\begin{equation}
 d_{\mathrm{out}} ({\color{red} r})
  =
   F_{\rho} ^{-1}  \left[
                p + (1-p) \,  F_{\rho} (r; \Nv)
            \right] .
    \label{eq:ppf:dist:out}
\end{equation}
Finally, the corresponding point can be obtained simply as
\begin{equation}
 \x_{\mathrm{out}} = d_{\mathrm{out}} \; \pmb{s}.
 \label{eq:IS:scaling}
\end{equation}
If a~sample of $n$ points all at once is needed, it is advisable to cover the distances from the origin uniformly with respect to probability, and therefore a~set of even space sampling probabilities $\pmb{p}: \{ p_i= (i-0.5)/n, \; i=1,\ldots, n \}$ is recommended.

\subsection{Interior between two $\Nv$-balls in standard Gaussian space
\label{sec:annulus}
}

A rotationally symmetrical region between two different radii, $r<R$, forms an annulus (a ring), $B_{\mathrm{ann}}$; see the red domain in Fig.~\ref{fig:chidist} top right. Its probability density for a~point $\x$ is again just a~scaled standard Gaussian density
\begin{align}
    h_{\mathrm{ann}} \left( \x; r , R\right)
    &=
        \displaystyle
        \frac{\Pi_{v=1}^{\Nv}  \varphi (\x_v)}
             {p_{\mathrm{ann}} }
    =
        \displaystyle
        \frac{\Pi_{v=1}^{\Nv}  \varphi (\x_v)}
             {p_R - p_r }
    \label{eq:sampl:dens:ann:x}
\end{align}
and zero otherwise. In this equation, the scaling denominator
$p_{\mathrm{ann}} = p_R - p_r$
is formed by the difference between two probabilities, $p_R = p_{\mathrm{in}} (R,\Nv)$ and $r_r = p_{\mathrm{in}} (r,\Nv) $.
Again, this density can be rewritten in terms if the radial distance
    $\rho \equiv   \lVert  x \rVert $.
Whenever $r < \rho < R $, the probability density becomes nonzero, and it reads
 \begin{align}
    h_{\mathrm{ann}} \left( \rho; r , R\right)
     &=
        \displaystyle
        \frac{ f_{\rho} (\rho; \Nv) }
             {p_{\mathrm{ann}}} .
    \label{eq:sampl:dens:ann:rho}
\end{align}

The cumulative distribution for this distance reads
$p(\rho) = \int_r^\rho f_{\rho} (t; \Nv)  / (p_R - p_r ) \dd t
=
\left( F_{\rho} (\rho; \Nv) - p_r \right)/ (p_R - p_r )
$ and its inversion can be used to obtain distances corresponding to selected sampling probability
$p \in (0,1)$
\begin{equation}
 d_{\mathrm{ann}}
 =
F_{\rho}^{-1}  \big(
                p \, ( p_R  - p_r )   +  p_r ; \Nv
            \big).
    \label{eq:ppf:dist:ann}
\end{equation}

Sampling random points from the annulus can be performed in the same fashion as described above, i.e., by using a~set of random unit directions from Eq.~\eqref{eq:rad:direction} and scaling them via Eq.~\eqref{eq:IS:scaling} by using the distance $d_{\mathrm{ann}}$ from Eq.~\eqref{eq:ppf:dist:ann}.

A natural question arises: why would we sample points from the annulus between two balls? In the proposed algorithm, the Gaussian space can conveniently be divided into the interior of a~ball of radius $r$, and its exterior. If we are no longer interested in exploring the ball interior (e.g., we are sure no failure occurs below distance $r$), and we believe that the most central failure point is roughly at a~distance $r$ from the origin, it makes no sense to consider very remote locations to estimate the failure probability by integrating the Gaussian density (the exterior of the \Nv-ball extends to infinity). It rather makes sense to focus only on regions that contribute considerably to the exterior probability, i.e., to limit the outer radius to a~value $R$. In this way, the described annulus, or  \emph{important ring}, is formed.

Assume now the worst possible scenario in which the failure region is exactly the exterior of the ball with radius $r$. In this case we know the corresponding failure probability $(1-p_r)$ via knowing the probability inside the ball $p_r = F_\rho^{-1}(r;\Nv)$ (safe state).
Note that such a~failure domain represents an upper bound of failure probability because any other failure surface will have a~smaller failure probability. It then makes sense to exclude regions outside a~greater ball with radius $R$ by which only a~negligible fraction of $1-p_r$ is excluded. For example, consider excluding the probability of $(1-p_r)/10^4$ only. The outer diameter $R$ then obeys $1-p_R = (1-p_r)/10^4$ meaning that
$p_R = 1 - (1-p_r)/10^4$. Therefore, the outer radius can be computed via Eq.~\eqref{eq:radius} as
\begin{equation}
\label{eq:estim:R}
    R = F_{\rho}^{-1} \big( 1 - (1-p_r)/10^4; \Nv
                   \big).
\end{equation}

\section{Sequential adaptive \emph{extension} of the experimental design
\label{sec:AdaptSeq:ED}
}

The premise behind the development of the presented algorithm is that each evaluation of the system performance (limit state function) is very expensive. Crude Monte Carlo type sampling strategies simply throw sampling points ($=$ integration ``nodes'') independently of each other, and do not reflect the structure of the problem being solved at all. Some advanced sampling techniques learn from previously obtained information but may not be able to fully exploit its potential.

\begin{figure*}[h]
	\centering
	\includegraphics[width=18cm]{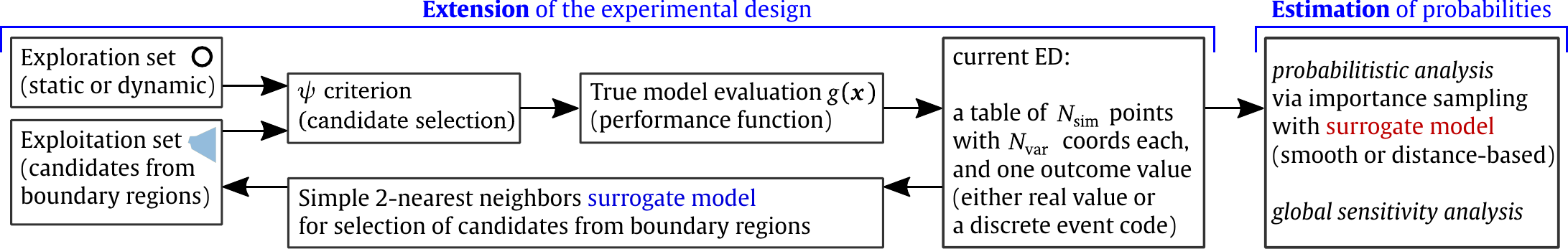}
	\caption{Flowchart of the proposed methodology consisting of two basic tasks: extension and estimation.}
	\label{fig:flowchart}
\end{figure*}

The set of points at which the true performance function has been evaluated so far will be referred to as the \emph{experimental design} (ED), and their number will be denoted as \Ns. The ED is simply a~table of \Ns\  points each with \Nv\  coordinates and one additional vector of $\Ns$ results obtained from the performance function. These results can be indices of one of the \emph{discrete states}, including failure, success, no result, or a~failure type code. The algorithms proposed in this paper are formulated to operate with such limited information. In particular, no other information is necessary for either of the two fundamental steps, namely (i) \emph{extension} of ED via the selection of the best candidate point for subsequent evaluation of the expensive performance function, and (ii) \emph{estimation} of the current failure probability using a~temporary binary surrogate model (or a~surrogate with discrete states as listed above). These two steps are recognized in Fig.~\ref{fig:flowchart}, which presents an overview of the proposed method. It involves the construction of two different surrogate models: one for the identification of domains for local refinement, and the other for fast numerical integration via sampling.

The existing methods for the extension of the ED often employ \textit{learning functions} \cite{TeixOconn:AdaptiveMetamodel:Review:SS2021}. Learning functions are convenient mathematical functions that weigh the metamodel properties to seek the best candidate to extend the current ED.
They evaluate a~set of candidates with criteria that are built on considerations of uncertainty in the model approximation and the proximity to the failure region and select the new most promising candidate.
\citet{Bichon2008} and \citet{Echard2011} introduced two of the most relevant works in this context:  \citet{Bichon2008} presented ``Efficient Global Reliability Analysis'', proposing the use of the ``Expected Feasibility Function'' to extend the ED, and \citet{Echard2011} developed a~method combining Kriging with Monte Carlo Simulation while using what they called a~U-function based on the probability of misclassifying a~candidate in order to extend the ED.
The literature on learning functions is quite rich. \citet{Lv2015} proposed the H learning function, which is built on entropy consideration, and combined Kriging with Line Sampling, \citet{Sun2017} defined a~learning function named the Least improvement function (LIF), which combines Kriging-based statistical information and the joint PDF of basic random variables. \citet{Zhou2019} proposed a~learning function in Polynomial Chaos Expansion that models uncertainty with a~Bayesian approach.  Quite recently, \citet{Zhang2019} proposed the Reliability Expected Improvement Function (REIF), which is to be combined with an adaptive Kriging surrogate, which relates to the expected improvement (EI) of \citet{Jones1998}.
Indeed, when the performance function is smooth and not just binary, its value is deemed to provide an additional measure of the distance from the failure surface. In such cases, this information can be used to increase the effectiveness of the algorithm in various ways. One of the possibilities is to construct a~surrogate model based on the existing ED. The extension of the ED can balance between the exploration and exploitation of sampling using conditional probability \cite{BAO2021107778}.
All of these approaches use the numerical values of the true performance function \gx\ and its surrogate.

In the present paper, we consider that the only pieces of information to be supplied for the selection are: (i) the position of the points in ED, (ii) the classification of the model output (failure, failure type, success, no information, etc.) and (iii) the joint probability density function.
In order to discover new failure domains and, at the same time, to refine the existing approximation of the failure surface (boundary), we prepare two sets of candidate points: the \emph{exploration set} (Sec.~\ref{sec:exploration:set}) and the \emph{exploitation set} (Sec.~\ref{sec:exploitation:set}), combine them together and perform the selection of the best candidate using the proposed $\psi$ criterion (Sec.~\ref{sec:criterion}).

\subsection{Exploration set (global)
\label{sec:exploration:set}
}

\begin{figure}[!bh]
    \centering
    \includegraphics[width=0.9\linewidth]{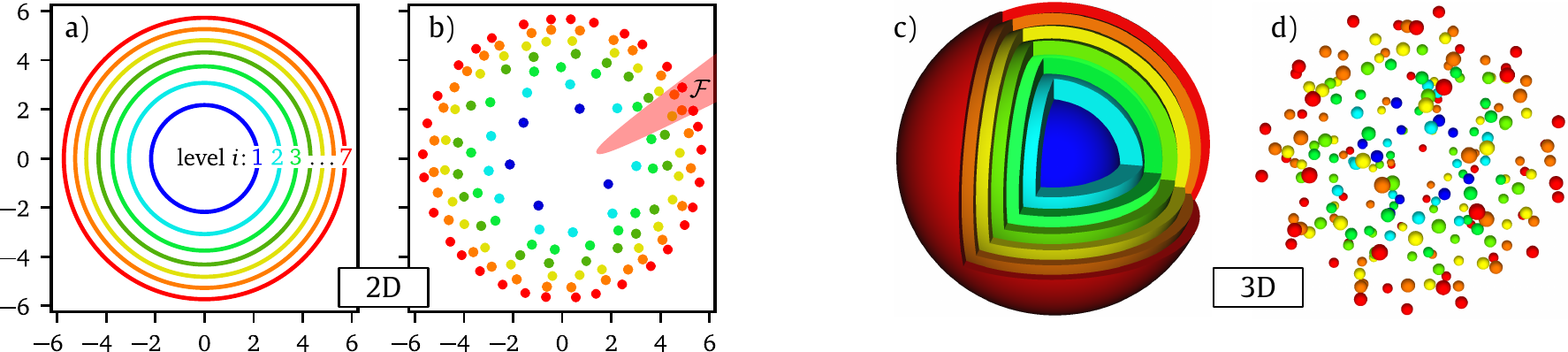}
    \caption{The first seven levels of exploration sets (layers) in $\Nv=2$ and $\Nv=3$ dimensional standard Gaussian space. Panels a,c: nested \Nv-balls. Panels b,d: realizations of random exploration sets.}
 \label{fig:ExplorationSet}
\end{figure}

We expect that a~rare event (failure) region can appear anywhere. The local probability density of any potential failure region size decreases with increasing distance from the origin in the standard Gaussian space. Therefore, in a~global search for failure domains, it makes sense to explore the space in a~controlled fashion by checking various directions with an increasing distance from the origin. Therefore, we propose the construction of a~sequence of nested \Nv-dimensional balls
in the Gaussian space whose radii increase towards infinity and which occupy the probability in  prescribed levels. In other words, we suggest preparing an \emph{exploration set} for ``onion-like'' layers; see Fig.~\ref{fig:ExplorationSet} left. Each layer is numbered by a~level number $i$, and the corresponding \Nv-ball is to be covered by a~prescribed number of points $n_i$ corresponding to this layer.
The boundary of each layer is an \Nv-ball, and the $n_i$ points selected from the surface of this ball should be uniformly dispersed. Moreover, it is preferable to avoid directionally collapsible candidates:  the candidates corresponding to level $i$ should have different radii vectors than the points from the preceding level $i-1$.

The probability content occupied by an \Nv-ball corresponding to layer $i$ is selected with regard to what level of failure probability the ball is roughly associated with. More exactly, the radii of the balls can be selected
using Eq.~\eqref{eq:radius} such that the exterior of that ball has a~prescribed probability, $p_\mathrm{out}$, see Eq.~\eqref{eq:pout}.

There is freedom in the proposed method to select: (i) probabilities $p_{\mathrm{out},i}$ corresponding to individual levels and (ii) the numbers of points $n_i$ corresponding to these individual levels.
The selection of these parameters is an important step. If the exploration algorithm is too \emph{aggressive}, meaning that the ball radii are rapidly increasing while they are each covered by only a~small number of points, there is a~risk of overlooking some localized failure domain \revA{\Fexact\ }(regions which are narrow at small radii; see the illustration in Fig.~\ref{fig:ExplorationSet}b in which the first four levels of exploration points miss the tip of the failure domain). In such a~case, it takes several other ``onion layers'' and expansive \gx\ evaluations to hit the failure domain for the first time. As will be shown below, once a~distant failure is hit, the algorithm will automatically back-trace the failure surface down the high probability region; however, many unnecessary calls to the performance function will be made. If, on the other hand, the ``onion layers'' are too densely packed and/or each covered by too many exploration points, it will take many calls to hit the failure domain for the first time and start the exploitation phase.
In the simplest situation, the exploration set is pre-generated in advance for a~given problem dimension. We remark, however, that the exploration set can be modified anytime during the run of the proposed method; see below. For example, suppose the exploration phase has already reached a~rather high radius, and the analyst suspects that a~narrow convex failure domain might have been overlooked. The exploration set can be enriched with additional layers or candidates at any time, and the selection criterion described below will automatically select the most relevant candidate(s) to check for potential failures.

Once the decision regarding the directions/radii coverage is made, it is not difficult to select $n_i$ points for each level approximately uniformly distributed on the unit sphere and scale them to the desired radius $\rho_i$. We refer to \ref{sec:rand:dir} and \ref{sec:unif:dir} for a~simple algorithm for the selection of the set of $n_i$ points from the surface of \Nv-dimensional ball.

Based on our experience, it is a~good practice to select the \Nv-balls so that each level leaves an unexplored exterior with probability $p_{\mathrm{out},i} = 10^{-i}$, where $i=\{1,2,3,\ldots   \}$. This means that the first ball consumes 90\% and leaves only a~probability content of 0.1, the second ball only leaves 0.01, the third 0.001, etc. Given these levels, the surfaces of these balls are sufficiently covered by exploration points when their number is selected as
\begin{equation}
  \label{Eq:ni}
  n_i = -\Nv \ln \left(  \frac{p_{\mathrm{out},i}}{\Nv} \right).
\end{equation}
These point counts, e.g., rounded down to integers using the floor function $n_i = \floor{n_i}$, will be used in all the numerical examples presented in this paper. With this rule, the number of evaluations to cover a~certain safety level is known in advance, and so is the number of function calls for a~given failure probability. Of course, any prior knowledge about the failure probability can be used to adjust the radii.
To get an idea about the point counts proposed in Eq.~\eqref{Eq:ni} for dimensions ranging between two and twenty, see Fig.~\ref{fig:ExplorationSet:Numbers} and Tab.~\ref{tab:ExplorationSet} in \ref{sec:explor:table}.

Until a~rare event is encountered in the ED for the first time (be it a~failure or a~non-result), the candidate set for selection via the $\psi$ criterion is formed by this \emph{exploration set} only. Once the desired event is hit, the surroundings of the point must also be considered in a~search for the most informative candidate. In other words, an \emph{exploitation set} with appropriate candidates must be constructed and included when selecting the next ED point.

\subsection{Exploitation set (local)
\label{sec:exploitation:set}
}

Assume now that the previously evaluated performance function confirmed at least one rare event (e.g., failure) at any point of the ED. It is clear that the failure surface has been crossed, and the location of the boundary should be refined locally. In the case of a~binary result, it is reasonable to test a~location somewhere close to existing failure and success points. More exactly, we propose searching for the boundary at locations (candidates) whose \emph{two nearest neighbors have different classifications} (failure, success, no result or simply \emph{different states}). From all candidates, we propose that only those having two \emph{different} classifications of their two nearest neighbors be retained. The exploitation set is prepared sequentially, i.e., after each extension of the ED, and so it is based on current information.
The preparation of the candidate set works in two steps:
  (i) the generation of a~large pool of ``dots'' in a~sufficiently large neighborhood of each existing rare event point (e.g., a~failure point) and
  (ii) the selection of only those candidates that potentially refine the boundary between different states (the two nearest ED neighbors signalling a~different state). The censored list of points is the \emph{exploitation set}.

The pool of candidates can be pre-generated at the beginning and kept constant throughout the whole analysis.
We, however, propose generating them after each discovery of a~new rare event (e.g., evaluation of the limit state function returning failure). A~set of ``candidate dots'' is simply generated in the spirit of importance sampling with Gaussian sampling density $h_i(\x)$ centered at each rare-event ED point, and the standard deviation is driven by the problem dimension:
\begin{align}
  \label{Eq:dots}
    \mu_i & = \x_i, \quad i=1,\ldots,\NT \\ \nonumber
    \sigma_i & = \sqrt{\Nv-1},
\end{align}
which spreads the dots sufficiently away from the existing point but at the same time focuses the dots sufficiently.
Since the censoring that extracts (masks) the candidates for the exploitation set is fast (based merely on the computation of distances), it is quick to generate and censor $10^6$ candidates or more.
Moreover, we remark that many operations can be pre-computed for the potential classifications (failure, success, etc.) while the true performance function is still being evaluated. Once the performance function \gx\ has returned the result, one of the pre-computed scenarios is used to call \gx\ with the best candidate immediately.

\begin{figure}[!t]
    \centering
    \includegraphics[width=\linewidth]{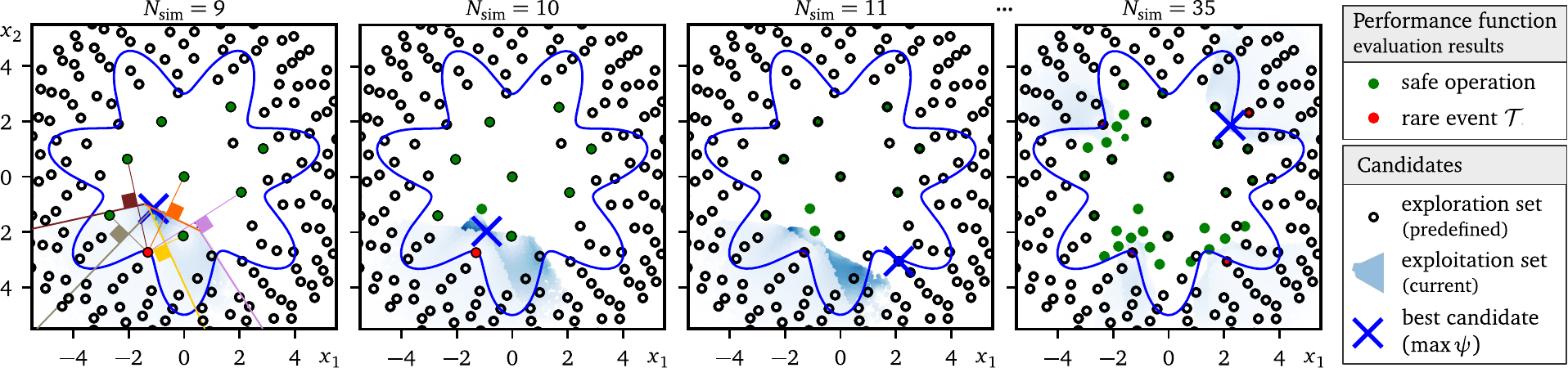}
    \caption{Two-dimensional illustration of the combined set candidates and the selection process (standard Gaussian space) for a~binary performance function dividing the space into a~``flower-shaped'' safe domain and the surrounding failure domain.  At $\Ns=9$ model evaluations, the rare event is hit for the first time. At $\Ns=10$ an \emph{exploitation} candidate is selected, while at $\Ns=11$ the \emph{exploration} candidate wins using the highest value of the $\phi$ criterion expressed by the blue color saturation.
    A~complete evolution of the extension of the ED using the $\phi$-based selection process for another run of the same problem is available as a~\href{https://vutbr-my.sharepoint.com/:v:/g/personal/vorechovsky_m_vutbr_cz/EaWDuW0xQYhAlhwqtnOww8kBRmGSLw4LhI-aejNjK-kT0w?e=zNdhaE}{\textcolor{blue}{\textsf{video}}}.
    }
 \label{fig:combined:set}
\end{figure}

Fig.~\ref{fig:combined:set} presents an illustration of the exploitation sets.
The figure shows the situation at four different stages after $\Ns$ evaluations of a~binary performance function. The true failure surface is plotted via the blue curve (a complicated shape with seven ``design points''). The algorithm started with the exploration candidates (empty circles), and \rev{the results obtained from evaluations of the limit state function} are \rev{visualized by \emph{filling} the circles} (either green  for ``success'' or red for ``failure'' event). Once the first red point was discovered, the algorithm started proposing \emph{exploitation}  candidates (the results obtained in \rev{ \rev{these} exploitation candidates can be distinguished from the exploratory ones by the fact that the colored circles do not have a~black ring around them}). The boundaries of this region are, in fact, axes of lines connecting ED points with different outcomes. These lines are plotted via solid lines of various colors. Note, however, that neither these lines nor the corresponding boundaries are constructed. The boundaries are naturally formed by censoring candidates with two different nearest ED neighbors. The boundaries in higher dimensions become formed by planes in 3D and hyperplanes in higher dimensions. No computational geometry is needed, and generalization to high dimensions is straightforward.

Now that the combined exploitation and exploration set has been prepared, the last step is the selection of the best candidate. In the situation depicted in the figure, the best candidate (blue cross) is selected from the exploration set, but this selection depends on the local situation and balances between the local and global sets to maximize the potential gain from the next evaluation of the performance function. The selection criterion is described in the following section.

\subsection{$\psi$ criterion for candidate selection
\label{sec:criterion}
}

Suppose the combined exploration-exploitation set of candidates is available. The task is to select the candidate in which the (supposedly costly) evaluation of the performance function delivers the maximum gain in terms of probability.
%
The new point will support the geometrical interpretation of the failure domain which is being constructed. The structural information about the shape and location of the failure surface is important, e.g., for sensitivity analysis \cite{Papaioannou2021}, see also Sec.~\ref{sec:sens}, and may also be useful for other approximation techniques (such as the Taylor expansion at ``design points'' etc.).

One of the main motivations behind developing the proposed algorithm is that
    no two ED points may be required close to each other unless they deliver a~significant amount of information in terms of probability content (that is, refine the boundary approximation between territories corresponding to different event types -- such as the failure surface \FS).

We propose a~novel $\psi$ criterion for the selection of the best candidate. Its maximization leads to the (approximate) maximization of the instantaneous gain in terms of probability content because the meaning of the criterion is the approximate amount of probability content being classified by evaluating \gx\ in the candidate. Indeed, the
values of $\psi$ can be viewed as ``bites'' of probability with a~clear geometrical meaning: each candidate represents a~certain region (its neighborhood) in the design space and the volume of this region can be multiplied by the average probability density to obtain the corresponding probability content:
\begin{equation}
    \label{Eq:PsiCrit:c}
    \psi_\cand
    =
    \underbrace{
                \sqrt{f_\cand f_\neigh}
                }
                _{\mathrm{ave \; probability}}
    \,
     \underbrace{
                \left( l_{\cand,\neigh} \right)^{\Nv}
                 }
                 _{\propto \mathrm{vol.}},
\end{equation}
where the term  $\sqrt{f_\cand f_\neigh}$ accounts for the (Gaussian) probability density of both the candidate, $\cand$, and the density in its nearest ED point, $\neigh$, (i.e., the existing = previously evaluated point in the ED). The term represents the geometrical average probability density of the two. The geometric mean between $n$ numbers $x_i$ is defined as $(\Pi_{i=1}^n x_i)^{1/n}$, and therefore the proposed term features $\sqrt{f_\cand f_\neigh}$. Why do we use the geometric mean instead of the simple arithmetic mean? The arithmetic mean is not a~suitable measure as it would favor infinitely distant candidates. The problem is that very distant candidates in Gaussian space that have zero probability density would still form a~nonzero average density with an existing ED point.  Therefore, very remote candidates would have a~high $\psi$ criterion and become selected, which is not preferable \revA{-- the expansion to distant locations is already controlled by the exploration set}.

\begin{figure}[!tb]
	\centering
	\includegraphics[width=16cm]{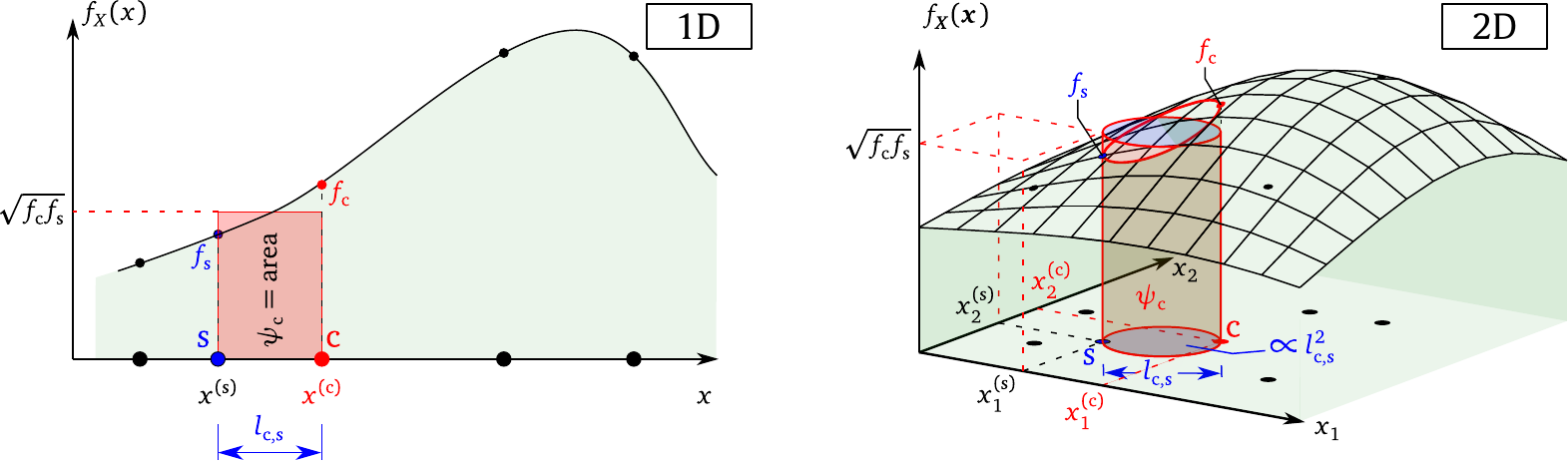}
	\caption{Geometrical meaning of the proposed $\psi$ criterion for a~candidate ``$\cand$'' in $\Nv=1$ and $\Nv=2$ dimensions. The black solid circles are the existing ED points and the blue circle ``$\neigh$'' is the nearest neighbor to candidate ``$\cand$''. $\psi_c$ is therefore proportional to the probability content of the red domain close to the candidate.}
	\label{fig:2D-phi}
\end{figure}

The term $l_{\cand,\neigh}$ is the distance between a~candidate, $\cand$, and its nearest existing ED point, $\neigh$. Simply, the distances between all candidates and all \Ns\ currently existing ED points are evaluated, and each candidate is paired with its nearest point. To compute the \Ns\ distances from each candidate, a~suitable metric must be selected. It this paper, we consider Euclidean distance between the candidate and its nearest neighbor as
\begin{equation}
    \label{eq:metric}
    l_{\cand,\neigh}
    =
    \sqrt{\sum_{v=1}^{\Nv}  \big|\x_v^{(\cand)}-\x_v^{(\neigh)}\big|^2  }.
\end{equation}
When this distance is raised to the problem dimension:  $\left( l_{\cand,\neigh} \right)^{\Nv}$, it becomes proportional to the volume of a~geometrical object selected from the design domain. This geometrical object, be it an \Nv-ball, \Nv-hypercube, or some other shape, has a~volume that quantifies the size extent of the neighborhood of the candidate $c$. We do not need to specify the exact shape of the volume as all the geometrically similar volumes for various candidates under comparison would share the same positive multiplier related to the specific geometric shape. This multiplier can be dropped from the definition as it does not alter the ranking of compared candidates. The geometrical interpretation of the criterion is illustrated in Fig.~\ref{fig:2D-phi}.

The criterion is constructed as a~product of two independent terms:
\begin{itemize}
    \item
        the volume term, which favors rapid \emph{exploration} of previously unexplored regions by expansion, and
    \item
        the probability term, which favors the \emph{exploitation} of high probability contents occupied by a~candidate near the origin.
\end{itemize}
The balance between exploration and exploitation is thus maintained automatically: a~candidate located close to an existing ED point can only be selected if the corresponding probability density is sufficiently high so that the gain is greater compared to other more distant points. And vice-versa, overly distant candidates are not preferred because their densities are low.
The criterion favors candidates leading to a~fast classification of the space because, the $\psi$ criterion at a~given stage of the computation is the maximum value of all $\psi_{\cand}$ selected among all candidates under comparison
\begin{equation}
    \label{Eq:PsiCrit}
    \psi
    =
    \max_{\cand} \left( \psi_\cand \right).
\end{equation}

The illustrative example in Fig.~\ref{fig:combined:set} reveals that the criterion automatically switches between the exploration and exploitation sets. Points selected from the predefined exploration set have a~black circumference and are filled with either red or green color. Filled circles without the black circumference have been selected from the exploitation sets, and these points could have appeared only after the discovery of the first rare event point (red). All candidates having two different types of two nearest neighbors are considered to be exploitation candidates and ranked together with the exploration candidates using the proposed $\psi$ criterion. The maximum $\psi$ criterion values always correspond to the blue crosses.
During the exploitation steps, the algorithm tends to alternate between safe and failure domains by refining the approximation of the failure domain.

It is useful to record the history of the $\psi$ criterion throughout the sequential extension of the ED. The criterion provides a~rough orientation as to how much the next candidate can modify the current estimation of \pF, and thus \rev{these two pieces of information} may serve \rev{for formulation of} a~\emph{stopping criterion}. If the value of $\psi$ drops below a~certain limit (say $\pF/100$), it might be reasonable to quit the algorithm and/or switch to another type of analysis. Experience has shown that $\psi$ will keep decreasing if the computation continues, and typically the rate of decrease slows down as further refinements of the boundaries cause only minor changes to the estimations.

This formulation of the stopping criterion does not guarantee that the algorithm has already discovered all important regions. It can happen that further computation will lead to the discovery of new, previously overlooked regions.
In such cases, the estimation of underestimated $\pF$ may suddenly increase. Analogously, it may also happen that the $\pF$ estimation is exaggerated, especially in high dimensions. If the ED is very small for a~problem dimension,
it is possible that the discovery of a~failure point is not followed by refinement of the boundary, and an overly rough approximation of the true failure surface via an oversimplified surrogate model may then estimate the extent of the failure domain as being too large, leading to the overestimation of the true $\pF$.

\subsection{Discussion on the $\psi$ criterion and the assumptions behind the proposed algorithm}
\label{sec:discussion:assumptions}

\paragraph{High dimensions and the hyperspherical method}
As pointed out by \citet{KATAFYGIOTIS2008208}, the geometry of the reliability problem in high dimensions is challenging: the volume of the probability space grows with the number of random variables. In such a~large volume, most of the contributions to the failure probability come from failure regions with small probability densities but large volumes. \citet{KATAFYGIOTIS2008208} discuss the significance of the ``design point'' (the most central failure point) in the case of strongly nonlinear problems, and they show that the selection of an appropriate importance sampling density is practically impossible in these cases. Indeed, high-dimensional reliability problems are hard to treat using importance sampling (no matter if the sampling density is centered in the origin or multiple densities are used centered in the identified design points). When the form of the importance sampling density is not appropriate, the variance of importance sampling estimation is known to explode with domain dimension \cite{AU2003139}.
\citet{KATAFYGIOTIS2008208} introduced a~concept of an Important Ring, which is a~rotationally symmetrical region between radii $\Nv \pm \epsilon$ centered at the origin of the design space, and explained why \IS\ with density centered around the ``design point'' becomes impractical. These geometric insights were exploited by \citet{WangSong2018}, who presented a~hyper-spherical extrapolation method for high-dimensional problems. Indeed, the shape of an annulus, Fig.~\ref{fig:chidist} left, and Eq.~\eqref{eq:mo:rho} reveal the fact that most of the probability is associated with a~thin layer around an \Nv-ball with a~radius of approximately $\sqrt{\Nv-1}$. The traditional assumption that the region in the vicinity of the most central point (a.k.a. the design point) has a~prevalent contribution to the failure probability may not be valid, especially for very high dimensions $\Nv$.
The present method solves this problem automatically by employing the proposed $\psi$ criterion, which automatically favors the largest probability contributions of candidates, no matter where they are.

\paragraph{Assumptions and limits}
As \citet{Rackwitz2001} pointed out, an important step in the development of methods is to \emph{show} where they do \emph{not} work, i.e., to find the limits of the applicability of a~concept and to construct counterexamples.
The assumption behind the developed algorithm in its simplest version is that the corresponding set (territory) forms an \emph{open domain}. As has now become clear, cases with scattered localized \emph{closed} rare event domains may not be treated well; it can happen that localized domains will be missed by the exploration set and encapsulated by the application of the proposed $\psi$ criterion during the refinement.
A separate numerical example featuring the ``Modified Rastrigin function'' (see Fig.~\ref{fig:Overview}) is devoted to this \revA{weakness} and is presented in Sec.~\ref{sec:rastrigin}.

\paragraph{Incorporation of a~priori knowledge}
Sometimes, the performance function represents a~problem for which the outcome can be known \emph{a priori}, i.e., without the need to run an expensive computer simulation. For example, suppose a~specific region is known to be a~safe region. This information can be passed to the algorithm easily without modifying the proposition of candidates or modifying the selection criterion. Whenever a~candidate is selected to become an ED point and a~\gx\ call is required, the point location is checked  first, and if the point belongs to the described region, the call is simply bypassed by associating the ED point with the known classification.

\section{Failure probability \emph{estimation} via sampling analysis
\label{Sec:surrogate:pf}
}

At any stage of the ED extension process, the desired probabilities can be estimated based on  point-wise information, i.e., the current ED with known \gx\ outcomes; see the right-hand part of Fig.~\ref{fig:flowchart}. Indeed, a~true computationally expensive model (a~simulator) can be substituted by a~computationally cheaper model. The surrogate model is constructed in a~solely non-intrusive way with respect to the original simulator, i.e., it is purely data-driven.
When the performance function \gx\ returns continuously distributed, trustworthy, and well-behaved output, traditional surrogate models such as PCE, Kriging, radial basis functions, etc., can be employed for fast sampling analysis.
Such a~surrogate model has the potential to improve the accuracy of the importance sampling estimation presented in this section because the failure surface (the boundary \FS{}) may be approximated more accurately. It is guaranteed that the above-described extension of the ED was performed in such a~way that the surrogate model was well supported, especially when close to the failure surface.

In this paper, however, we focus on cases in which the original simulator is trusted to provide categorical information only, i.e., one of a~finite set of classifications. Similarly, the surrogate will simply be a~finite-state classifier.

\subsection{Rare event surrogate via the nearest neighbor classifier}

\begin{figure}[!htb]
    \centering
    \includegraphics[width=17cm]{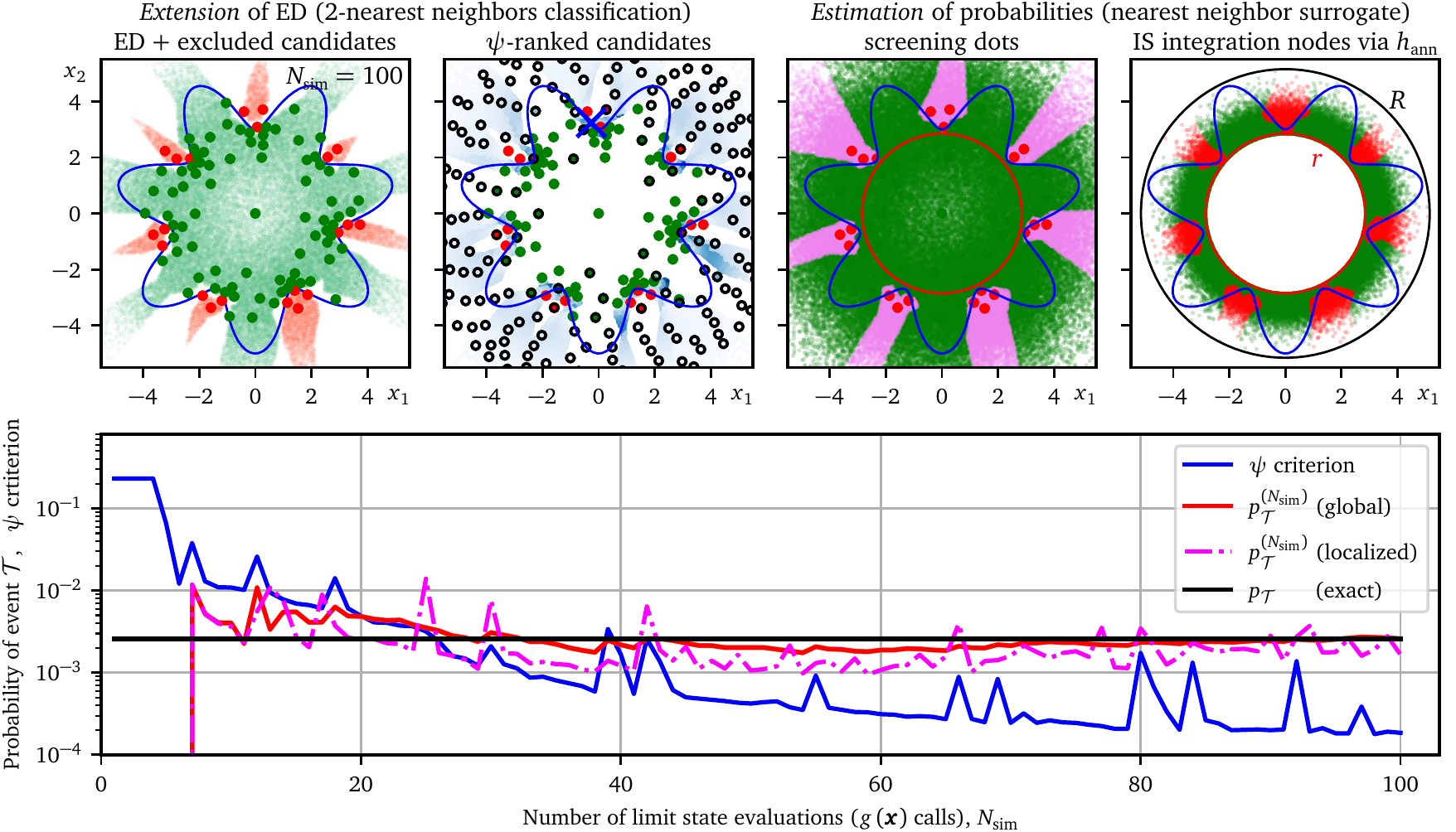}
    \caption{Top row: on-the-fly classification of territories.
      Top left panels: two nearest neighbors for the classification of boundary regions -- red and white candidates are censored out, and the retained (blue) candidates are considered for the \emph{extension} of ED.
      Top right panels: classification of dots and nodes via the nearest neighbor; used for screening and cubature.
      Bottom: history of the rare event probability \emph{estimation} and the $\psi$ criterion quantifying the approximate ``probability bites'' by selected candidates.
      The complete evolution of all panels is demonstrated in the \href{https://vutbr-my.sharepoint.com/:v:/g/personal/vorechovsky_m_vutbr_cz/EaWDuW0xQYhAlhwqtnOww8kBRmGSLw4LhI-aejNjK-kT0w?e=zNdhaE}{\textcolor{blue}{\textsf{Wavy circle video}}}.
    }
 \label{fig:NearestNeigh}
\end{figure}

We propose the construction of an intermediate discrete-valued surrogate at any location \x\ based simply on its nearest ED neighbor at which the performance of the system has already been evaluated. For example, an integration node \x\ will only be considered as a~``failure location'' if the nearest neighbor from the ED signaled failure. In this way, an effective approximation indicator function \IFN\ of the true indicator function \IF\ is obtained at any stage with \Ns\ evaluated points.
Analogously, a~general event $\pazocal{T}$ is estimated to occur if the known model response in the nearest neighbor is $\pazocal{T}$ and a~collection of such points becomes associated with a~nonzero indicator value $\ITX$. This corresponds to the division of the design domain \DD\ into a~finite number of territory types (see the two top right parts of Fig.~\ref{fig:NearestNeigh}) with just a~binary classification.
The surrogate must decide about the previously ``undecided'' regions visible in the top left part of the figure, i.e., region considered in the ED extension as potential boundary regions between two different classes of model outputs.

Making a~response prediction based on its nearest neighbor is equivalent to constructing a~\vor\ diagram, i.e., Dirichlet tessellation, which is a~standard way to partition a~domain into disjoint cells called Dirichlet regions ($=$Thiessen polytopes $=$ \vor\ polygons).
The \Ns\ points from the current ED with known classifications play the role of the \emph{seeds} (sometimes called  sites or generators).
This \vor-type approximation may seem to be too crude, but as will be shown below, when the classification is built using points obtained via the proposed sequential adaptive algorithm based on the $\psi$ criterion, \ITN{}  quickly tends to the true $\ITx$  as \Ns\ increases. An important aspect is that no computational geometry is involved; the only operation needed is the comparison of distances.

We remark that the described distance-based surrogate model is not restricted to the Gaussian space only. The surrogate model and also the estimation of \pF{} can be made either in the standard Gaussian space or, it can be computed directly in the original space, standardized to zero mean and unit variance. If, however, any of the variables are bounded, then experience has shown that the proposed sequential refinement and classification becomes less efficient.

\subsection{Importance sampling estimation}

Assume the surrogate model for event $\pazocal{T}$ is ready in the form of indicator function $\ITX$. The related event probability defined by Eq.~\eqref{eq:pf:definition2}
can today be estimated very efficiently, e.g., via importance sampling (\IS).
In \IS\ the original joint density $f_{\X}(\x)$ for sample selection is replaced by the importance sampling density $h(\x)$, which we consider to be jointly Gaussian with independent marginals. Generating millions of realizations of Gaussian random variables takes less than a~second on contemporary computers, and implementations are provided ready to use in various free numerical packages (for example the \lstin{NumPy} package \cite{harris2020array} for powerful operations over multidimensional dimensional arrays in combination with vectorized functions and routines \revA{\cite{ChuSadRypVor:CPC:Spirrid:13}} from \lstin{scipy} available in \lstin{Python}).

The probability of an event type $\pazocal{T}$ occurring when the indicator function $\ITX$ signals it is defined as the expectation:
$
\pT
= \mathbb{E} \left[ \ITX \right]
=
\idotsint_{\DD} \ITx \, f_{\X}(\x) \, \dd \x
$.
Let $h(\x)$ be the \IS\ density which is positive wherever event $\pazocal{T}$ occurs.  The probability of event $\pazocal{T}$ can be rewritten as
\begin{align}
 \label{eq:IS:anal}
    \pT =
    \int_{\DD}
    \frac{ \ITx \,f_{\X}(\x) } {h(\x)}
    h(\x)
    \dd \x
    =
    \mathbb{E}_h \left[ \frac{ \ITX \,f_{\X}(\X) } {h(\X)} \right],
\end{align}
where  $ \mathbb{E}_h \left[ \cdot \right] $ denotes the expectation for $\X$ being distributed according to $h$: $\X \sim h$. The \IS\ estimation of $\pT$ based on the current approximation of $\ITN{}$ of the true indicator function $\ITx$ is made with $n_{\IS}$ integration nodes  via the arithmetic average
\begin{align}
\label{eq:IS:estim}
    \PT{\Ns}
    &\approx
    \frac{1}{n_{\IS}} \sum_{i=1}^{n_{\IS}}
     \ITNi{}
         \frac{ f_{\X}  (\x_i) }
              { h       (\x_i) },
    \quad
    \X_i \sim h.
\end{align}
The \IS\ estimator is unbiased by construction. The variance of the \IS\ estimator follows the standard definition
\begin{align}
 \label{eq:IS:var:anal}
    \text{Var}_h[\pT ]
    &=
    \mathbb{E}_h
    \left[
    \left(
    \frac{ \ITx \,f_{\X}(\x) } {h(\x)}
    \right)^2
    \right] - \pT^2
  =
    \int_{\DD}
    \left(
    \frac{ \ITx \,f_{\X}(\x) } {h(\x)}
    \right)^2
    h(\x)
    \dd \x -\pT^2
\end{align}
Its estimation using $n_{\IS}$ integration nodes $\X_i $ sampled from $h$, $i=1,\ldots,n_{\IS}$, can again be performed using an arithmetic average
\begin{align}
\label{eq:IS:var:estim}
    \text{Var}_h[ \PT{\Ns} ]
    &\approx
    \frac{1}{n_{\IS}}
    \Bigg\{
    \left[
     \frac{1}{n_{\IS}}
    \sum_{i=1}^{n_{\IS}}
     \ITNi{}
         \frac{ f^2_{\X}  (\x_i) }
              { h^2       (\x_i) }
    \right]
    -
    \left(\PT{\Ns}\right)^2
    \Bigg\}
\end{align}
and finally, the coefficient of variation of the estimation reads
\begin{align}
\label{eq:IS:cov:estim}
    \text{CoV}_h[ \PT{\Ns} ]
    &\approx
    \frac{\sqrt{\text{Var}[ \PT{\Ns} ]}}
    {\PT{\Ns} }
\end{align}
A~straightforward application of \IS\ estimation would be to use Gaussian sampling density centered at the origin (zero mean values) and a~magnified standard deviation which is somehow based on the smallest distance between the failed ED point and the center point. Though this is a~robust strategy, it is not very efficient, as many integration nodes fall outside the territories corresponding to the desired event type. Another disadvantage is the need to compute the weight ratio for each integration node.

It is known that the optimal choice of the importance sampling density for event $\pazocal{T}$ (e.g. a~failure) is proportional to the original density, but defined only over the importance (failure) region in which the event $\pazocal{T}$ occurs:
 $h_{\text{opt}}(\x) =  \ITx \,f_{\X}(\x) / \pT$.
Such an optimal density leads to zero variance of the \IS\ estimator Eq.~\eqref{eq:IS:estim}.
Generally, using such an optimal density is not feasible for two reasons: the probability $\pT$ being estimated is not known,
and even if the density $h_{\text{opt}}(\x)$ can be constructed, a~method to efficiently sample from such a~density is generally not available.

We argue that the crucial component in formulating the best \IS\  density is to determine territories for exclusion from the sampling because the sampling of points from the important region is then performed proportionally to the original density.
Therefore, we argue that it is worthwhile to obtain the precious description of the important region, i.e., the territory indicating the target event $\pazocal{T}$ via $\ITx$. As described in the preceding subsection, its point-wise approximation can be obtained as $\ITN$. Therefore, we propose starting with a~simple importance sampling run to obtain ``screening dots'' providing good geometrical information about the territory of event $\pazocal{T}$.

\paragraph{Screening via a~set of localized importance sampling runs}
Given the point-wise information from ED, we expect that the target territory is in the vicinity of points associated with $\pazocal{T}$. From all the \Ns\ points obtained so far, we select only those \nT points which were classified as $\pazocal{T}$-events.
In order to map their surroundings and describe the corresponding \vor\ cells, we propose the performance of such a~set of \NT localized \IS\ runs around each such ED point that the Gaussian sampling densities $h_i$ are always centered at one of the $\pazocal{T}$-classified points, and $\sigma_i = \sqrt{\Nv-1}$. These are the same kind of dots as were used as candidates in the \emph{extension} task, see Eq.~\eqref{Eq:dots}.
It suffices to throw roughly hundreds or thousands of screening dots around each of the $\nT$ points if one wishes to sample the territory fairly accurately.
Based on the nearest neighbor classification, the territory belonging to $\pazocal{T}$ will now be described by a~high number of retained screening dots and not only the \nT points. Even though the \vor\ cells are not constructed explicitly via computational geometry (such as the QuickHull algorithm), their descriptors, such as the locations of vertices or volumes, can be estimated very accurately \cite{masvor:ADES:VoronoiApproxErr:19}.

\paragraph{Cubature via a~set of \emph{local} \IS\ runs}
One can use the retained screening dots as cubature nodes and estimate the probability of event $\pazocal{T}$ directly in the fashion of \IS. Each such point has its likelihood ratio computed as the fraction $f_X(\x) / h_i(\x)$, where $h_i(\x)$ can be different for dots obtained from sampling around a~different $\pazocal{T}$ point.
This direct approach is referred to as ``localized \IS'' from here on and will be reported in the convergence diagrams; see the magenta points and line in Fig.~\ref{fig:NearestNeigh}. Such a~systematic cubature can be quite accurate (provided the most important failure regions are well covered by ED). The disadvantage is the need to compute the weight ratio for each integration node (the ratio between the original density and the local sampling density) and also the number of nodes that are not classified as $\pazocal{T}$-event tends to be high, which degrades the results.

\paragraph{Cubature via a~\emph{global} \IS\ via rotationally invariant density in an important ring}
Suppose the screening dots were not localized in a~very small region for which the localized importance sampling would be efficient. We propose the use of use a~large pool of ``integration nodes'' selected from sampling density $h_{\mathrm{ann}}(\x)$ which excludes the useless \Nv-ball of the radius $r$ ($=$ the distance of the most central ``screening dot'' from the origin; see the magenta points in Fig.~\ref{fig:NearestNeigh}). Additionally, we propose that the density also excludes the exterior of the \Nv-ball with the radius $R>r$. The outer radius can initially be calculated using Eq.~\eqref{eq:estim:R}, which excludes only a~negligible fraction of the contents of the exterior of the $\Nv$-ball of radius $r$; see also Fig.~\ref{fig:chidist} top.
Once a~previous estimation of the rare event probability exists, such as the preceding $\PT{\Ns-1}$, the outer radius should be based on this information, i.e., the excluded probability should form a~negligible portion of it, say
$ \PT{\Ns-1} / 10^4$. Using Eq.~\eqref{eq:radius}, the outer radius is obtained as
\begin{equation}
\label{eq:radiusR:pT}
    R (\PT{\Ns{\color{red}-1}}; \Nv)
    =
    F_{\rho}^{-1} \left(  1-\frac{\PT{\Ns{\color{red}-1}}}{10^4}; \Nv \right).
\end{equation}
Based on the information from the screening dots, we presume that the interior of the \Nv-ball of  radius $r$ does not contain any failure event.

All numerical examples in this paper are analyzed using this strategy. It is our experience that using \IS\ with density $h_{\mathrm{ann}}(\x)$, hereinafter called ``global \IS'', provides a~good balance between robustness and efficiency.
\rev{We remark that in a~hypothetical case, in which the rare event occurs exactly outside a~circle/ball/hyperball with radius $r$ in the standard Gaussian space,  the sampling density $h_{\mathrm{ann}}(x)$ becomes an \emph{optimal} \IS\ sampling density (with zero variance of the estimator). The reason is that $h_{\mathrm{ann}}(x)$ is proportional to the original standard Gaussian density and yet it removes completely the contribution from the safe region (\Nv-ball with radius $r$).}

Let us now consider the sampling density $h_{\mathrm{ann}} \left( \x; r , R \right)$ introduced in Eq.~\eqref{eq:sampl:dens:ann:x} and substitute it into Eq.~\eqref{eq:IS:estim}
\begin{align}
 \label{eq:ptau}
    \PT{\Ns}
    &\approx
     \frac{1}{n_{\IS}} \sum_{i=1}^{n_{\IS}}
     \ITNi
         \frac{  f_{\X}(\x_i)}
              { h_{\mathrm{ann}} (\x_i) }
    =
     \frac{p_{\mathrm{ann}} }
          {n_{\IS}}
          \underbrace{\sum_{i=1}^{n_{\IS}} \ITNi}_{n_{\IS,\pazocal{T}}}
    =
    p_{\mathrm{ann}}
    \frac{{n_{\IS,\pazocal{T}}} }
    {n_{\IS}}.
\end{align}
In other words, computation of the likelihood ratio ($f_{\X}/h$) at each node is \emph{not} needed as the Gaussian densities cancel out. This is because the samples generated via the simple procedure from Sec.~\ref{sec:annulus} have their  density proportional to $f_{\X}$.
It suffices to simply compute the proportion of \IS\ nodes that signaled event $\pazocal{T}$ (e.g., failure) and multiply it with  $p_{\mathrm{ann}}$. The closer the true event domain is to the annulus, the closer
    $ h_{\mathrm{ann}}$ is to the optimal
\IS\ density, and so the estimation variance vanishes.
With this strategy, the annulus between the two radii $r$ and $R$ is effectively examined.
Whatever event occurs outside the outer radius $R$ is associated with a~negligible probability.

The variance of such an \IS\ estimation is also simple to obtain. By substituting
$n_{\IS,\pazocal{T}}$ as the number of samples that lead to the event $\pazocal{T}$ and $p_{\mathrm{ann}} $ as the constant likelihood ratio into Eq.~\eqref{eq:IS:var:estim}, the estimation variance becomes
\begin{align}
\label{eq:IS:var:estim:C}
    \text{Var}_h[ \PT{\Ns} ]
    \approx
    \frac{1}{n_{\IS}}
    \Bigg\{
    \left[
        \frac{  n_{\IS,\pazocal{T}} }{  n_{\IS} }
        p^2_{\mathrm{ann}}
    \right]
    -
    \left(\PT{\Ns}\right)^2
    \Bigg\}
    =
    \frac{1}{n_{\IS}}
    \Bigg\{
        \left[
     \PT{\Ns}
     p_{\mathrm{ann}}
    \right]
    -
    \left(\PT{\Ns}\right)^2
    \Bigg\}
    =
    \frac{ \PT{\Ns} }
         { n_{\IS}  }
        \left(     p_{\mathrm{ann}}     -     \PT{\Ns}
        \right),
\end{align}
and therefore, by using Eq.~\eqref{eq:ptau}, the general coefficient of variation in Eq.~\eqref{eq:IS:cov:estim} simplifies to
\begin{align}
\label{eq:IS:cov:estim:C}
    \text{CoV}_h[ \PT{\Ns} ]
    &\approx
    \frac{1}{\sqrt{ n_{\IS} } }
    \sqrt{
    \frac{ p_{\mathrm{ann}}}{\PT{\Ns} } -1
    }
    =
     \frac{1}{\sqrt{ n_{\IS} } }
    \sqrt{
        \frac{n_{\IS}}
             {n_{\IS,\pazocal{T}}
        } -1
    }
\end{align}
There can be two reasons for this coefficient of variation of the estimator potentially being unacceptably high:
 (i)  the number $n_{\IS}$ of integration nodes is small (the square root in the first fraction), and
 (ii) the failure region is highly localized due to which which the important ring contains only a~small proportion of nodes associated with the rare event (the fraction in the second square root is much greater than one).
In the first case, the relevant measure is to increase $n_{\IS}$, while in the second case, it pays to switch to the set of local \IS\ runs around various $\pazocal{T}$ points (see above), or generally around clusters of rare-event clouds which can be localized using the ``k-means clustering'' algorithm.

Fig.~\ref{fig:NearestNeigh} presents all the important information about the process of \emph{extension} of ED (top left) and the \emph{estimation} of probabilities which can be performed at any time during the process. The evolution of both is shown in the  \href{https://vutbr-my.sharepoint.com/:v:/g/personal/vorechovsky_m_vutbr_cz/EaWDuW0xQYhAlhwqtnOww8kBRmGSLw4LhI-aejNjK-kT0w?e=zNdhaE}{\textcolor{blue}{\textsf{Wavy circle video}}} covering the history from the very first limit state function evaluation up to $\Ns=200$. The video frames, which can be displayed one by one, correspond to individual stages of the process, thus enabling a~detailed inspection of the process.
The accuracy of the estimation is excellent already at $\Ns \approx 70$ when all of the seven ``failure regions'' become discovered. The blue line plotted in the bottom diagram shows the amount of the ``probability bite $\psi$'' occupied by the neighborhood of the selected candidate. It can be seen that once about $\Ns=30$ points have been evaluated, the $\psi$ contributions become smaller than the rare event probability itself. The decrease in $\psi$ with an increasing $\Ns$ can be used, along with the stabilization of probability estimations, to decide the profitability of further \gx\ function evaluations (stopping criterion).

The red line, such as the one visible in Fig.~\ref{fig:NearestNeigh}, is always accompanied in this paper by a~scatterband of $\pm$ one standard deviation obtained from Eq.~\eqref{eq:IS:cov:estim:C}. In particular, the two lines parallel the estimate ${\PT{\Ns} } $ are computed as ${\PT{\Ns} } \left( 1 \pm  \text{CoV}_h[ \PT{\Ns} ] \right)$. We remark that such a~scatterband accounts for the variance in the \IS\ estimator
for a~given classification surrogate only. Since we used large numbers of integration dots (many thousands),  the scatterband is very narrow (almost invisible).
  However, it must be stressed that Eq.~\eqref{eq:IS:var:estim:C} does \emph{not} account for the variability associated with the varying surrogate classifier \ITN, which evolves differently in various runs of the ED extension algorithm.
This means that another run of the proposed extension algorithm may result in different evolution of the rare event probability estimations. However, our experience is that in almost all tested cases, the differences in the evolution of the classifier are negligible and they quickly diminish as the number of points $\Ns$ increases. The only exception is the ``Modified Rastrigin'' counterexample (Sec.~\ref{sec:rastrigin}), in which the surprise discoveries of various closed parts of the failure domains are very much randomly variable.


\section{Global sensitivities and importance measures of individual variables
\label{sec:sens}}

\begin{figure}[!htb]
    \centering
    \includegraphics[width=\textwidth]{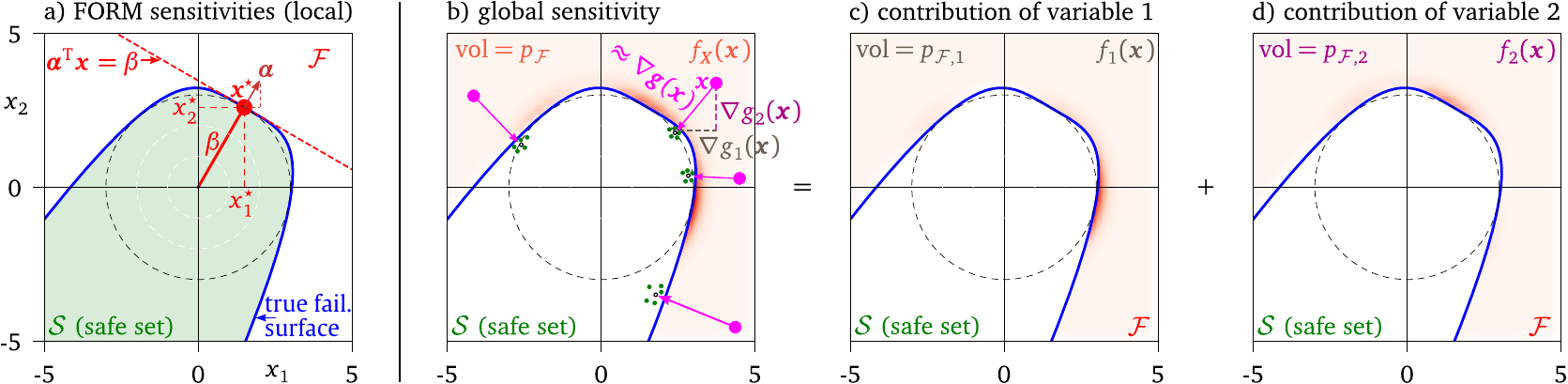}
    \caption{Measures of importance with respect to failure probability.
    a) illustration of FORM $\alpha$-sensitivities constructed using linearization of a~failure surface in the most central failure point $\xDP$.
    b)-c) illustration of the proposed global sensitivity measures $s_v^2$ by  decomposition of $\pF$ into additive contributions by individual variables.
    }
 \label{fig:global:sens}
\end{figure}

One of the by-products of the design point search for FORM/SORM is the simple evaluation of $\alpha$-sensitivities \cite{HohenRackw:Sens:86,Madsen1988}.
These values have a~clear geometrical meaning, and they have immediate application in the partial safety factor method in design codes.
Suppose we have a~linear performance function \gx{} with one distinct design point
    $\xDP = \arg\,\min \{ \lVert \x \rVert \; | \; \gx = 0 \}$.
The Euclidean distance of this most central failure point to the origin of the standard Gaussian space is the safety index
    $\beta = \lVert \xDP \rVert
           = \sqrt{ \xDP \tran \cdot \xDP }
           = \sqrt{ \sum_{v=1}^{\Nv} (x_v^{\star})^2 }
    $.
The distance can also be written using the vector of $\alpha$-sensitivities, $\balpha$, which has the unit size:
$
 \lVert \balpha \rVert
    = \sqrt{ \sum_{v=1}^{\Nv} \alpha_v^2 }
    = 1
$.
The linear and normalized approximation $M(\x)$ to the safety margin can thus be expressed as
$M(\x) = \balpha \tran (\xDP - \x) = \beta - \balpha \tran \x$. The variance of $M$ is $\sqrt{ \sum_{v=1}^{\Nv} \alpha_v^2 }
    = 1$ \cite{Madsen1988}.
Vector $\balpha$ is the negative of the gradient of the performance function at the origin of the standard normal space, i.e., $\balpha$ points in the \emph{important direction} (see Fig.~\ref{fig:global:sens}a), and the coordinates of the design point can be obtained as $\xDP = \beta \balpha$. The components of $\balpha$, i.e., the direction cosines of \balpha\ are called the $\alpha$-sensitivities (or $\alpha$-factors), and they are regarded as measures of the sensitivity of the reliability index to inaccuracies in the value of \x\ at the design point
\begin{align}
    \alpha_v
    =
    \left.
        \frac{\partial \beta}{\partial x_v}
    \right
    \rvert_{\xDP}
    .
\end{align}
The squared components of vector $\balpha$ can be obtained directly from the coordinates of the design point:
\begin{align}
    \label{eq:pf:definition:clas:alpha}
    \alpha_v^2
    =  \frac{(x_v^{\star})^2}
            {\beta^2}
    =  \frac{(x_v^{\star})^2}
             { \sum_{v=1}^{\Nv} (x_v^{\star})^2}
    ,
    \quad
    v = 1,\ldots, \Nv
    .
\end{align}

The relative importance measures by FORM/SORM are highly affected by the accuracy of FORM/SORM solutions, and they may be inadequate for component or system reliability problems featuring multiple design points or surfaces with high curvatures. Yet, good importance measures are needed to identify critical random variables concerning the reliability of a~product or a~process and enable focusing on critical uncertainties (disregarding unimportant ones) in the optimization of the product reliability.

In this paper, we introduce a~new generalized importance measure that considers all (known) points contributing to failure, weighted by the original density. 
\revA{This measure may use the the performance function's gradient when available, but can be defined also for discrete-state performance function for which the gradient is locally zero.}
The proposed measure comes as a~by-product of the above-proposed technique for reliability estimations.
Consider that the total unit contribution of \emph{any} point $\x=\{ x_1, \ldots, x_v,  \ldots, \x_{\Nv} \}$  \rev{to an event of interest} can be decomposed into individual coordinates in the spirit of the $\alpha$-sensitivities
\begin{align}
 \label{eq:alpha:share}
    \alpha_v^2(\x)
    =  \left(
            \frac{\color{red} \nabla g_v(\x)   }
                 {\color{red} \lVert \nabla \pmb{g}(\x) \lVert}
        \right)^2
    =   \frac{                  \color{red} g_v^2(\x) }
             { \sum_{v=1}^{\Nv} \color{red} g_v^2(\x) }
    ,
    \quad
    \text{i.e.}
    \;
    \sum_{v=1}^{\Nv} \alpha_v^2(\x) = 1
    \;\; \text{for any} \; \x
    ,
\end{align}
\revA{where $ g_v(\x)$ is the projection of the local gradient $\lVert \nabla  \pmb{g}(\x) \lVert$ 
onto direction $v$. This gradient is perpendicular to contour lines of the performance function $\gx$. In FORM approximation, the assumption of linearity of $\gx$ guarantees that the direction of the gradient is identical for all points in the design domain and can be taken as the radius vector connecting the origin with the design point.
When \gx\ is a~general smooth function, the gradient and its components can be computed by evaluating the partial derivatives (either analytically or numerically). For discrete-state performance functions treated in this paper, such gradient does not provide any meaningful information about the steepest descent and must be replaced by another information available from the previously evaluated points in the ED. A~reasonable way to replace the gradient in an existing (supposedly) failure point is the direction to the \emph{nearest safe point}. These directions are perpendicular to the failure surface, see the magenta arrows in Fig.~\ref{fig:global:sens}b which illustrate the local gradients at various points in the failure set \Fexact. Technically, these directions are not difficult to compute. Assume the extension algorithm has already discovered at least one failure point. The nearest-neighbor classification therefore divides a~number of integration nodes into safe and failure groups. In each failure integration node, one can now approximate the direction perpendicular to the failure surface by taking $K$ nearest neighbors from the group of safe nodes. The line connecting the failure node with the centroid of $K$ nearest safe nodes is a~good approximation of the ``gradient'' direction, see the empty circles Fig.~\ref{fig:global:sens}b representing such centroids. An efficient way to perform the computations is to employ K-d tree space-partitioning data structure for organizing points. This data structure enables classification used for fast identification of the $K$ nearest neighbors from the group of safe nodes. For \lstin{Python}, the functionality is readily available via \lstin{sklearn.neighbors.KNeighborsClassifier} in {scikit-learn} library \cite{scikit-learn}. 
The function allows for vectorized call which quickly finds the $K$ nearest neighbors for a~large set of nodes from the failure set at once.
}

The resulting \revA{squared projections} $\alpha_v^2(\x)$ can be viewed as importance measures of the individual dimensions and can be used as additive shares of probability density at any point $\x$. This choice is natural as
$ \alpha_v$s  weigh \revA{locally} the individual increments \revA{needed to change the system state}:
$
    \alpha_v(x)
    =
    {\color{red}
        \frac{\partial g(\x)}{\partial x_v}
    }
$.
Using these shares, the total failure of probability defined in Eq.~\eqref{eq:pf:definition1} can be rewritten as
\begin{align}
    \label{eq:pf:definition:sens}
    \pF
    \equiv
    \idotsint_{\Fexact}  \sum_{v=1}^{\Nv} \alpha_v^2(\x) f_{\X}(\x)
    \; \dd \x
    =
    \sum_{v=1}^{\Nv}
    \idotsint_{\Fexact}   \alpha_v^2(\x) f_{\X}(\x)
    \; \dd \x
    =
    \sum_{v=1}^{\Nv}
    \pFv
    .
\end{align}
In this way, the probability of failure (or analogously any event type) is obtained as a~sum of contributions $\pFv$ of individual variables.
Each variable contributes to \pF\ by
\begin{align}
    \pFv
    =
    \idotsint_{\Fexact}   \alpha_v^2(\x) f_{\X}(\x)
    \; \dd \x
    =
    \idotsint_{\Fexact}
    f_v(\x)
    \; \dd \x
    ,
\end{align}
where we define a~part of the standard Gaussian density ascribed to a~single variable $v$ as (see also the illustrations in Figs.~\ref{fig:global:sens}c and d)
\begin{align}
    f_v(\x) =  
    {\color{red} \alpha^2_v(\x)}
    f_{\X}(\x)
    .
\end{align}
This local contribution to $\pFv(\x)$ is easy to evaluate as it is dependent only on the \revA{local gradients at} point $\x$ and its standard Gaussian density.
Finally, the shares $\pFv$ can be standardized by $\pF$ to form the proposed global importance measures $s_v^2$ associated with individual variables
\begin{align}
    \label{eq:pf:s}
    s_v^2
    =
    \frac{\pFv}{\pF},
    \quad
    \text{i.e.}
    \;
    \sum_{v=1}^{\Nv} s_v^2(\x) = 1
    .
\end{align}
An important aspect is that the proposed $s_v^2$s are not based on values of the performance function, as only the binary information indicating an \emph{event} is needed. This is an important property because the importance measures should \emph{not} be dependent on the way a~performance function is defined if it provides the same failure boundary. A~robust importance measure for sensitivity to an event should be invariant under reformulations or reparametrizations of the underlying problem.

\begin{figure}[!tb]
    \centering
    \includegraphics[width=\textwidth]{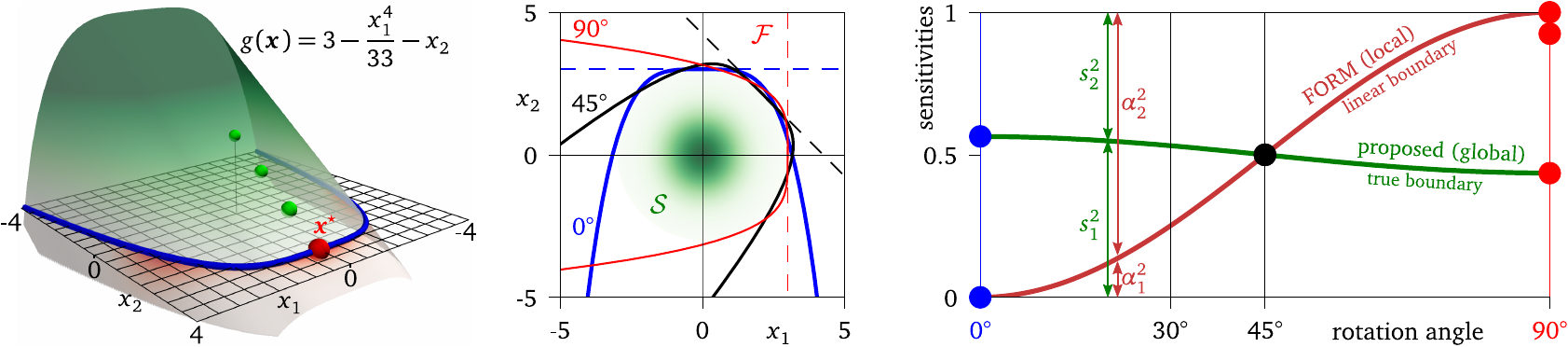}
    \caption{Left: the nonlinear performance function and the design point search.
            Middle: failure surfaces for various rotation angles $\gamma$.
            Right: Comparison of the classical $\alpha^2$-sensitivities with the proposed importance measures $s_v^2$ for various rotations of the problem by angle $\gamma$.
    }
 \label{fig:sensi-quadr}
\end{figure}

The numerical values of $s_v^2$ generally do not match with the classical factors $\alpha_v^2$ defined in Eq.~\eqref{eq:pf:definition:clas:alpha}. The reason is that the proposed measures $s_v^2$ weigh the contributions to \pF\ from the whole failure set \Fexact, while the classical $\alpha$-factors, in fact, consider only a~single design point. The dark red line in Fig.~\ref{fig:sensi-quadr} right presents the classical $\alpha$-sensitivities for a~linear failure surface in two dimensions in dependence on its rotation around the origin (see the dashed lines in Fig.~\ref{fig:sensi-quadr} middle). 
The meaning of the proposed measure is different from the $\alpha$-sensitivities. We argue that it provides meaningful results also for complicated failure domains (nonlinear, non-smooth, disconnected failure sets, etc.).

To show an example in which the results are quite different from $\alpha$-sensitivities, we consider the following nonlinear bivariate performance function:
 $\gx = \beta - x_1^4/c - x_2$,
where $\beta = 3$. In order to have a~unique most central failure point, the constant $c$ must be greater than 32; we take $c=33$.
Since \gx{} is linear in $x_2$ and constant in $x_1$ when $x_1=0$, the gradient point search initiated at the origin easily finds the unique design point $\xDP=\{0,3\}$; see the point trajectory in Fig.~\ref{fig:sensi-quadr} left. The classical $\alpha$-sensitivities are thus $\balpha = \{ 0, 1\}$. These are the sensitivities related to the unique $\beta$ point: the failure surface \FS\ is linear at $\xDP$, and the SORM correction fails because the curvature in the $\beta$-point vanishes.
However, it is evident that there are many highly probable regions with nonzero $x_1$ which considerably contribute to the failure probability.
In particular, there are two symmetrically distributed regions around points
  $x_1 = \pm {\color{red}2.84}$, $x_2=1.02$; see the red regions in Fig.~\ref{fig:sensi-quadr} left).
Indeed, the proposed global importance measures are \revA{not much different from each other}:
$s_1^2 \approx {\color{red}0.57}$ and $s_2^2 \approx {\color{red}0.43}$. The fact is that $s_1^2>s_2^2$ may be seen as contradicting the FORM sensitivity $\alpha^2_1 = 0$, i.e., the zero importance of variable $x_1$.

Consider now a~clockwise rotation of the coordinate system by an angle $\gamma$. The failure surface revolves as a~rigid body and moves the $\beta$-points along a~circle of radius $\beta=3$; see Fig.~\ref{fig:sensi-quadr} middle, and also the case of $\gamma = 30 \degree$, which is visualized in Fig.~\ref{fig:global:sens}.
The classical $\alpha$-sensitivities become strongly dependent on the angle: $\balpha = \{ \sin^2(\gamma) , \cos^2(\gamma)\}$; see Fig.~\ref{fig:sensi-quadr} right. The same figure documents that the proposed global importance measures are almost insensitive to the angle of rotation\revA{, because it considers all three failure regions}.

An extreme case documenting the fundamental differences between the local and the proposed global importance measure is the case of a~rotationally symmetric failure domain in standard Gaussian space (the exterior of the \Nv-ball)  has all importance measures identical: $\pFv = 1/\Nv$, which expresses the fact that no variable is more important than any other for the achievement of failure. The classical $\alpha$-sensitivities would, in this case, take arbitrary values in order to make their squared sum equal to one, i.e., any point on the failure surface \FS{} of the \Nv-ball can be selected.

The evaluation of individual $\pFv$ (and therefore also $s_v^2$) is very cheap as it can be seen as a~by-product of the sampling analysis employed to deliver an estimation of $\pF$.
Suppose we have an existing set of points that were sampled proportionally to the standard Gaussian density $f_{\X}(\x)$. It can be the global importance sample with \Ns{} points that was obtained outside the \Nv-ball or in the annuloidal $h_{\mathrm{ann}}(\x) \propto f_{\X}(\x)$.
From this sample, we only select a~vector $\x$ containing $\nF$ points marked as ``failure'' (or any other event type, based on the nearest neighbor surrogate). All these points are equally probable, and they each represent the same share of the estimated failure probability $\pF(\x_i) = \pF / \nF$, $i=1,\ldots,\nF$. This share can be further split into individual directions in the spirit of Eq.~\eqref{eq:alpha:share} \revA{by selecting the nearest safe point to the failure point $\x_i$, and computing the $v$ squared standardized projections of such a~pair at point $\x_i$: ${\color{red} \alpha^2_{v}(\x_i)}$. 
In this way,} the contribution of the $i$th point in the $v$th direction reads
\begin{align}
    \pFv (\x_i) =
    \pF(\x_i)
    {\color{red} \alpha^2_{v}(\x_i)}
    =
    \frac{ \pF}{\nF}
    {\color{red} \alpha^2_{v}(\x_i)}
    ,
    \quad
    \text{i.e., each point contributes }
    \;
    \pF(\x_i) = \sum_{v=1}^{\Nv} \pFv (\x_i) = \frac{ \pF}{\nF}
    \; \text{for any} \; \x_i ,
\end{align}
where $x_{i,v}$ is the $v$th coordinate of point $\x_i$.
The desired estimation of importance measures $s_v^2$ of individual directions can be obtained by summation over $\nF$ failing nodes with a~fixed direction index $v$ and dividing by the failure probability
\begin{align}
    \label{eq:estims:sv}
    s_{\pazocal{F},v}^2
    \approx
    \sum_{i=1}^{\nF} \frac{\pFv (\x_i)}{\pF}
    =
    \frac{ 1}{\nF}
    \sum_{i=1}^{\nF} 
    {\color{red} \alpha^2_{v}(\x_i)}.
\end{align}
Sensitivity to any other event type $\pazocal{T}$ is obtained analogously by retaining only $\nT$ points corresponding to that event and computing the average:
$   s_{\pazocal{T},v}^2
    \approx
    \sum_{i=1}^{\nT} 
        \left[ 
            {\color{red} \alpha^2_{v}(\x_i)}
        \right] / {\nT}
$. Therefore, the proposed sensitivity measure is just a~cheap by-product of the proposed method.
In cases when the sampling probability is not proportional to the standard Gaussian density, it is no longer true that all points have the same contribution of $1/{\nT}$, and therefore straightforward re-scaling in a~similar manner to importance sampling must be employed
\begin{align}
    s_{\pazocal{T},v}^2
    \approx
     \frac{1}{p_\pazocal{T}}
     \frac{1}{n_{\IS}}
    \sum_{i=1}^{n_{\IS}}
    \ITNi{}
         \frac{ f_{\X}  (\x_i) }
              { h       (\x_i) }
              {\color{red} \alpha^2_{v}(\x_i)}
\end{align}

\section{Numerical examples}
\label{sec:numex}
\begin{figure}[!tb]
    \centering
    \includegraphics[width=18cm]{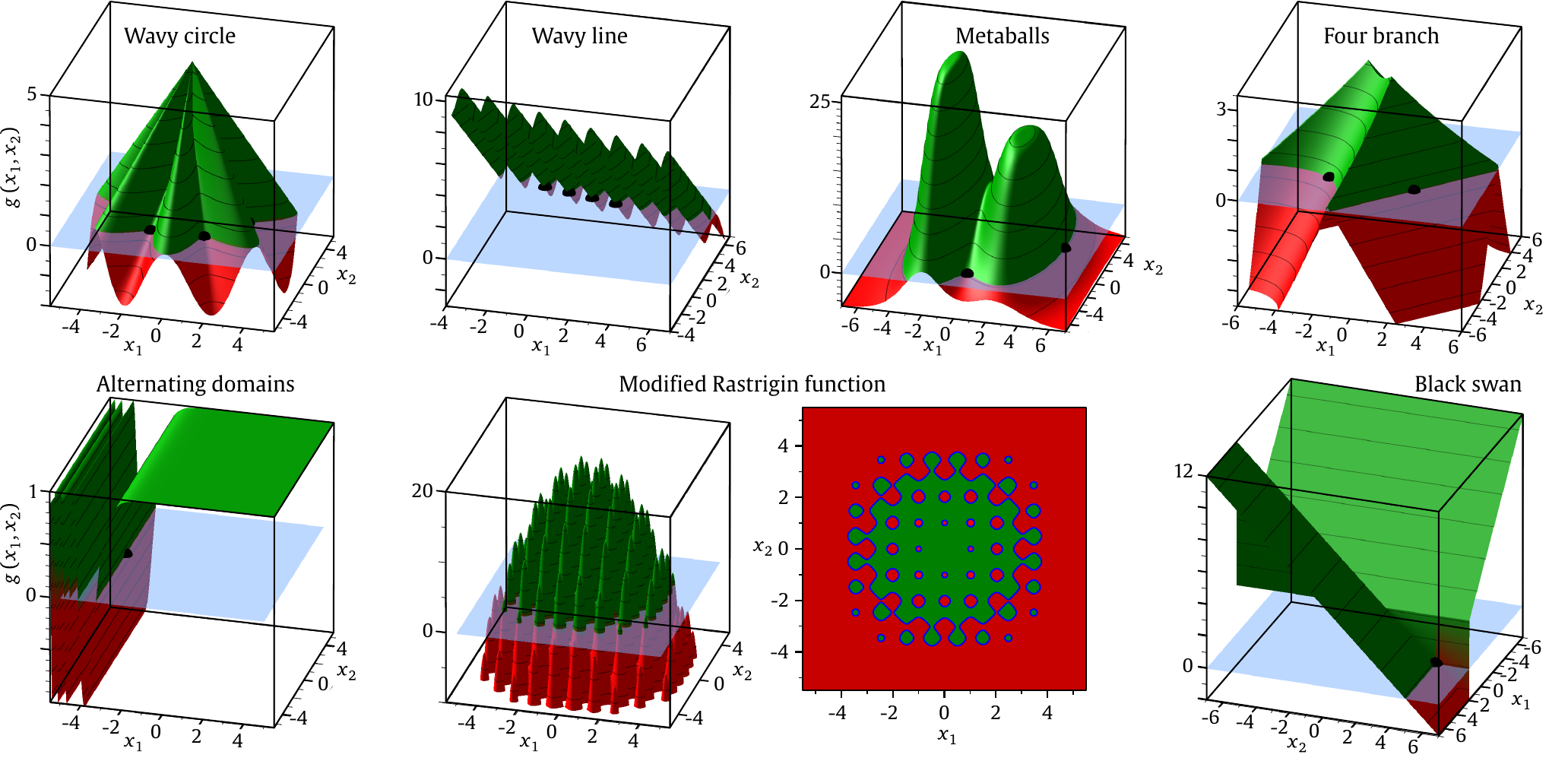}
    \caption{Overview of seven two-dimensional examples used for demonstration of the proposed method. Design points are visualized using solid black balls.}
 \label{fig:Overview}
\end{figure}

We present a~variety of numerical examples which have been selected to explore different classes of problems posing unique challenges. In the first seven examples, two-dimensional problems are defined in the space of independent standard Gaussian random variables; see Fig.~\ref{fig:Overview} for a~quick overview of the selected functions. \rev{After that, an~engineering example with Gaussian inputs shows the applicability of the proposed method to real nonlinear computational mechanics problem. The next example remains bivariate, however, it documents the straightforward applicability of the approach to problems with correlated non-Gaussian variables.}
Then, problems in higher dimensions are analyzed.
In all definitions of the functions, we present the expressions that return smoothly \rev{or non-smoothly} varying output variables. However, the proposed extension algorithm receives categorical information only (such as binary ``failure-success'' codes). The same holds for the estimation, which uses only the indicator functions signaling an event. The only exceptions are ``Four Branch'' and ``Metaballs'' examples for which we also examine the degree of improvement in probability estimation when a~smooth interpolation of the point-wise information in the ED via the Radial Basis Function is employed as the classifier.

\subsection{Wavy circle}

The failure surface used to demonstrate the \emph{extension} and \emph{estimation} steps in Fig.~\ref{fig:NearestNeigh} is a~sine wave of amplitude $a=1$ superposed onto a~circle of average radius $r_{\mathrm{ave}}=4$. The corresponding limit state function can be formulated as, e.g.,
\begin{align}
    \label{eq:SinWave}
     g\left( x_1, x_2 \right)
    &=
     r_{\mathrm{ave}} + a \sin(k \varphi_{\x} )   - \lVert \x \rVert ,
\end{align}
where $ \lVert \x \rVert = \sqrt{ x_1^2  + x_2^2 } $ is the Euclidean distance from the origin,
      $  \varphi_{\x} = \atantwo(x_2,x_1) = \arctan (x_2/x_1)$ is the angle of vector $\x$ from axis $x_1$, and
      $k=7$ is the number of design points. The sum of the first two terms represents the maximum safe distance from the origin, and the rare event (failure) occurs ``behind'' the wavy boundary, i.e., when $g \leq 0$. The presented algorithm uses only binary information (failure or success).
The seven most central failures are located at an identical distance of $ r_{\mathrm{ave}} - a = 3$.
Fig.~\ref{fig:NearestNeigh} presents the situation after $\Ns = 100$ evaluations of the limit state function, and the full history starting with the initial evaluation at the origin is included in the
\href{https://vutbr-my.sharepoint.com/:v:/g/personal/vorechovsky_m_vutbr_cz/EaWDuW0xQYhAlhwqtnOww8kBRmGSLw4LhI-aejNjK-kT0w?e=zNdhaE}{\textcolor{blue}{\textsf{Wavy circle video}}}.
The video captures the progressive (i) discovery of all the design points and (ii) the refinement of the failure surface, both alternating based on ranking via the $\psi$ criterion to occur proportionally to the Gaussian density $f_{\X}$. The estimation of failure probability converges towards the exact failure probability result $\pF \approx 2.582 \cdot 10^{-3}$.  

The \emph{global} \IS\ performed on the nearest-neighbor surrogate provides stable estimations in the ring between $r = 3$ and $R \approx 5.2$, see Fig.~\ref{fig:NearestNeigh}.
The proposed global sensitivities estimated via Eq.~\eqref{eq:pf:s} are all identical
    $ s_{\pazocal{F},1}^2  = s_{\pazocal{F},2}^2  =  0.5$, which corresponds to the distribution of seven failure regions, while the local FORM-style values of $\balpha$ sensitivities are very different and depend on which design point is considered (e.g., $\balpha = \{ 0, 1\}$ for the point  $\xDP = \{ 0, 3\}$).

We used \textsf{OpenTURNS} software \cite{Baudin2016} to run various techniques for failure probability estimation.
The gradient-based design point search for FORM analysis initiated at a~random location needs, after some help, 33 model evaluations to discover one of the design points, and thus FORM approximates the probability incorrectly as $\Phi(-3) \approx 1.35 \cdot 10^{-3}$. The SORM analysis increases the number of limit state function calls because it computes the failure surface curvature about the design point and decreases the estimation by about $ 3.2 \cdot 10^{-4}$ (the values slightly differ depending on the method used: Tvedt, Hohenbichler, and Breitung).


The results obtained with SuS implemented in \textsf{OpenTURNS} very much depend on the computational budget. If the number of limit state function evaluations reserved for each probability level is sufficiently high (a few thousand), the stochastic gradient optimization is able to correctly locate all seven failure regions and estimate the failure probability accurately. When, however, the total number of function calls drops below roughly one thousand,  the estimate becomes incorrect. Similar statements are true about results obtained from the Adaptive Directional Stratification Algorithm in \textsf{OpenTURNS}: the number of function evaluations must be in the thousands for the method to provide sufficiently good results. For this function, the proposed technique provides better efficiency than the above method\revA{s} because it provides more accurate estimations with considerably fewer function calls.

\begin{figure}[!tb]
    \centering
    \includegraphics[width=17cm]{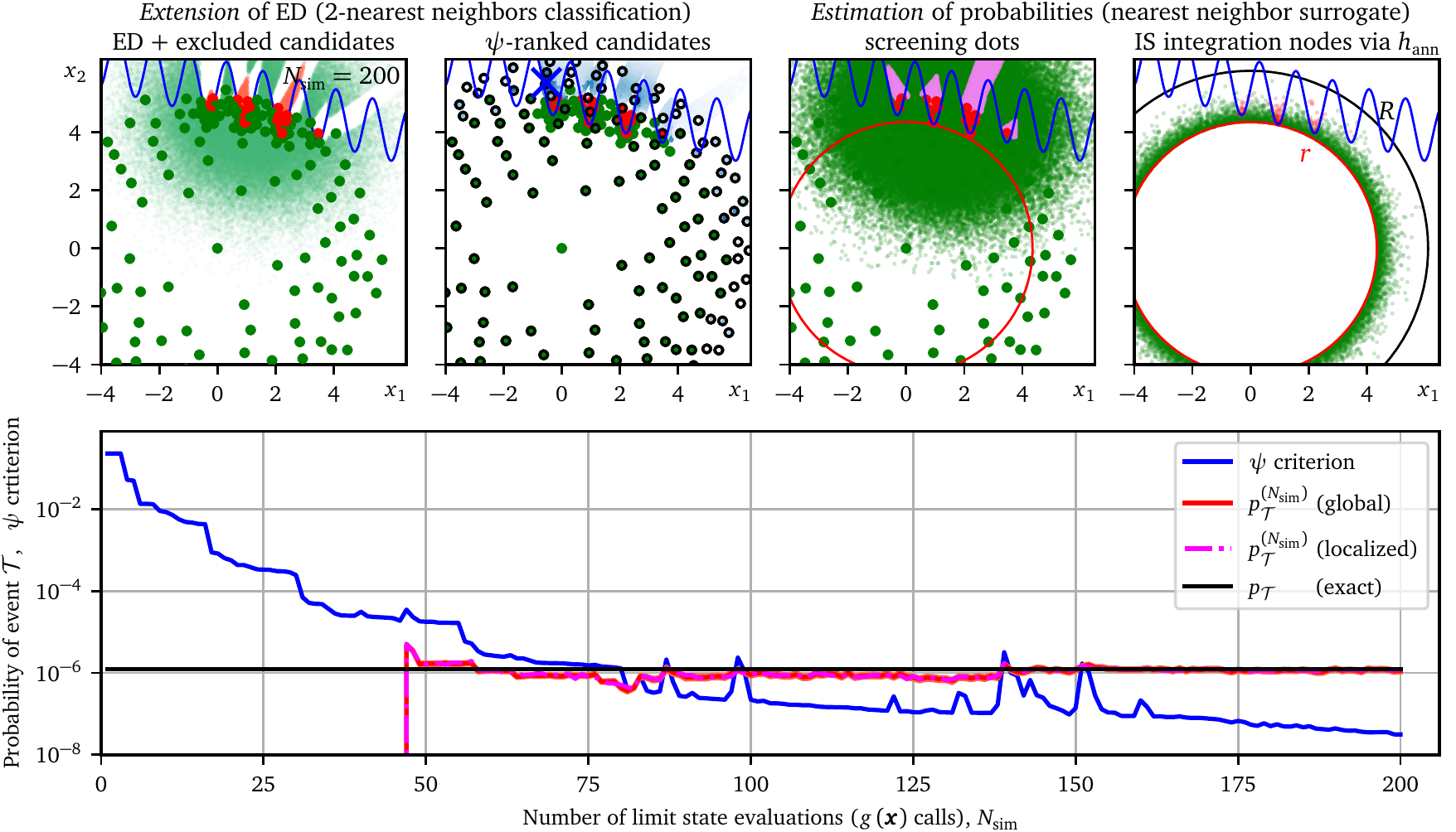}
    \caption{The ``Wavy line'' problem in Eq.~\eqref{eq:LinSin}.
      The complete evolution of all panels is available as \href{https://vutbr-my.sharepoint.com/:v:/g/personal/vorechovsky_m_vutbr_cz/ERcZ-RGf9Y5ArClnBENCqO0BXYTvF53H6FY4aieMspm4FQ?e=fgTk1E}
        {\textcolor{blue}{\textsf{Wavy line video}}}.
        }
 \label{fig:LinSin}
\end{figure}

The picture changes completely when the methods are supplied with binary information about \gx\ only. Design point search, SuS, directional sampling, etc., break down completely, and no efficiency/accuracy comparison is possible with the proposed technique. The same is true also for various tail approximation methods, which are based on fitting various distribution functions to the supposedly smooth output variable. This is true for all functions used in the remainder of this paper, and therefore, a~comparison of the presented algorithm with SuS or directional sampling will not be presented \revA{in most of the examples}.

The only method class from the one\revA{s} available from the wide selection in \textsf{OpenTURNS} is the class of pure sampling methods, such as crude Monte Carlo, or \IS\ around the mean values. These methods are able to operate given the dichotomous nature of information: they count the number of failed samples. Of course, the efficiency is much worse compared to both proposed \IS\ techniques (global and local), which involve sampling such that the fraction of failure samples of all function evaluations is much higher. For example, a~single run with $\Ns=1\,000$ of the standard \IS\ around mean values, and with a~Gaussian sampling density whole standard deviations are equal to three, the $\pF$ estimation is around 0.003, and the cov of the estimator is 0.15, which is quite high.


\subsection{Wavy line}

The previous example revealed that the algorithm distributes attention evenly among all seven failure regions, which share the same distance from the origin.
 Another test featuring multiple design points documents the ability to treat multiple ``design points'' with different contributions to the failure probability.
Inspired by \citet{Sundar2016}, who presented a~similar problem to illustrate their construction of a~surrogate model based on an artificial neural network, we define the performance of the system as
\begin{align}
 \label{eq:LinSin}
     g\left( x_1, x_2 \right)
    &=
    -\frac{1}{4} x_1  - x_2 + \sin (5 x_1) + 5.5,
\end{align}
where the constants were selected such that the failure surface is a~sine function (with its frequency controlled by the multiplier 5) superposed onto linearly decreasing function (with a~slope of $-1/4$), which is shifted in the direction of $x_2$ by a~constant 5.5.
The example is a~low-dimensional problem with a~very strongly nonlinear limit state possessing multiple effectively disjoint failure domains, see Fig.~\ref{fig:LinSin}.

FORM analysis is based on the single $\beta$-point $\xDP = \{0.943626,4.26411\}$, whose distance from the origin is
$\beta = 4.3672719$. Therefore, the FORM solution estimates the failure probability as $\Phi (-\beta) = 6.29 \cdot 10^{-6}$. The proposed method quickly converges to the correct solution $\pF = 1.217 \cdot 10^{-6}$, see the \href{https://vutbr-my.sharepoint.com/:v:/g/personal/vorechovsky_m_vutbr_cz/ERcZ-RGf9Y5ArClnBENCqO0BXYTvF53H6FY4aieMspm4FQ?e=fgTk1E}
        {\textcolor{blue}{\textsf{wavy line video}}} associated with Fig.~\ref{fig:LinSin}.
        
The estimated global sensitivities \revA{highlight the role of variable $x_1$:}
    $ s_{\pazocal{F},1}^2  \approx   0.829$ and 
    $ s_{\pazocal{F},2}^2  \approx  0.171$. 
    \revA{This is counter-intuitive and very much different from the reasonable }
 FORM sensitivities based on $\xDP$ and thus highlighting the role of variable $x_2$:
$\balpha = \{ 0.0467,     
              0.9533 \}$.  
\revA{
The reason for this discrepancy is that the performance function is taken as binary and therefore the information about the linear gradient of \gx\ in Eq.~\eqref{eq:LinSin} is not available. The algorithm approximates the local gradient at any point as the direction to the nearest safe point, i.e., perpendicularly to the wavy failure surface. Most of these directions are close to horizontal (almost parallel with $x_1$) and due to this prevailing horizontal projection, the sensitivity to the first variable appears to be dominant.
}

\subsection{Metaball function -- complicated topology}
\label{sec:MetaBalls}

\begin{figure}[!bt]
    \centering
    \includegraphics[width=17cm]{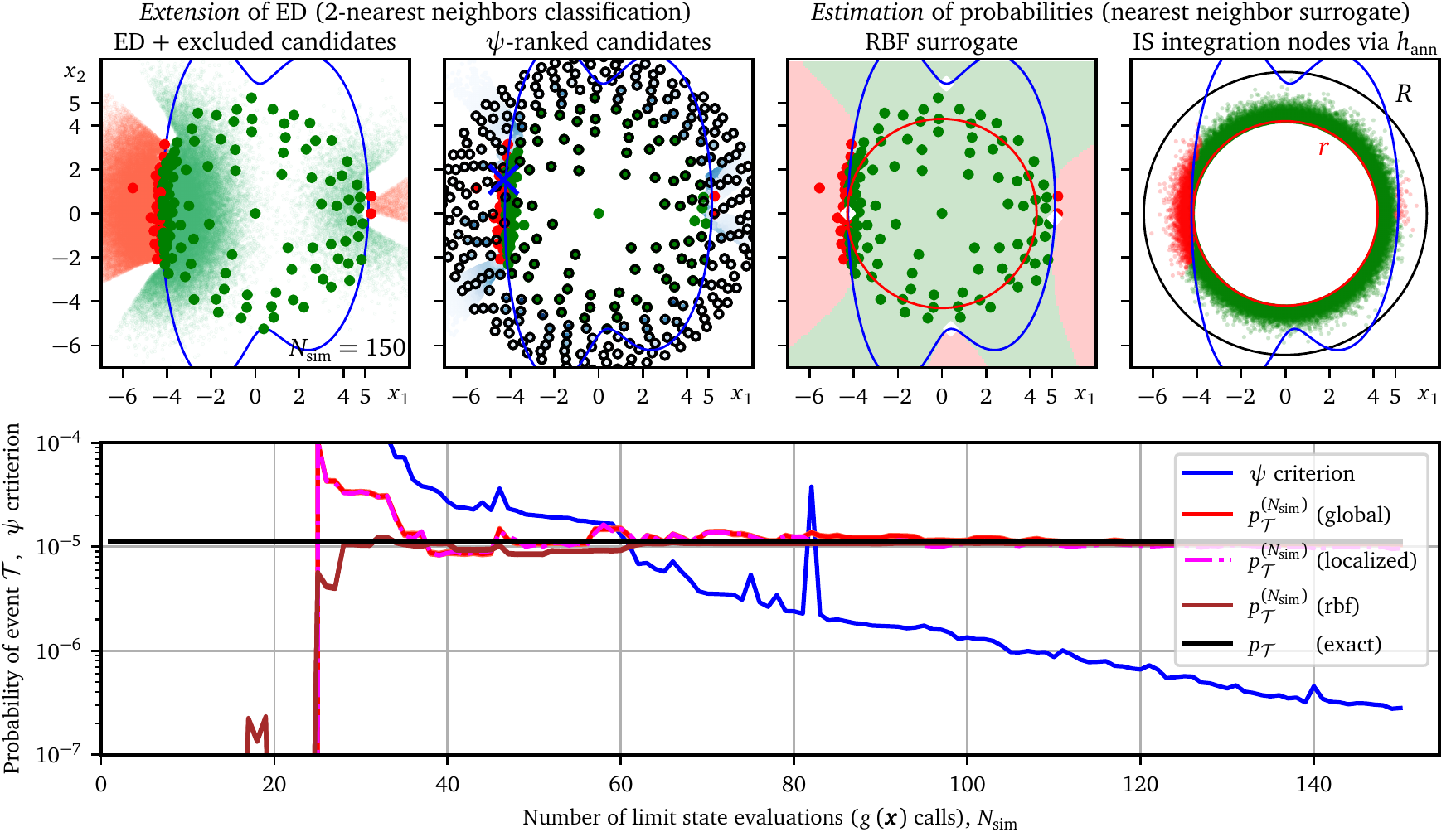}
    \caption{The ``MetaBalls'' problem in Eq.~\eqref{eq:MetaBall}.
      The complete evolution of all panels is available in the \href{https://vutbr-my.sharepoint.com/:v:/g/personal/vorechovsky_m_vutbr_cz/EcqNjLpEokZLjo1FmBqTXR8BYWl62ZkXpYrwYdUT0phb0w?e=kxraG3
        }
        {\textcolor{blue}{\textsf{MetaBalls video}}}.
        }
 \label{fig:MetaBall}
\end{figure}

\citet{Breitung:19:RESS}, inspired by \cite{wiki:Metaballs}, used the Metaball example to document the risk that SuS does not move its stochastic gradient optimization towards the region with the highest contribution to failure probability. The particular definition of the Metaball function in \cite{Breitung:19:RESS} reads
\begin{align}
\label{eq:MetaBall}
     g\left( x_1, x_2 \right)
    = \frac{30}{
        \left(
            \frac{4\left( x_1 + 2\right)^2 }{9} +
            \frac{x_2^2}{25}
        \right)^2 + 1
    }+
    \frac{20}{
        \left(
            \frac{\left( x_1 - 2.5\right)^2 }{4} +
            \frac{\left( x_2 - 0.5\right)^2 }{25}
        \right)^2 + 1
    } - 5
\end{align}
Failure occurs when $\gx<0$ and the associated failure probability reads $\pF \approx 1.12857 \cdot 10^{-5}$.
The proposed method quickly converges to this result once the first failure is hit, see Fig.~\ref{fig:MetaBall}. When the ED contains about 40 sequentially added points via the $\psi$ criterion, the estimation of failure probability using the crude nearest neighbor surrogate is sufficient. The ``probability bites'' approximated by the $\psi$ criterion become more than one order of magnitude less than $\pF$ when $\Ns>100$, see the blue line. Further extension of the ED leads to even better refinement of the true boundary and, therefore, also more accurate probability estimation; see Fig.~\ref{fig:MetaBall} and its evolution captured by the  \href{https://vutbr-my.sharepoint.com/:v:/g/personal/vorechovsky_m_vutbr_cz/EcqNjLpEokZLjo1FmBqTXR8BYWl62ZkXpYrwYdUT0phb0w?e=kxraG3
        }
        {\textcolor{blue}{\textsf{MetaBalls video}}}.

The estimation method presented in this paper uses the simple nearest neighbor classification surrogate.
However, other classifiers such as Support Vector Machine, Radial Basis Function, Gaussian process, Random Forest, Neural Networks, Naive Bayes, AdaBoost, Quadratic Discriminant Analysis, Polynomial Chaos, etc., can be employed to increase the accuracy of the estimation. We now examine how the estimation accuracy increases by employing
radial basis function (RBF) interpolation built using the true \gx\ values, not just the binary output. We have tested the simplest RBF with default parameters (class \lstin{Rbf} from the \lstin{scipy.interpolate} sub-package \cite{2020SciPy-NMeth}) to perform classification into
failure and safe domains based on the sign of the RBF surrogate. For each sample size, the RBF interpolator was set up based on the current ED. The global \IS\ estimation was then applied for each sample size to estimate \pF. It leads to a~stable almost exact estimate already for $\Ns \geq 60 $; see the brown line in Fig.~\ref{fig:MetaBall}. The third panel from the left in the top row of the figure and also in the associated \href{https://vutbr-my.sharepoint.com/:v:/g/personal/vorechovsky_m_vutbr_cz/EcqNjLpEokZLjo1FmBqTXR8BYWl62ZkXpYrwYdUT0phb0w?e=kxraG3
        }
        {\textcolor{blue}{\textsf{MetaBalls video}}} captures the evolution of classification based on smooth RBF interpolation.
The involvement of the smooth interpolator indeed slightly improves the accuracy and stability of the \IS\ estimate. However, in this particular example, the estimate based on the simple nearest neighbor surrogate provides almost equally good results, despite the fact that only the binary information about $\gx$ is being used.

We now discuss the role of the complicated contour of this particular function (see Fig.~\ref{fig:Overview}) for gradient-based optimization methods such as FORM and SuS.
There are four distinct ``design points'' for this function \revA{(see the white diamonds in the third panel from the left in Fig.~\ref{fig:MetaBall}}):
the most central failure point is
$\xDPi[1] \approx \{ -4.26376, -0.00067 \}$
with the distance from the origin $\beta_1  \approx 4.26376$,
and the other three read:
$\xDPi[2] \approx \{ 5.1265, -0.2256 \}$
with $\beta_2  \approx 5.131$, 
$\xDPi[3] \approx \{  0.399, -5.241 \}$
with $\beta_1  \approx 5.257$, 
and
$\xDPi[4] \approx \{  0.201, 5.895 \}$
with $\beta_4  \approx5.898$. 
The corresponding four FORM approximations are
$p_1 = \Phi (-\beta_1) = 1.005 \cdot 10^{-5}$,
$p_2 = 1.437  \cdot 10^{-7}$,
$p_3 = 7.336  \cdot 10^{-8}$, and
$p_4 = 1.838  \cdot 10^{-9}$,
which reveal that the first design point is the only decisive one, and its discovery and usage in FORM approximation leads to very accurate probability estimate (12\% error only).

However, the most central failure point
$\xDPi[1]$ is hard to discover because the path to it from around the origin leads over a~``high peak''. Therefore, the standard gradient-based search algorithms do not discover it, and instead slide towards $\xDPi[3]$. They then estimate the \pF\ three orders of magnitude wrong as $p_3$.
As argued by \citet{Breitung:19:RESS}, SuS, which is in fact a~stochastic gradient method, also struggles to move the point cloud towards $\xDPi[1]$. Indeed, we confirm that using excessively small sample sizes in individual levels of the SuS algorithm implemented in \textsf{OpenTURNS} \cite{Baudin2016} makes the point cloud descend from the saddle point corresponding to the coordinate origin down the depression towards $\xDPi[3]$. The probability estimate is wrong because the SuS algorithm was unable to navigate the cloud towards $\xDPi[1]$ when the number of \gx\ evaluations dropped below several thousand. This behavior was also visually reported by \citet{Breitung:19:RESS}, who used it to show that decisive regions, located relatively close to the origin, i.e., with high failure probability content, can be overshadowed by other, less important regions in SuS. The reason is that for small sample sizes, the sequences of point clouds orient the downhill search according to basically local information only.

The Directional Sampling implemented in \textsf{OpenTURNS} was also unable to provide a~correct result with a~small sample size. The method provides good and stable results, but the number of limit state evaluations must be in the thousands.

\revA{The} proposed global sensitivities \revA{equal}
    $ s_{\pazocal{F},1}^2 =  0.994$ and 
    $ s_{\pazocal{F},2}^2 =  0.006$ 
    \revA{underlying the fact that all the ``gradients'' in high density regions of \Fexact\ (i.e. in the vicinity of points $\xDPi[1]$ and $\xDPi[2]$) are almost perfectly aligned with $x_1$.}
    \revA{The FORM sensitivities based on $\xDPi[3]$ are $\balpha \approx \{ 0.0058, 0.9942 \}$ incorrectly putting significance to variable $x_2$. This is a~consequence of not discovering the much more important point $\xDPi[1]$ for which the FORM-style sensitivities would be $\balpha \approx \{ 1, 0 \}$.}

\subsection{Four Branches problem}

\begin{figure}[!tb]
    \centering
    \includegraphics[width=17cm]{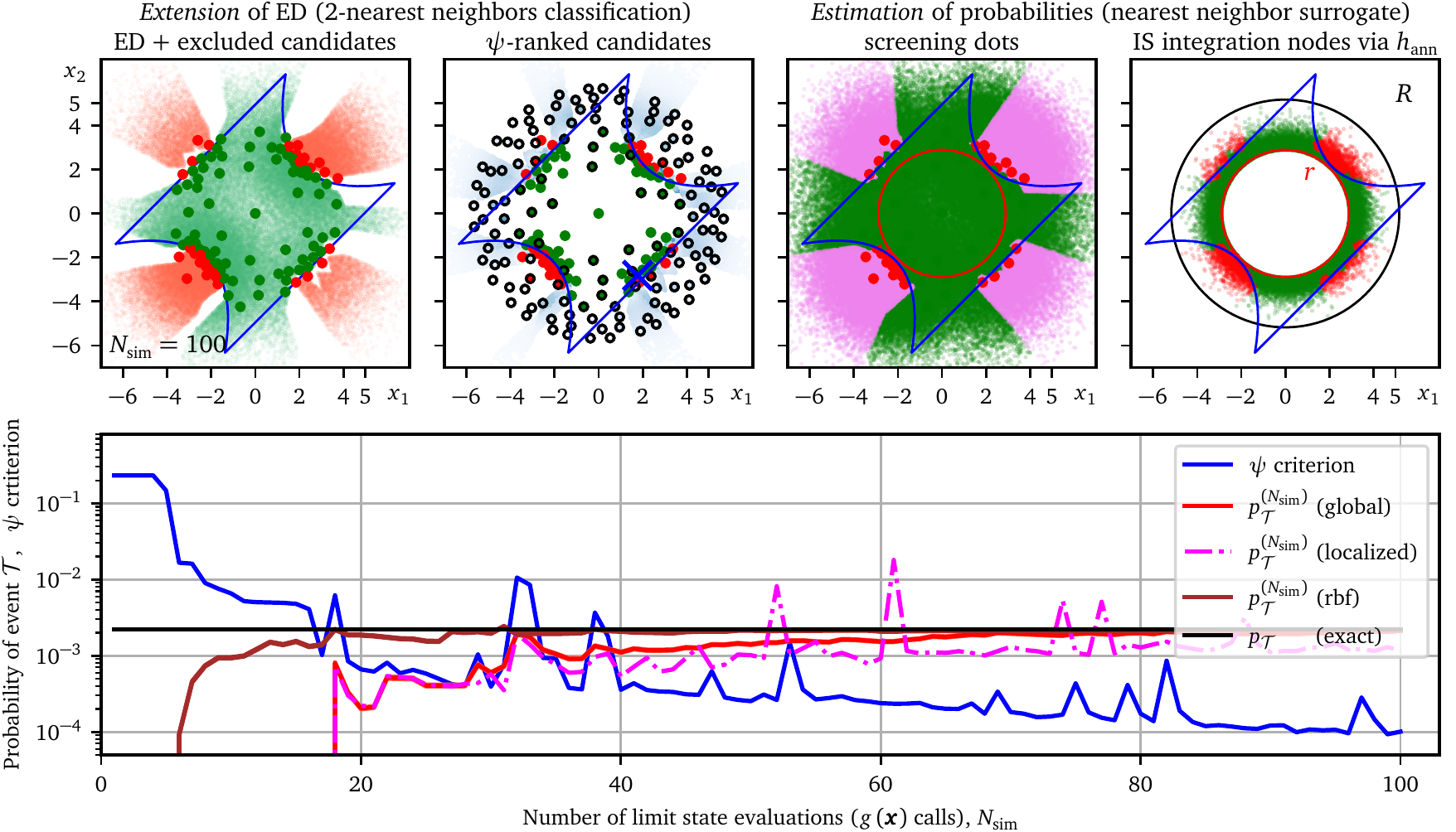}
    \caption{The ``Four Branches'' problem in Eq.~\eqref{eq:FourBranch}.
      The complete evolution of all panels is shown in the  \href{https://vutbr-my.sharepoint.com/:v:/g/personal/vorechovsky_m_vutbr_cz/EWynt1wUgFRGrBMdIM9ypzkBZeTIlkDm9L96AGrFdJvVNQ?e=k6Mj8c}
        {\textcolor{blue}{\textsf{Four Branches video}}}.
    }
 \label{fig:FourBranch}
\end{figure}

The following 2-dimensional ``Four Branch function'' \cite{Borri1997} is a~common benchmark problem in reliability analyses; see, e.g., studies with various parameter settings \cite{Schueremans2005,Echard2011,Papaio:Papa:Straub:SeqIS:SS:16,Schbi2017}. The function describes the failure of a~\emph{series system} with four distinct limit state components: two linear and two nonlinear branches of the failure surface. The limit state function reads
\begin{align}
 \label{eq:FourBranch}
    g \left( x_1 , x_2 \right)
    &= \min
     \left\{\begin{array}{ll}
        3 + 0.1 \left( x_1 - x_2 \right)^2  - {\left(x_1+x_2\right)}/{\sqrt{2}} \\
        3 + 0.1 \left( x_1 - x_2 \right)^2  + {\left(x_1+x_2\right)}/{\sqrt{2}} \\
        x_1 - x_2 + 7/\sqrt{2}\\
        x_2 - x_1 + 7/\sqrt{2}
        \end{array}\right.
        .
\end{align}
The failure event is defined as $\gx \leq 0$. Various authors have used various parameters replacing the number $7$ (originally 3.5, sometimes 6, or 11). In our definition, there are two pairs of design points:
    two points when $x_1= x_2 = \pm 3\sqrt{2}/2 $ at a~distance $\beta_{1,2} = 3$ and another two when $x_1=-x_2 = \pm 7\sqrt{2}/4 $ at a~distance $\beta_{1,2} = 3.5$.
The exact result failure probability is $\pF \approx 2.222\cdot 10^{-3}$,
and the proposed global sensitivities \revA{are identical}
    ($ s_{\pazocal{F},1}^2  = s_{\pazocal{F},2}^2  =  0.5$) \revA{due to symmetry reasons}.

Fig.~\ref{fig:FourBranch} shows the four blue lines forming the failure surface. The associated
    \href{https://vutbr-my.sharepoint.com/:v:/g/personal/vorechovsky_m_vutbr_cz/EWynt1wUgFRGrBMdIM9ypzkBZeTIlkDm9L96AGrFdJvVNQ?e=k6Mj8c}
        {\textcolor{blue}{\textsf{Four Branches video}}} demonstrates the behavior of the proposed method by adding points one  by one.
The extension algorithm refines the boundary proportionally to the probability density featured in the $\psi$ criterion. The consequence of this is that the classification close to the four remote intersections of failure surfaces is not performed correctly for small designs. While this is no problem for the global \IS\ probability estimation, which focuses on the high-density regions, it is a~source of erroneous estimation for the local \IS. As can be seen, the magenta dash-dot line in Fig.~\ref{fig:FourBranch} represents wrong and unstable results, which are degraded due to the accentuation of the inaccurately classified regions (heavily covered by the local \IS\ integration nodes).
The global \IS\ applied to the same binary surrogate classifier provides accurate and stable results for EDs as small as $\Ns \approx 80$. The top right panel in Fig.~\ref{fig:FourBranch} shows that the corner regions are not important for the global \IS.

We have examined the simplest Gaussian RBF-based classifier in the same way as we did in the preceding example. Using the global \IS\ estimation leads to stable, almost exact results which are already in the range of $\Ns \in (20,80)$, see the brown line in Fig.~\ref{fig:FourBranch}. The improvement is due to the more accurate classification of the failure surface in the vicinity of the four design points; the ``corners'' were not classified correctly neither by RBF nor by the binary surrogate.
The efficiency of the RBF classifier is as good as the efficiency of the best methods known in the literature.
Tab.~\ref{tab:comparison} compares many advanced techniques applied to the Four Branches example, along with the numbers of limit state evaluations (\Ns) and the rare event probability estimates.
Many sampling techniques, such as the sequential importance sampling employed in \cite{Papaio:Papa:Straub:SeqIS:SS:16},
use too many limit state function calls, and good results are obtained only when building a~surrogate model.
All the presented techniques, however, use the numerical values for the \gx. We repeat again that the proposed algorithm achieves almost the same efficiency; however, it does so using only the categorical information about \gx, which makes the proposed method very robust.

\begin{table}
\caption{Summary of results for the Four Branches problem obtained with other methods from the literature.}
\centering
\label{tab:comparison}
\begin{tabular}{llrll}
  \toprule
  Year & Method & $\Ns$ & $\PF{\Ns} \,(\cdot 10^3)$ & $\mathrm{cov} \left( \PF{\Ns} \right) \left[ \% \right]$ \\
  \midrule
2000 & DS~\cite{Waarts2000:fourbranch} & 227 & $2.2556$ & 37\\

2011 & AK-MCS+U~\cite{Echard2011} & 96 & $2.233$ & -\\
2011 & AK-MCS+EFF~\cite{Echard2011} & 101 & $2.232$ & -\\
2011 & ${}^2$SMART~\cite{BOURINET2011343:fourbranch} & 1035 & $2.21$ & 1.7\\


2013 & CE-AIS-GM~\cite{KURTZ201335:fourbranch} & 3~943 & $2.15$ & 3\\
2014 & MetaAK-IS$^2$~\cite{CADINI2014109:fourbranch} & 48+90 & $2.22$ & 1.7\\
2016 & AK–SS~\cite{HUANG201686:fourbranch} & 45 & $2.233$ & 4.94\\
2017 & KRA~\cite{XUE20171:fourbranch} & 116 & $2.22$ & 4.7\\
2017 & ASVM-MCS~\cite{Pan2017} & 89 & $2.13$ & 2.2\\

2018 & iRS~\cite{GUIMARAES201812:fourbranch} & 33 & $2.24$ & -\\
2019 & AKEE-SS~\cite{ZHANG201990:fourbranch} & 41.7 & $2.20$ & 3.07\\


2020 & BSC+RLCB~\cite{yi2020efficient:fourbranch} & 37 & $2.240$ & 4.89\\

2020 & DRL~\cite{XIANG2020106901:fourbranch} & 2597 & $2.314$ & -\\


2021 & ABSVR1~\cite{WANG2021114172:fourbranch} & 30 & $2.214$ & -\\
2021 & ABSVR2~\cite{WANG2021114172:fourbranch} & 43 & $2.222$ & -\\

2021 & SuS+K~\cite{RePEc:fourbranch} & 24 & $2.234$ & 2.05\\

2021 & RVM~\cite{LI2021381:fourbranch} & 73 & $2.218$ & 2.14\\
2021 & SVM~\cite{LEE2021107481:fourbranch} & 92 & $2.13$ & 10\\
2022 & RVM-MIS~\cite{WANG2022108287:fourbranch} & 161 & $2.252$ & 2.759\\
2022 & CE-DIS~\cite{ZHANG2022108306:fourbranch} & 275 & $2.20$ & 4.79\\
2022 & APCK-PDEM~\cite{ZHOU2022108283:fourbranch} & 35 & $2.226$ & 0.32\\

\bottomrule
\end{tabular}
\end{table}

\subsection{Black Swan}

Consider now a~highly localized failure region characterized by the \emph{simultaneous} violation of two thresholds: failure $\pazocal{F}$ occurs when $x_1>2$ and $x_2>5$. This failure region emerges for the following simple function studied in chapter 5 of a~book \cite{au_engineering_2014}, where the authors discuss the difficulties of SuS reaching the failure domain:
\begin{align}
 \label{eq:blackswan}
     g\left( x_1, x_2 \right)
    &=
     \left\{\begin{array}{ll}
        5 - x_1, & \text{for } x_1 \leq 2 \\
        5 - x_2, & \text{for } x_1 > 2
        \end{array}\right.
\end{align}
With this definition of the limit state function, the failure region described above corresponds to $ \gx <  0 $.
    The exact solution is the product of the probabilities obtained from two simple FORM-like  solutions:
$\pF = \Phi(-5) \Phi(-2) \approx 6.52136 \cdot 10^{-9}$.
In this example, we show that it can be very hard to hit the ``black swan'' event. The exploration set must have at least one point located in the rare event domain (failure region); without it, the discovery of the event is not possible. However, once a~failure event is localized, the proposed extension algorithm is very effective in the refinement of the small part of the failure surface which is associated with a~high probability.

\begin{figure}[!htb]
    \centering
    \includegraphics[width=17cm]{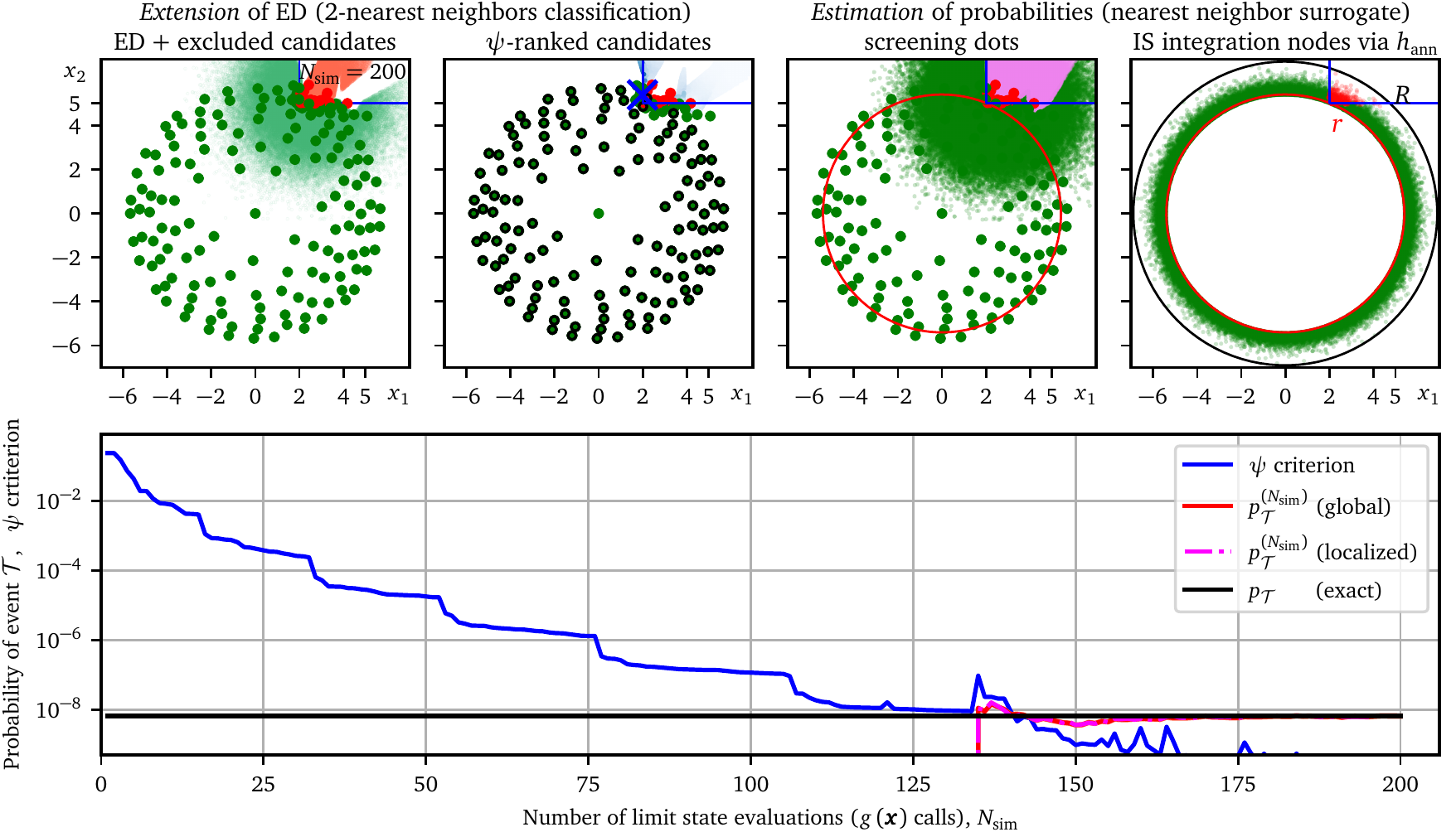}
    \caption{The ``Black Swan'' problem in Eq.~\eqref{eq:blackswan}.
      The complete evolution of all panels is shown in the  \href{https://vutbr-my.sharepoint.com/:v:/g/personal/vorechovsky_m_vutbr_cz/EZBQLNIqHopBkbIZXGOqTz0BJkr-EZ_ik5qhVH5oa4vkNg?e=LxssLe}
        {\textcolor{blue}{\textsf{Black Swan video}}}.
        }
 \label{fig:BlackSwan}
\end{figure}

We generally recommend generating many layers of the predefined exploration set, covering remote territories in the input space. Unnecessarily distant points will not be selected while refinement is in progress exploiting more important regions. However, without offering remote exploration points, the discovery of new disjoint failure regions would not be possible.

The estimated global sensitivities
    $ s_{\pazocal{F},1}^2  \approx  {\color{red} 0.3189}$ and 
    $ s_{\pazocal{F},2}^2  \approx  {\color{red} 0.6811}$
\revA{proposed in this paper are somewhat different from} the FORM-style values $\balpha$:
    $2^2/(2^2+5^2) \approx 0.1379$ and
    $5^2/(2^2+5^2) \approx 0.862$, 
\revA{which are based on the single} most central failure point $\xDP = \{2,5\}$.
\revA{The proposed sensitivities are based on the whole failure region in which the directions to the nearest failure surface approximating the missing gradient information are taken as either parallel or perpendicular to $x_1$. In this way, the failure region $\Fexact: \{x_1>2 \text{ and } x_2>5\}$ becomes split into two triangles separated by a straight line $x_2=x_1+3$. These two triangles contribute separately to $\pF=p_1+p_2$, where $p_1 = \int_{x_1=2}^{\infty} \left( \int_{x_2=x_1+3}^{\infty} f_{\x}(x_1,x_2) \dd x_2 \right) \dd x_1 \approx 2.0794 \cdot 10^{-9}$, leading to $s_{\pazocal{F},1}^2 = p_1/\pF$
and $s_{\pazocal{F},2}^2 = (\pF-p_1)/\pF$.
}

\subsection{Modified Rastrigin -- scattered small closed failure regions}
\label{sec:rastrigin}
As discussed already in Sec.~\ref{sec:discussion:assumptions}, the proposed algorithm may struggle when the rare event domain is formed by one or more small, closed regions, which are scattered over the design domain. A~2-dimensional modified Rastrigin function used by \citet{Echard2011} was selected to document it. It is a~modification of the Rastrigin function, which is a~standard benchmark for optimization algorithms \cite{Muhlenbein1991-xc}. This modification features regions with both positive and negative values of the function, which are interpreted as safe and failure events, respectively. It features a~highly non-linear limit state function with non-convex and disjoint failure domains. The function is a~rotationally symmetrical paraboloid with a~superposed wavy cosine function with a~large amplitude:
\begin{align}
 \label{eq:rastrigin}
     g\left( x_1, x_2 \right)
    &=
     10 - \sum_{v=1}^{2} \left(  x_v^2 - 5 \cos (2 \pi x_v) \right)
    .
\end{align}
The function is pictured in Fig.~\ref{fig:Overview} together with a~top view showing the twenty closed failure regions and one open failure region spreading to infinity.
Due to the bidirectional symmetry, the exact global sensitivities are both equal $ s_{\pazocal{F},1}^2  = s_{\pazocal{F},2}^2  =  0.5$.

\begin{figure}[!bt]
    \centering
    \includegraphics[width=17cm]{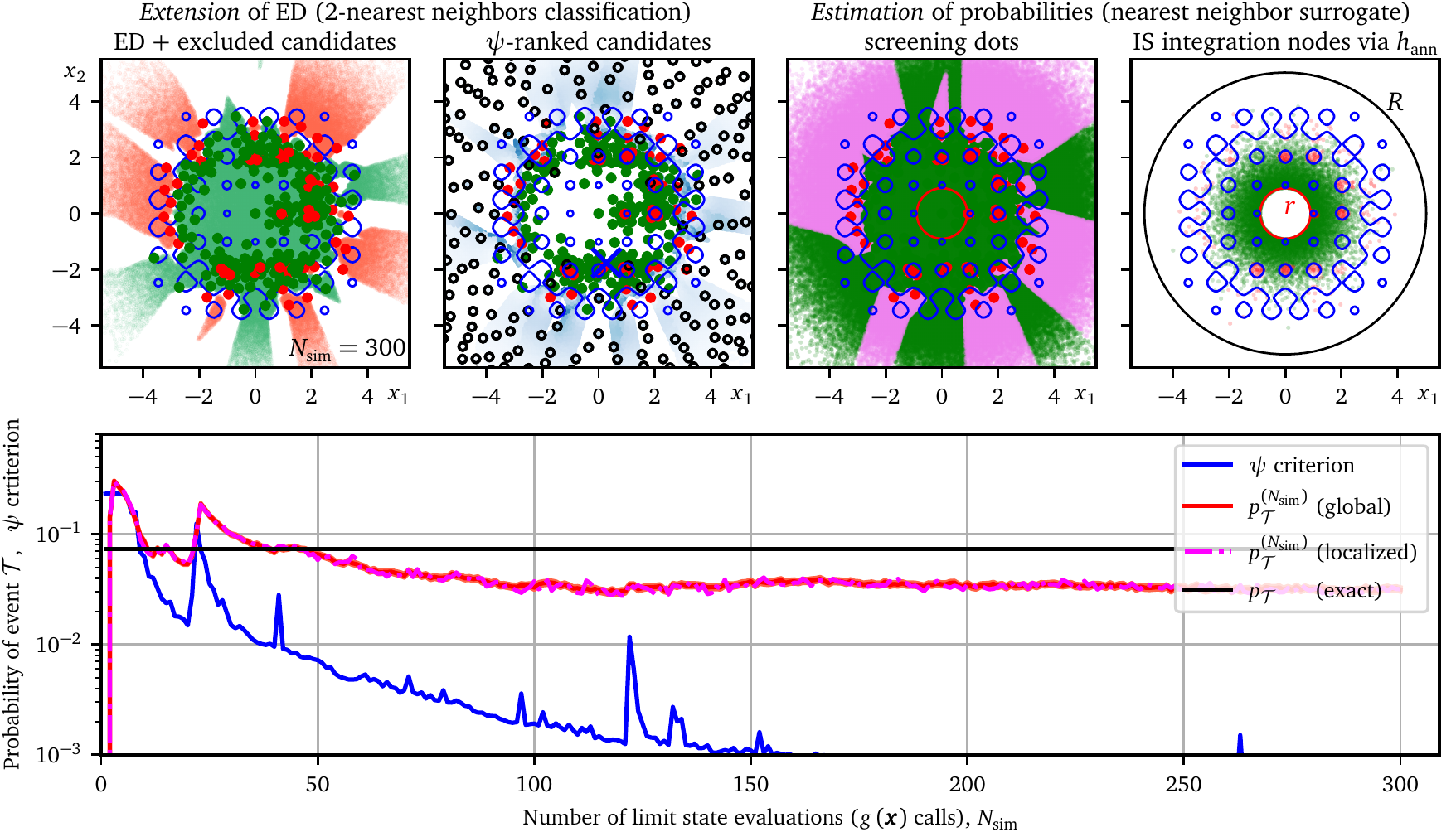}
    \caption{The ``Modified Rastrigin'' problem in Eq.~\eqref{eq:rastrigin}.
      The complete evolution of all panels is shown in the  \href{https://vutbr-my.sharepoint.com/:v:/g/personal/vorechovsky_m_vutbr_cz/EYkoii0qwBZMjyNq5Pdt1lYBZBQDELrQhscDgyTkLcExtg?e=AmuprD}
        {\textcolor{blue}{\textsf{Rastrigin video}}}.
        }
 \label{fig:Rastrigin}
\end{figure}

The \IS\ analysis of the nearest neighbor surrogate tends to produce an incorrect \revA{\pF\ }estimation of $0.0336$, while the same \IS\ with the true function leads to a~much higher value of $ \pF \approx 0.072986$. 
The reason is visible in Fig.~\ref{fig:Rastrigin} and in the attached  \href{https://vutbr-my.sharepoint.com/:v:/g/personal/vorechovsky_m_vutbr_cz/EYkoii0qwBZMjyNq5Pdt1lYBZBQDELrQhscDgyTkLcExtg?e=AmuprD}
        {\textcolor{blue}{\textsf{Rastrigin video}}}:
\rev{
the algorithm completely misses several important failure regions. Candidates located at small, closed failure regions are incorrectly classified as points with the same type of event as their two nearest neighbors. While such a~decision saves effort in other examples, it yields to the encapsulation of domains that are then never hit (if they were hit later, it would lead to an automatic fix for the problem as such regions get automatically backtracked and refined). In the algorithm run presented, only eight out of twenty closed failure regions were hit in stages $\Ns < 132$, and then one more region (apart from the open failure region) was discovered. It is a~matter of chance how many regions are discovered, and the very small region at $\{1, 0 \}$ was just a~matter of great luck.}
 Unfortunately, multiple runs of the proposed algorithm with the predefined density of the exploration set used in this example do \emph{not} remedy the problem completely: the predefined \rev{exploration} sets are too coarse to hit all of the eight small circular regions at coordinates as small as $x_1, x_2 \approx \pm 1$.

One can think about two different measures available to remedy the problem; one in the \emph{extension} step and the other in the \emph{estimation} step.

A straightforward solution on the \emph{extension} side of the algorithm is to use additional exploration candidates that are automatically included in the ranking procedure via the $\psi$ criterion. The refinement of the exploration set is simple: it can be achieved by, e.g., adding a~new \Nv-ball radius (Fig.~\ref{fig:ExplorationSet}) covered by (many) evenly distributed candidates. Once at least one point from the unsafe region is discovered, the method automatically refines the failure surface in its vicinity via exploitation candidates. However, this measure contradicts the central narrative of the article, namely the minimization of the number of evaluations of the function $\gx$.

Another possibility is to improve the \emph{estimation} step by training an advanced surrogate model on the $\Ns$ data point evaluated so far. As exemplified previously in the MetaBalls example, see Sec.~\ref{sec:MetaBalls}, if the numerical values of $\gx$ are usable to construct a~close-fitting approximation via, e.g., Kriging, RBF or PCE, the classification into event types can be much better than the poor nearest-neighbor classification used in this example. This measure assumes that the values of the function \gx\ are not just a~discrete classification of the state but that they suggest something about the shape of the function, including an estimate of where the function is negative. In such a~case, however, it is possible to use the information also for the selection of suitable extension candidates in the same way as other learning functions available in the literature. This paper focuses mainly on categorical functions.

Finally, we remark that if the closed area corresponds to an ``unknown result'' event, i.e., in the case of no relevant model response, then such a~response cannot be used trivially, e.g., as in the Kriging or PCE smooth approximations. \rev{However, the} proposed extension method treats it \rev{naturally} as a~new event type and tries to encapsulate it \rev{by refining the boundary around it, and also providing the associated probability of such an event}.

\subsection{Alternating domains -- a~noisy function with open failure regions}

To demonstrate the resistance of the proposed method to noise in the limit-state functions, we constructed a~limit-state function with alternating safe and failure domains, separated by a~sequence of parallel linear boundaries.
 The rare event $\pazocal{F}$ occurs whenever the function
\begin{align}
 \label{eq:CosExp}
     g\left( \x \right) = g\left( x_1 \right) = \cos\left[x_1 \exp \left( - x_1 - 4  \right)  \right]
\end{align}
fulfils $\gx<0$.
The boundaries between the alternating states are solutions satisfying
    $\left(2k + 1 \right)\pi = -2 b_{k} \exp (- b_{k} - 4)$, for $k=0,1,\ldots, \infty$.
    The $x_1$ coordinates of the boundaries are
    $\pmb{b} \approx
     \{ -3.267544, $ 
    $    -4.13154$,  
    $   -4.5466$,   
    $    -4.8239$,   
    $    -5.03$,     
    $    -5.20$,     
    $    -5.34$,     
    $    -5.46$,     
    $    \ldots \}$.
The exact failure probability can be calculated using a~series of simple FORM-style solutions with the above-listed ``design points'' $\{b_k,0\}$: $\pF = \sum_{k=0}^{\infty} \left( -1  \right)^{k}  \Phi \left( b_{k}  \right)  \approx  5.266 \cdot 10^{-4}$. 

\begin{figure}[!tb]
    \centering
    \includegraphics[width=17cm]{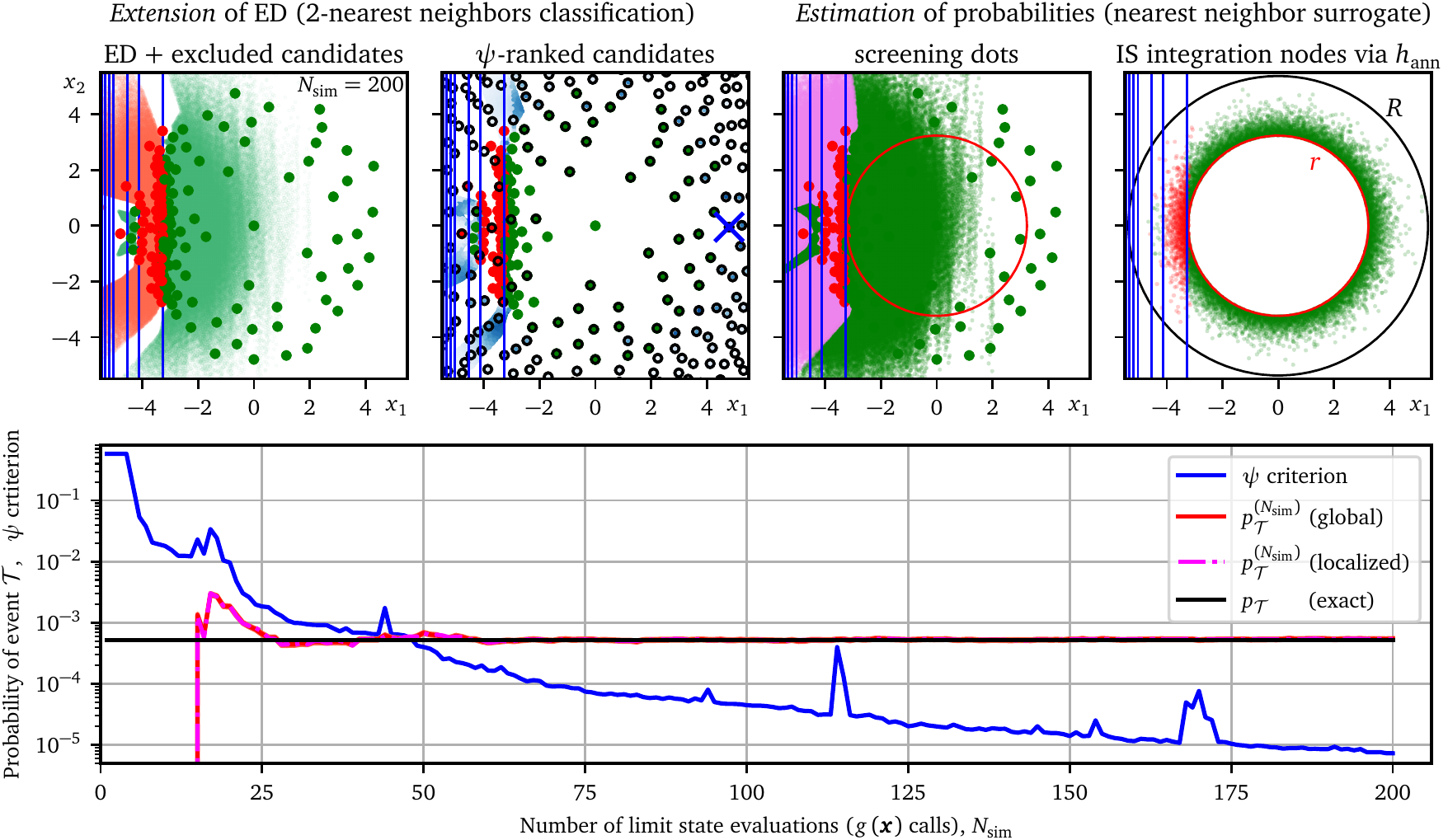}
    \caption{The ``Alternating Domains'' problem in Eq.~\eqref{eq:CosExp}.
      The complete evolution of all panels is shown in the  \href{https://vutbr-my.sharepoint.com/:v:/g/personal/vorechovsky_m_vutbr_cz/EfyQCyDw1UtCk7PHF_ZXml4BDblyYGKQsJUsfqu0IaJdBw?e=Rid9HD}
        {\textcolor{blue}{\textsf{Alternating Domains video}}}.
        }
 \label{fig:CosExp}
\end{figure}

The robustness of the proposed method is demonstrated through the discovery and subsequent refinement of the boundaries, which occur proportionally to the probability content $\psi$, see Fig.~\ref{fig:CosExp} and the associated \href{https://vutbr-my.sharepoint.com/:v:/g/personal/vorechovsky_m_vutbr_cz/EfyQCyDw1UtCk7PHF_ZXml4BDblyYGKQsJUsfqu0IaJdBw?e=Rid9HD}
        {\textcolor{blue}{\textsf{Alternating Domains video}}}.
At $\Ns \approx 60$, the first boundary is well refined, and the probability of failure is estimated very accurately. The estimate is not influenced much by the discovery of a~safe region placed in between two failure domains ($\Ns=117$). Due to the low influence of the secondary, tertiary, and other more distant boundaries, it takes many function calls to discover and refine them. The $\psi$ criterion correctly predominantly favors refinement of the first boundary over mapping the more distant regions.

\revA{The} estimated values of the proposed global sensitivities \revA{equal}
    $ s_{\pazocal{F},1}^2 =  1$ and
    $ s_{\pazocal{F},2}^2 =  0$, \revA{underlying the fact that all the ``gradients'' based on the direction to the nearest safe state are aligned with $x_1$. They match the FORM sensitivities based on $\xDP = \{ b_0, 0\}$: $\balpha = \{ 1, 0 \}$}.

\begin{figure}[!tb]
    \centering
    \includegraphics[width=19cm]{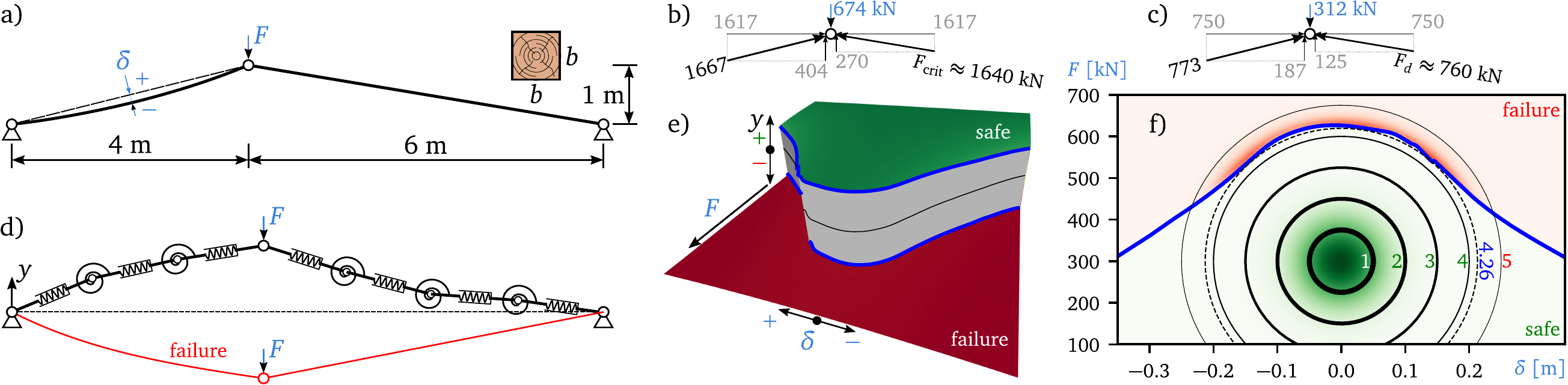}
    \caption{
    \rev{
    Unsymmetrical shallow von Mises truss.
    a) Initial geometry with two random variables $F$ and $\delta$;
    b,c) Simple static estimations of forces derived from the right hand side bar which either attains the critical Euler's force or the design compressive buckling force;
    d) Sketch of the discrete dynamical model and the final failure state.
    The course of dynamical response obtained by the FyDiK code is shown in  \href{https://vutbr-my.sharepoint.com/:v:/g/personal/vorechovsky_m_vutbr_cz/EbTKnyEyM1NPrm1ypgiJfkoBpMz4pu9crGVFeFhrhPHOVA?e=6Y0ExN}
        {\textcolor{blue}{\textsf{FyDiK animation}}};
    e) Limit state function computed for a~fine grid of combination ($F,\delta$) showing a~sudden jump between positive (green) and negative (red) final vertical coordinates $y$ of the loading point,
    f) Bivariate Gaussian density with contribution to $\pF$ shown via red color.
    }
    }
 \label{fig:vonMisesSketch}
\end{figure}

\rev{
\subsection{Unsymmetrical von Mises truss}
 \label{sec:vonMises}
In this section, we demonstrate the relevance of the proposed method for an engineering problem.
Consider a~shallow two-bar planar truss subjected to a~vertical load at its top joint. This is no doubt the most popular example of a~structural system where snap-through is the prevailing form of instability. It is associated to the name of von Mises \cite{Mises:23,Mises:25}, who first used this model to explore kinds of structural instability more general than Euler's buckling of a~single bar. It is an example of bistable shallow structures, which has been used for many years as a~benchmark in the numerical analysis of nonlinear structures \cite{Crisfield:91}. The stability analysis of these truss systems has been extensively studied since they display a~wealth of bifurcation phenomena. In recent years, a~growing interest has been observed also on the dynamic behavior of multi-stable structures, including the dynamic buckling of structures \cite{Orlando2018,FrantikvonMises}. The motivation for this interest stems also from the need to develop deployable space structures. We use this type of structure and the associated nonlinear dynamic solution of it as an engineering example for which the solution of the performance is quite complex and where the proposed technique for analysis of reliability provides a~robust solution very efficiently.
}

\rev{The particular initial geometry of the unsymmetrical planar timber von Mises truss spanning 10~m is displayed in Fig.~\ref{fig:vonMisesSketch}a. The two bars are made of a~hard wood (density $800$~kg/m$^3$, modulus of elasticity $E=12$~GPa), and they have an identical constant square cross-section with edge length $b=0.28$~m. The area of the cross-section is
    $A=7.84\cdot10^{-2}$~m$^2$ and the  moment of inertia
    $I=5.124\cdot10^{-4}$~m$^4$. If the two bars were ideal (no imperfections), a~slowly increasing vertical loading force $F$ first attains the Euler's critical force in the right hand side bar
    ($F_{\mathrm{crit}} = \pi^2 EI/l^2_{\mathrm{r}} \approx 1640$~kN). At that moment, the opposing horizontal projections of forces in the two bars are equal and therefore, the left  bar has its axial force of about 1667~kN. The sum of the vertical projections of the forces in static equilibrium equals $F\approx670$~kN, see Fig.~\ref{fig:vonMisesSketch}b. A~more refined calculation takes into the account the normal strain in the two bars, which results in the displacement of the loading point and change in the geometry. Such a~solution with the same critical force $F_{\mathrm{crit}}$  in the right bar corresponds to the vertical force of $644.2$~kN only. 
    To consider initial imperfections of the bars and determine the design resistance of the system, one can use the European standard Eurocode 5 \cite{EC5}. The planar buckling resistance of a~bar in compression in the absence of bending moment basically reduces the compressive resistance of a~cross section by employing a~buckling reduction factor $k_c$. The design compressive strength of the solid timber along the grains ($f_{d,0,d}=20.9$~MPa) is obtained from its characteristic counterpart $f_{c,01,k}=34$~MPa by applying the modification factor $k_{\mathrm{mod}}=0.8$ (load duration and moisture content) and the partial safety factor for a~material property $\gamma_{M} = 1.3$; see e.g. \cite{Hassan2019}. The right hand side bar has a~greater slenderness than the left bar ($51 < 75.3$), so it suffices to focus on the right bar and determine its buckling reduction factor $k_c = 0.46$. The corresponding design force in the bar becomes $760$~kN. The same analysis of static equilibrium of the undeformed configuration yields the vertical force corresponding to the design resistance of the truss $312$~kN, see Fig.~\ref{fig:vonMisesSketch}c. 
    This simple analysis provides an orientation about the order of loading forces relevant to the real structure.
}

\begin{figure}[!tb]
    \centering
    \includegraphics[width=17cm]{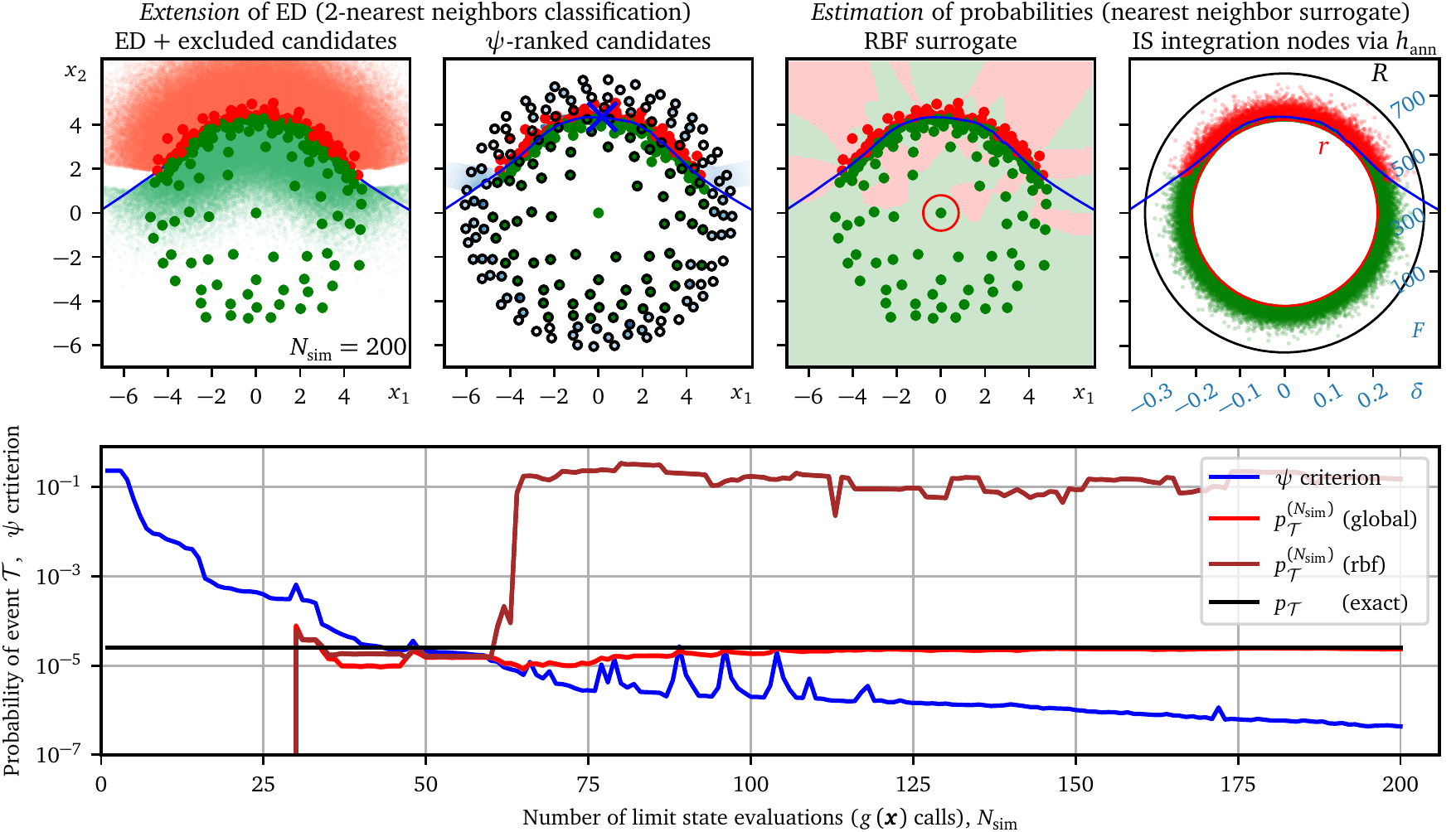}
    \caption{
    \rev{The ``von Mises'' problem sketched in Fig.~\ref{fig:vonMisesSketch}a.
      The complete evolution of all panels is shown in the  \href{https://vutbr-my.sharepoint.com/:v:/g/personal/vorechovsky_m_vutbr_cz/ER-mAY66_KRHi8PMHXvN8QQB8b9TJJG9_nvgT_LlpdNHpg?e=zYpat8}
        {\textcolor{blue}{\textsf{von Mises video}}}.
        }
        }
 \label{fig:vonMises}
\end{figure}

\rev{
Assume now that there are two random variables in the studied von Mises truss: (i) a~random loading vertical force $F$, and (ii) a~random imperfection $\delta$ of the left hand side bar in the form of one sine half-wave measured perpendicular to the straight line, see the sketch in Fig.~\ref{fig:vonMisesSketch}a. We are interested in computation of the probability that the truss will fail to keep its upright shape, that is, the loading point drops below the horizontal line connecting the supports. Therefore, the limit state function $\gx = y$, where $y$ is the final vertical coordinate of the loading point, and $\pF = \mathrm{P}\left( y<0\right)$.
We model the random amplitude of the sine imperfection $\delta$ by  Gaussian distribution with zero mean and standard deviation $\sigma_{\delta} = 50$~mm. The random force $F$ follows Gaussian distribution with the mean value of $\mu_F = 300$~kN and standard deviation $\sigma_F = 75$~kN. Since the structure is very shallow, it can happen that after a~sudden application of the vertical force, the vibrating bars can actually compress and deform in such a~manner that the snap-through process occurs and the loading point drops down. In such a~failure event, both bars have tensile stresses in the final stable state. To decide whether the initial configuration switches into the failure one is probably impossible by using an analytical expression. Therefore, we employ the explicit dynamics solver \textsf{FyDiK} \cite{fydik} to model the complex nonlinear dynamic behavior; see the sketch in Fig.~\ref{fig:vonMisesSketch}d and the attached \href{https://vutbr-my.sharepoint.com/:v:/g/personal/vorechovsky_m_vutbr_cz/EbTKnyEyM1NPrm1ypgiJfkoBpMz4pu9crGVFeFhrhPHOVA?e=6Y0ExN}
        {\textcolor{blue}{\textsf{FyDiK animation}}}. 
The animation compares two solutions with a~small difference in the loading force magnitude: the top truss keeps its upright configuration while the bottom truss is loaded by a~larger force leading to snap-through. The changing colors of individual segments of the bars represent the pulsating normal forces. The approach to modelling the dynamical response von Mises beams using \textsf{FyDiK} software was previously presented in \cite{FrantikvonMises}. In the present application, each bar was discretized into 10 elements connected via joints with rotational springs representing the bending stiffness of the bar. The normal stiffness is reflected via the attached normal springs.
 The linear viscous damping coefficient was set 11~Nsm$^{-1}$kg$^{-1}$ approximating the relative damping of about 3\%.
}

\rev{ The proposed adaptive sequential sampling calls the \textsf{FyDiK} solver with a~combination of the input  imperfection amplitude and the force, and the software is set to return the vertical coordinate $y$ of the loading point after the damped system converged a~stable state. The outcome of the solution is not smooth: the final stable states are either the ``safe'' configurations with the loading point $y \approx +1$~m above the supports or the ``failure'' configurations with the vertical coordinate about $y \approx-1$~m. The landscapes in the two domains are almost constant functions compared to the large difference of almost 2~m between them. In each domain, the output is a~smooth function with a~small gradient, but the derivative does not exist on the \emph{failure surface}, i.e., the boundary between the positive and negative outputs. Note that the failure surface is \emph{not symmetrical} with respect to the zero imperfection $\delta$ because the dynamical response evolves differently for concave and convex initial shapes of the left bar, even if the absolute  amplitudes were equal. The failure surface even exhibits small complex features which are sensitive to the particular proportions of the damping, inertia (density $\rho, A$) and stiffness ($E, A, I$), see Fig.~\ref{fig:vonMisesSketch}e.
}

\rev{
The physical space of Gaussian random variables is easily transformed into the standard Gaussian space by the component-wise linear mapping $(X_i-\mu_i)/\sigma_i$, $i=\{ \delta, F \}$. The joint probability density is visualized in Fig.~\ref{fig:vonMisesSketch}f along with a~color distinguishing between contributions to failure probability (red) and the prevailing safe (green) region. The region which contributes the most to \pF\ is in the vicinity of failure surface, which can roughly be approximated by a~quarter of a~circle with radius $\approx4.26$ in the standard Gaussian space in which the proposed algorithm operates.
Fig.~\ref{fig:vonMises} shows the evolution of sample and convergence of estimates for a~single run of the proposed method. At $\Ns =30$, the first failure configuration is hit, and since then the refinement of the boundary proceeds quickly. The proposed technique provides a~stable convergence of the estimate to the correct value $\pF = 2.557 \cdot 10^{-5}$, see the  \href{https://vutbr-my.sharepoint.com/:v:/g/personal/vorechovsky_m_vutbr_cz/ER-mAY66_KRHi8PMHXvN8QQB8b9TJJG9_nvgT_LlpdNHpg?e=zYpat8}
        {\textcolor{blue}{\textsf{von Mises video}}}.
   The \gx\ function is smooth in the two respective regions and when gradient methods originate their optimization in the safe domain, the $\gx$ function monotonically decreases towards the failure surface. 
    A~straightfoward use of FORM implemented in \textsf{OpenTURNS} provides the following results: after 122 calls to the limit state function in the Cobyla optimization solver, a~``design point'' is found at $(-1.898,3.949)$, which is at distance $\beta = 4.38$ from the origin so that the estimation of failure probability becomes incorrect: $5.9\cdot10^{-6}$. 
    SuS estimates the failure probability relatively well (depending on the particular setting such as the number of samples per probability level and the desired coefficient of variation), but the total number of limit state function calls must be greater than about $10\,000$, i.e., two orders of magnitude higher than in the proposed technique.
    A~good result is obtained with the standard importance sampling in \textsf{OpenTURNS}: setting the standard deviation of the Gaussian weighting function to about three and making 5\,000 calls to \gx\ provides $\pF$ estimates with the same accuracy as the proposed technique does after about 100 calls only.
    Directional sampling in \textsf{OpenTURNS} requires about 1,500 \gx\ calls to achieve the same accuracy level as the standard importance sampling.
    The proposed method uses the binary information only and yet it provides better results with less function calls. Moreover, it  can not be confused by the jump in the performance function $\gx$ between values for failure and safe states.
    This jump causes problems to some methods based on smooth surrogates. Smooth surrogate functions may provide wildly fluctuating interpolation which incorrectly classifies safe regions as failure events and vice versa. We document this by employing the RBF interpolation in the same manner as in the ``Four Branches'' example. The classification based on RBF is plotted in the third panel from the left in the top row of Fig.~\ref{fig:vonMises}. The brown line in the convergence panel shows that the corresponding estimation of failure probability ($\approx 0.8$) is almost five orders of magnitude wrong. 
The estimated global sensitivities defined in this paper are
    $ s_{\pazocal{F},\delta}^2  \approx  {\color{red}0.278}$ and
    $ s_{\pazocal{F},F}^2  \approx  {\color{red}0.722}$,
    and these numbers correspond to the relatively long part roughly the circular \revA{failure surface} approximating the Gaussian density isoline, see  Fig.~\ref{fig:vonMises} top.
}

\begin{figure}[!tb]
    \centering
    \includegraphics[width=17cm]{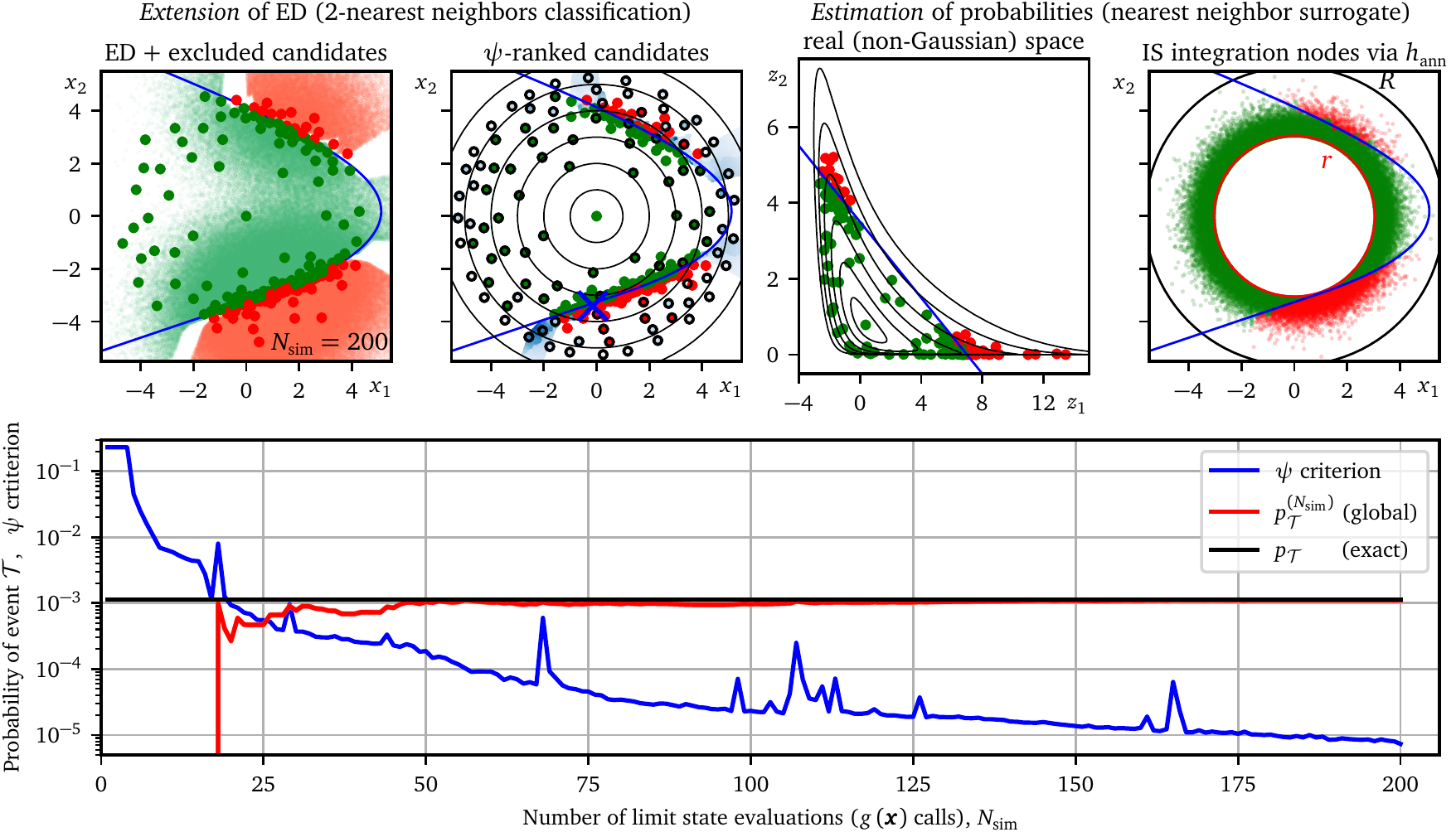}
    \caption{
    \rev{The ``Nataf'' problem in Eq.~\eqref{eq:nataf:g}.
      The complete evolution of all panels is shown in the  \href{https://vutbr-my.sharepoint.com/:v:/g/personal/vorechovsky_m_vutbr_cz/EY4NSKRC2TBJjRe_JYPElUwB8H5J6B_Ax0VXtEdMHvtugw?e=FO79Ba}
        {\textcolor{blue}{\textsf{Nataf video}}}.
        }
        }
 \label{fig:Nataf}
\end{figure}

\rev{
\subsection{Nataf example \label{sec:nataf:example}}
Until this point, all numerical examples were showing the proposed method on problems with bivariate Gaussian distribution. The purpose of this example is to verify that the method works for non-Gaussian input variables, provided the transformation to the standard Gaussian space is available. We named several popular options for this transformation and the relevant references in Sec.~\ref{sec:problemStatement}, and in the present example, we use the Nataf model. Consider a~linear limit state function involving two correlated non-Gaussian variables
\begin{equation}
    \label{eq:nataf:g}
    g \left( z_1 ,z_2 \right)
    = 7 - z_1 - 2z_2.
\end{equation}
The marginal $z_1$ has Gumbel (right-skewed, i.e. ``max'') distribution known also as type I Fisher-Tippett distribution,
with the scale parameter equal to 1 and zero mode (location parameter). The second marginal $z_2$ has Weibull (``min'') distribution with unit scale parameter and the shape parameter (exponent known as Weibull modulus) equal to $1.5$. This two-parameter Weibull variable is bounded from the left by the zero lower bound.
The Pearson correlation between these non-Gaussian variables equals $-0.708$ and a~failure event is signalled by $g<0$.
The true probability of failure can not be uniquely determined because the joint probability density $f_{\X}(\x)$ is not fully defined. However, we suppose that the joint density $f_{\X}(\x)$ constructed via the Nataf model is the true one.
The transformation to the standard Gaussian space makes the originally linear limit state function strongly nonlinear, see Fig.~\ref{fig:Nataf}. A~large-sample analysis in the standard Gaussian space provides the failure probability $\pF \approx 1.143 \cdot 10^{-3}$, a~value towards which the proposed technique quickly converges.
}

\rev{To apply the proposed method for extension of ED and estimation of $\pF$, it suffices to set up the transformation from the standard Gaussian space to the real space in which the limit state function is evaluated for each selected candidate, and the binary output about the system performance is associated with it. Otherwise, the technique works in the standard Gaussian space as described above. In Nataf model, the transformation to the real (correlated non-Gaussian) space proceed in two steps. First, a~coloring transformation, i.e., mapping to correlated Gaussian space is performed. There are infinitely many options to perform this mapping. Among the linear maps, Cholesky decomposition and eigendecomposition (known also as the Principal Component Analysis -- PCA, proper orthogonal decomposition, Karhunen-Lo\`{e}ve expansion, orthogonal transformation of covariance matrix) are the most famous ones and it can be shown \cite{NovVor:Trans:18} that they are special cases of a~general transformation pattern. This projection operation is non-separable meaning that generally all marginals are involved at once. The pairwise Pearson correlations between Gaussian marginals must be obtained
first by solving simple bivariate integrals \cite{DerKiureghian1986}. In our case, the underlying bivariate Gaussian marginals have correlation $\rho_{\mathrm{G}} = -0.8$. To perform the coloring transformation, we use PCA, which is a~preferred alternative as it provides an efficient way of reducing the dimension of the uncorrelated Gaussian space compared to the original dimension by ignoring the the components with the smallest variance contributions (eigenvalues)~\cite{NovVor:Trans:18,Vorech:08:CrossCorr}. The two eigenvalues form a~diagonal of matrix $\pmb{\Lambda} = \mathrm{diag}(1+\rho_{\mathrm{G}}, 1-\rho_{\mathrm{G}})$ and the corresponding orthonormal eigenvectors $\phi_1 = \left(c,c\right) \tran$ and $\phi_2 = \left(-c,c\right) \tran$, $c=1/\sqrt{2}$, form the square eigenvector matrix $\pmb{\Phi} = \left( \Phi_1, \Phi_2 \right)$. Given these matrices, any point $\x$ from the uncorrelated standard Gaussian space is transformed to the correlated standard Gaussian one via $\x^{(c)} = \pmb{\Phi} \pmb{\Lambda}^{1/2} \x$. The coordinates $\x_c$ are then mapped individually by component-wise \emph{memoryless isoprobabilistic transformation}: $z_v = F_v^{-1} \left[ \Phi \left( x^{(c)}_v \right) \right]$, where $\Phi$ is the standard Gaussian distribution function and $ F_v^{-1}$ is the inverse of the given non-Gaussian distribution function of variable $v$ (in our case Gumbel and Weibull). Finally, the coordinates $z_v$ are used in Eq.~\eqref{eq:nataf:g} to evaluate the system performance.
}

\rev{
The sampling in standard Gaussian space proceeds as usual and balances between the exploration and refinement of the (unknown) highly nonlinear failure surface; see the top left panel in Fig.~\ref{fig:Nataf}. The corresponding points transformed to the physical space are visualized in the third panel from the left, along with the probabilistic isolines and the original linear limit state function. The plot at the bottom of Fig.~\ref{fig:Nataf} shows the rapid convergence towards the exact \pF\ value. The complete evolution of all panels is shown in the  \href{https://vutbr-my.sharepoint.com/:v:/g/personal/vorechovsky_m_vutbr_cz/EY4NSKRC2TBJjRe_JYPElUwB8H5J6B_Ax0VXtEdMHvtugw?e=FO79Ba}
        {\textcolor{blue}{\textsf{Nataf video}}}.
We remark that the estimated global sensitivities of the underlying uncorrelated standard Gaussian variables do not provide the desired sensitivities of failure to the original variables. The underlying non-dimensional variables can live in subspace with reduced dimension and the projection removes the original meaning of the variables. The particular rotation in linear maps (such as the Cholesky- or eigen-decomposition or their generalization) is arbitrary \cite{NovVor:Trans:18}.
}


\subsection{Linear failure surface in higher dimensions}
 \label{sec:highdim}
The previous 2D examples revealed the robustness of the algorithm regarding the complicated failure surface and \gx\ function values. What remains a~question is how the algorithm efficiency scales with dimension. As clear already from Eq.~\eqref{Eq:ni}, a~higher dimension is increasingly hard to cover by a~small number of points. 

There are many classes of \revA{potential} problems to study in higher dimensions, and
\revA{one can distinguish between}  two extreme scenarios. In the first, (A), failure region is extremely unlikely and highly localized, such as in the ``Black Swan'' example. The number of function evaluations needed to hit the event using the exploration set becomes high in high dimensions; see Fig.~\ref{fig:ExplorationSet} and Tab.\ref{tab:ExplorationSet}. The table reveals that to explore and hopefully hit a~single ``Black Swan'' event at a~radius $\rho$ corresponding to a~probability as low as $10^6$ in $\Nv=10$ dimensions, the limit state function must have been evaluated for almost all previous levels of the ball, i.e. $(46+69+92+115+138+161)=460$ times. When, however, the event is discovered, the refinement of the failure surface around it does not cost many refinement steps. Localized \IS\ is the more suitable estimation procedure. In extreme scenario (B), the opposite limiting case is a~failure region in the shape of an $\Nv$-ball. While estimation using the global \IS\ is extremely effective because the sampling density $h_{\mathrm{ann}}$ is the optimal \IS\ density, the refinement of the failure surface is a~hard task. The reason is that the  $\Nv$-ball with a~given probability content has the greatest possible extent of the surface weighted by the probability density, meaning that all points close to the failure surface have high $\psi$ criteria, and accurate refinement necessitates many $\gx$ evaluations.

To present a~reasonable higher-dimensional example relevant to many practical problems, we use a~compromise: a~simple linear failure boundary (a line, a~plane, or generally a~hyperplane). \rev{In the following, we consider again that the input random variable\revA{s} are jointly Gaussian with independent standard marginals.} There is no reason to make the linear failure surface rotated in the space of input variables because the proposed framework is rotationally invariant in Gaussian space. Therefore, it suffices to make the limit state function simply depend on the first dimension only
\begin{align}
 \label{eq:linear}
     g\left( x_1,  x_2, \ldots  \right)
    &=
     \beta - x_1    .
\end{align}
Such a~problem has a~trivial analytical solution: $\pF = \Phi(-\beta)$. We set $\beta = 4.7534243$ to achieve
the failure probability $\pF = 10^{-6}$.

\begin{figure}[!tb]
    \centering
    \includegraphics[width=19cm]{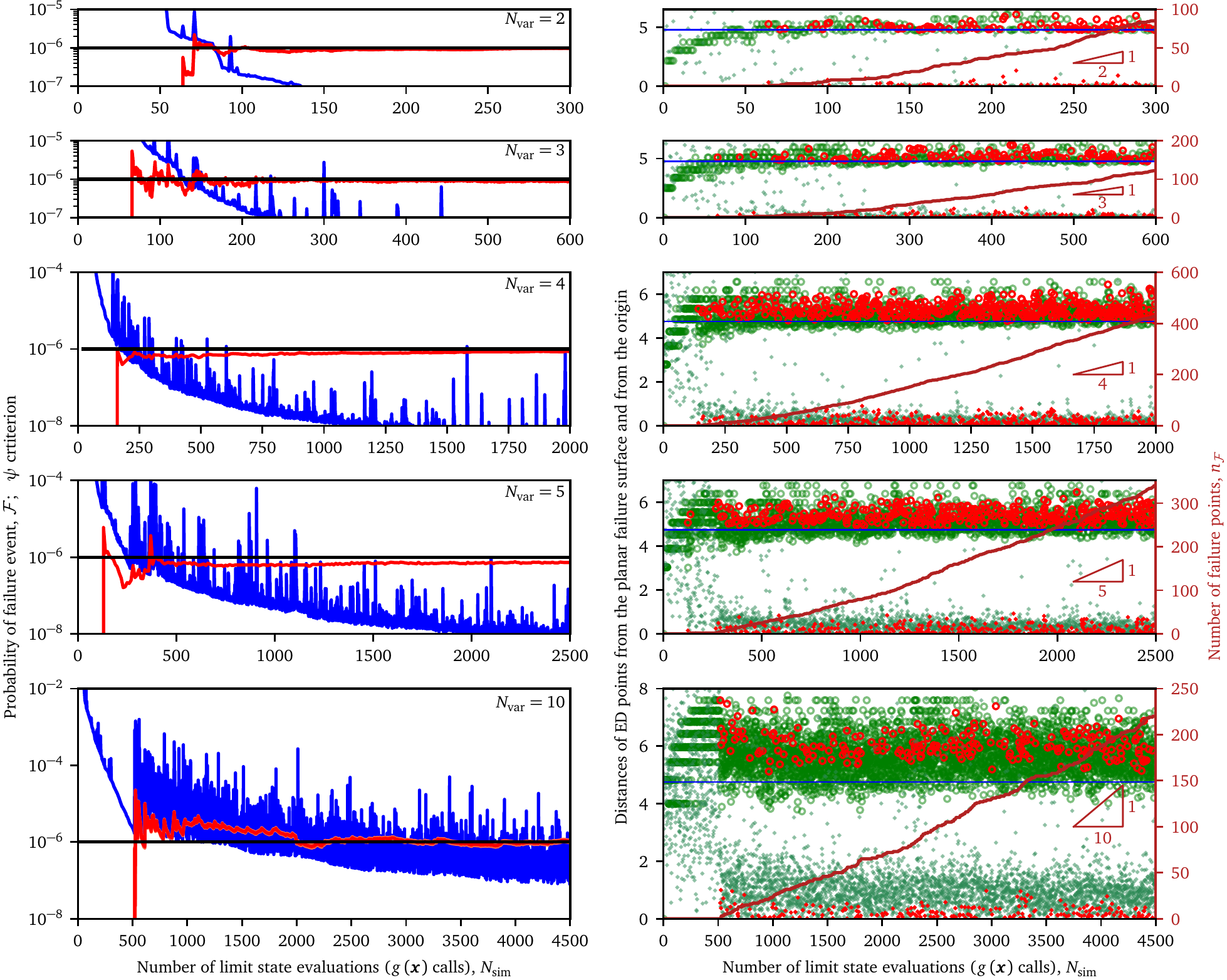}
    \caption{Recorded histories of exploration and estimation for the ``Linear Failure Surface'' problem in Eq.~\eqref{eq:linear} studied for $\Nv = 2$, 3, 4, 5 and 10 (rows). The left column reports histories of the $\psi$ criterion and the estimated $\PF{\Ns}$. The right column reports the radial distances of ED points from the origin (empty circles), as well as distances from the planar failure surface (small diamonds). The secondary vertical axis shows the number of failure  points.
      The complete evolution of ED extension in $\Nv=2$ and $\Nv=3$ dimensions is shown in  \href{https://vutbr-my.sharepoint.com/:v:/g/personal/vorechovsky_m_vutbr_cz/EbOMMFGxz-NEvawCP1_dSq4BFEgZuHBAcAWXbbi2MaMwPQ?e=16aPuW}
        {\textcolor{blue}{\textsf{video 2D}}}
        and
        \href{https://vutbr-my.sharepoint.com/:v:/g/personal/vorechovsky_m_vutbr_cz/EZEZ7UDaYn5BiHzwApQdhkABpJfF2Klaxao3ibs3GyjYhw?e=YtfeeR}
        {\textcolor{blue}{\textsf{video 3D}}}.
        }
 \label{fig:Lin}
\end{figure}

\begin{figure}[!tb]
    \centering
    \includegraphics[width=18cm]{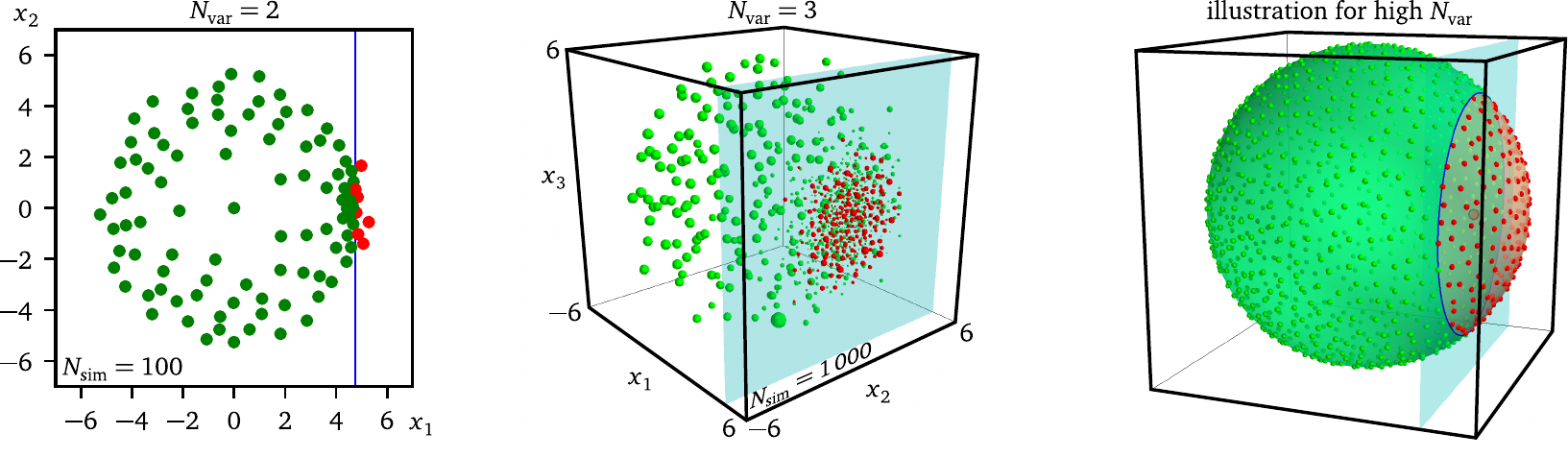}
    \caption{The ``Linear Failure Surface'' problem in Eq.~\eqref{eq:linear}.
      The complete evolution of $\Nv=2$ and $\Nv=3$ dimensions are depicted in \href{https://vutbr-my.sharepoint.com/:v:/g/personal/vorechovsky_m_vutbr_cz/EbOMMFGxz-NEvawCP1_dSq4BFEgZuHBAcAWXbbi2MaMwPQ?e=16aPuW}
        {\textcolor{blue}{\textsf{video 2D}}}
        and
        \href{https://vutbr-my.sharepoint.com/:v:/g/personal/vorechovsky_m_vutbr_cz/EQnTIkUOmepFhdftK4Ezx6oBCuAYl5D6hBsW-bVfVyulcg?e=2eEy4Q}
        {\textcolor{blue}{\textsf{video 3D}}}.
        The rightmost panel illustrates the situation for high $\Nv$, where all the probability is associated with a~thin layer of radius $r = \sqrt{\Nv-1}$, and the failure surface ($\Nv-1$ dimensional object visualized as a~plane) becomes effectively an $\Nv-2$ dimensional object visualized as the blue curve.
        }
 \label{fig:Lin:points}
\end{figure}

Fig.~\ref{fig:Lin} presents the results for $\Nv = 2,3,4,5$ and $10$ dimensions. The blue line in the left column is the estimated amount of probability resolved by evaluation of the corresponding \gx\ ($\psi$ criterion), and the red line plots the estimates of the failure probability.
It is clear that the purely ``exploratory phase'' with the expanding search until the first failure is hit consumes increasingly more $\gx$ function evaluations as the space dimension increases; compare the radial distances of points from the origin plotted as empty circles in the right column of Fig.~\ref{fig:Lin}, which are organized at individual distance levels. For $\Nv=10$, it takes about 500 \gx\ function calls to hit the failure event and begin refining the large failure surface. Stabilization of the probability estimates necessitates very fine refinement of the failure surface, which also consumes many function calls. The need to spend higher numbers of points in the purely exploratory  phase in higher dimensions is an inevitable consequence of the fact that the numerical value of $\gx$ cannot be used to orient the search, e.g., in the direction of the negative gradient, as SuS or methods building a~smooth surrogate do. The estimated $\pF$ for $\Nv=10$ shows quite a~serrated profile, although the coefficient of variation is very small due to the use of a~high number $n_{\IS}$ of integration nodes. The reason is that the boundary approximated via the \vor\ cells is also very serrated. Its extent is large, and the refinement would necessitate very many additional \gx\ calls.
One can also notice that the decrease in $\psi$ criterion is less rapid in high dimensions because the volume of the space simply increases with the space dimension. This is manifested via the increase of the extent of the failure surface part with a~high Gaussian density.
Apart from the radial distances plotted via empty circles in the right column of Fig.~\ref{fig:Lin}, we also
plot the distance of points from the planar failure surface as small diamonds (``safe'' green and ``failure'' red points).
Once the first failure event is hit, the extension algorithm primarily selects the points to refine the failure surface. However, as can be seen, their distance from the origin is considerably greater than the shortest distance of the plane $\beta$, which is marked by the horizontal blue line.
The fact that more points are needed in higher dimensions to refine the high-density part of the failure surface is also illustrated in Fig.~\ref{fig:Lin:points}. The complete evolution of the refinement process is captured point-by-point in individual frames of
    \href{https://vutbr-my.sharepoint.com/:v:/g/personal/vorechovsky_m_vutbr_cz/EbOMMFGxz-NEvawCP1_dSq4BFEgZuHBAcAWXbbi2MaMwPQ?e=16aPuW}
    {\textcolor{blue}{\textsf{video 2D}}}
and
    \href{https://vutbr-my.sharepoint.com/:v:/g/personal/vorechovsky_m_vutbr_cz/EZEZ7UDaYn5BiHzwApQdhkABpJfF2Klaxao3ibs3GyjYhw?e=YtfeeR}
    {\textcolor{blue}{\textsf{video 3D}}}.

The non-decreasing, maroon-colored line in the right column of Fig.~\ref{fig:Lin} also shows the number of failure points $\nF$. The secondary vertical axis and the small triangles reveal the ratio of failure points over all the limit state function evaluations. It can be seen that the proposed refinement algorithm tends to the ratio  $\nF / \Ns = 1/\Nv$, which is excellent in $\Nv=2$ dimensions where almost all limit state function evaluations refine the failure surface from both sides, but less efficient in higher dimensions in which increasingly more points are spent on the exploration of new territories and a~smaller share is devoted to boundary refinement.

When the value of the limit state function cannot be used to judge the direction of descent towards a~rare event because the output is just a~categorical variable, the solution becomes very hard in high dimensions. On the one hand, the dimension $\Nv$ of the input space becomes effectively $\Nv-1$ only because the whole probability content is limited to a~thin layer between (i) the inner ball containing almost no probability and (ii) the outer ball behind which there is again a~negligible probability content. This is illustrated in Fig.~\ref{fig:Lin:points} right, in which almost all points sampled from the high-dimensional Gaussian space reside on the green and red parts of the spherical ``important ring''.
Indeed, it is no longer true that the greatest contribution to the rare event (failure) probability resides in a~small region in the vicinity of the most central failure point (the big grey point under the failure spherical cap), as is the received wisdom based on low-dimensional problems.
The input space reduces to the thin ``important ring'', which virtually shrinks into just a~hypersphere with radius $r = \mathrm{mo}_{\rho} = \sqrt{(\Nv-1)}$; see Eq.~\eqref{eq:mo:rho}. When this radius is greater than the distance of the most central failure point from the origin, it makes no sense to perform density integration in the vicinity of the design point. The integration must be concentrated in the spherical cap only.
This reduction may seem to be of great help: instead of searching for rare events in the $\Nv$-dimensional Gaussian space, it suffices to explore the surface of an $\Nv$-ball with radius $r$. One can suggest that it suffices to cover the hypersphere uniformly by points and evaluate the limit state function there to estimate the probability of rare event $\pazocal{F}$ using the crude Monte Carlo estimator: $\nF/\Ns$. Unfortunately, the extent of this very flat hyper-surface becomes enormous as $\Nv$ grows large, which can be shown by substituting the radius $r$ from Eq.~\eqref{eq:mo:rho} into  Eq.~\eqref{eq:sur} and analyzing the leading terms
\begin{align}
\label{eq:sur:impring}
    \mathrm{Sur}\left[  B_{r} \right]
    =
    \frac{2 \pi^{\Nv/2}}{\Gamma\left(  \frac{\Nv}{2}\right)}
    \left( \sqrt{\Nv-1} \right) ^{\Nv-1}
    \approx
    \sqrt{2}
    \left(2 \pi \mathrm{e} \right) ^{ \left( \Nv-1 \right) /2}
    .
\end{align}
This expression underlines the exponential increase in the \Nv-ball surface with dimension \Nv.
Spreading points evenly on the surface of a~high-dimensional ball is itself a~challenging problem and cannot be achieved by transforming the known coordinates of well-distributed points covering a~unit hypercube \cite{VorMas:Nbody2:ADES:20,EliVorSad:miniMax:ADES:20,VorEli:Technometrics:20,Sobol:67,Sobol:76,Niederreiter:88,Tezuka:95,FanWan93,Owen:ScramblSobolXing:98,LEcuyerLemieux:02}.
However, even if we knew how to spread points very evenly on the hypersphere, thus reducing the variance of the Monte Carlo estimator
(for which the coefficient of variance is
   $ \sqrt{ \left({1-\pF} \right) / \left({\pF \; \Ns} \right)}$, i.e.,
   $\propto \Ns^{-1/2}$),
it would be of little help when there is a~need to decrease the number of points, \Ns.
One can argue that once at least one rare event location is discovered, it suffices to refine the failure surface on the hypersphere, i.e., just the blue curve in Fig.~\ref{fig:Lin:points} \revA{right}, which marks the intersection between the failure surface (visualized as via the blue plane) and the hypersphere.
However, either the failure probability is small (the rare event domain occupies a~small portion of the \Nv-ball surface) and therefore, it is difficult to discover it, or the failure probability is not that small, but then a~sufficient refinement of the failure surface on the hypersphere consumes many calls of the limit state function because the extent of the boundary is enormous despite the fact that it is an $(\Nv-2)$-dimensional object only. We conclude that the ``curse of dimensionality'' does not seem to have a~simple solution for categorical limit state functions \gx.

\revA{The proposed global sensitivity indices for the binary definition of the problem returns  $ s_{\pazocal{F},1}^2  \approx  1$, and the remaining sensitivities tend to zero because the direction to the nearest failure point is aligned with $x_1$.  This result matches the standard FORM sensitivities for the smooth definition of \gx\ in Eq.~\eqref{eq:linear}.}

\section{Conclusions}

This paper presents simple yet robust and efficient methods for the sequential \emph{extension} of experimental design and \emph{estimation} of rare event probabilities for computational models, which can be non-smooth, or can return only a~finite number of states or even have blind spots for which there is no result at all.
The \emph{extension} algorithm balances the \emph{gradual exploration} of new territories and \emph{refinement via the exploitation} of important regions by maximizing the proposed $\psi$ criterion. The criterion expresses the approximate amount of probability being classified by any proposed candidate for extension. The \emph{estimation} can be performed at any time during the extension process by quickly analyzing the point-wise information only. By obtaining data sequentially, it is possible to exploit the information from previous stages to inform the decision algorithm, minimize wasted resources, and continuously provide answers about the desired probabilities and sensitivities.
Two types of distance-based surrogate models are used to create a~quick and rough geometrical representation of the problem, particularly the partition of space into nonoverlapping subdomains of different event types.

The \emph{extension} of the experimental design makes no assumptions about the performance function and, therefore, is invariant with respect to its reparametrizations and reformulations, which do not alter the failure domain shape \& location, and the method is resistant against noise and jumps.

The proposed $\psi$ criterion for \emph{extension} involves a~set of primitive tools: evaluation of Gaussian density, computing distances among points,  masking (censoring), and the sorting of numerical arrays. The \emph{estimation} task uses two variants of standard importance sampling applied to a~surrogate model.

The proposed method combines the strengths of both sampling and approximation methods and keeps refining the local geometrical interpretation of the limit state function.

Simple yet apt \emph{global sensitivity measures} are proposed, which can be obtained for any rare event type as a~by-product of the presented method.

For the studied numerical examples, the existing methods that rely on the supposedly smooth contours of limit state functions are not competitors as they break down entirely for categorical functions. However, even if many of the examined functions are nicely smooth, many of the existing advanced methods still provide less accurate results based on a~higher number of function calls compared to the presented algorithm.
While the proposed  technique can be used for such smooth limit state functions too, and it provides fast convergence when combined with interpolating surrogate models, its strength and robustness are fully utilized where standard methods do not work: finite-state limit state functions which are expensive to evaluate.

The proposed method needs no fine-tuning of parameters, as there are no such variables that depend on the analyst. The only freedom is in the density of the initial \emph{exploration set} and with the option of refining it \rev{anytime} during the extension process. \rev{The prescribed numbers of exploration points for each radial distance directly control the convergence rate of the purely exploratory phase, i.e. until the discovery of the first failure point.}

The method can help in solving hard practical reliability problems for which the existing methods fail due to their strong assumptions about the performance function being well-formulated and well-behaved.

When the response of the limit state function is just categorical, the best candidate location (\emph{extension} of ED) and also the \emph{estimation} of probabilities can be pre-computed in advance for all potential outcomes while the expensive limit state function is still being computed. In this way, the wall time spent with the proposed algorithm can be \emph{de facto} decreased to zero.
Its applicability has been demonstrated for small to medium dimensions; high dimensions (several tens to hundreds of independent input variables) remain a~challenge.

\section*{Acknowledgment}

The  author  acknowledges the financial  support  provided  by  the Ministry of Education, Youth and Sports of the Czech Republic under project No. LTAUSA19058, and additionally by the Czech Science Foundation under project No. GC19-06684J.
\revA{The author thanks his colleagues Dr. Jan Eli\'{a}\v{s} for noticing an error in the original formulation of the proposed sensitivity measures, Dr. Frant\'{i}k for setting up the FyDiK solver needed to perform analyses of the von Mises truss numerical example and Dr. Sad\'{i}lek for help with Python programming.}



\bibliographystyle{elsarticle-num-names}
\bibliography{bibliography}

\normalsize

\appendix
\addcontentsline{toc}{section}{Appendices}

\section{Sampling a~unit random direction in \Nv\ dimensions
    \label{sec:rand:dir}
}

Sampling a~random unit direction from uniform distribution of directions in $\Nv$ dimensional space can be accomplished using a~simple procedure introduced in \cite{Muller1959}. It is achieved by generating a~random Gaussian point,
         $\pmb{m}=\{ m_1,\ldots,m_{\Nv} \}$, and scaling it onto a~surface of a~unit \Nv-ball $B_r$.
The Gaussian point has independent coordinates, each of which can be obtained via inverse transformation of the standard Gaussian distribution function, $\Phi^{-1}(p)$. Therefore, by choosing a~random sampling probability $p \in \left<0,1 \right)$, each coordinate can be obtained as $m_v=\Phi^{-1}(p)$. Once all the coordinates are obtained, a~random point $\pmb{s}$ on the surface of $B_r$ is
\begin{equation}
 \pmb{s} = \frac{\pmb{m}}{\lvert|\pmb{m} \rvert|}
 \label{eq:rad:direction}
\end{equation}
The normalization denominator is the Euclidean length $\lvert|  \pmb{m} \rvert| =  \sqrt{ \sum_{v=1}^{\Nv} m_v^2}$.
By scaling the point to a~unit length, the dimension of the problem gets reduced from \Nv{} to $(\Nv-1)$. The procedure is illustrated in Fig.~\ref{fig:test1} for a~set of 64 points in two dimensions.

\section{Sampling unit direction with uniform distribution \label{sec:unif:dir}}

If $n$ points are desired and generated according to a~procedure in \ref{sec:rand:dir} independently of each other, there is no guarantee that the points will be distributed evenly over the surface of a~unit ball.
There exist evenly distributed point sets on the surface of a~unit \Nv-ball for some point counts and dimensions \Nv; see e.g. \cite{Hardin1996}. There are also various classes of designs, such as the spherical t-design \cite{Delsarte1977,Hardin1996}, spiral point method \cite{Rakhmanov1994}, and others. Fekete points on a~unit sphere are points that minimize potential energy resembling the energy of a~system of $n$ repelling particles with unit charges according to Coulomb's law. Similar approaches have recently been developed for optimal designs from \Nv-dimensional unit cubes in periodic space \cite{VorMasEli:Nbody:ADES:19,VorMas:Nbody2:ADES:20,MasVor:ADES:CUDAparticles:18}.

\begin{figure}
\flushleft
\begin{minipage}{.5\textwidth}
  \centering
  \includegraphics[width=.9\linewidth]{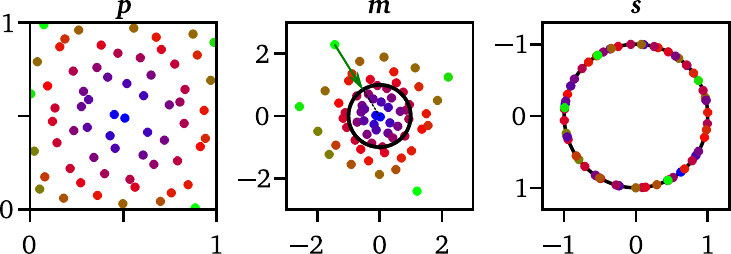}
  \captionof{figure}{Illustration of the Muller \cite{Muller1959} method for 64 points in $\Nv=2$ dimensions.}
  \label{fig:test1}
\end{minipage}%
\hspace{5mm}
\begin{minipage}{.45\textwidth}
  \centering
  \includegraphics[width=\linewidth]{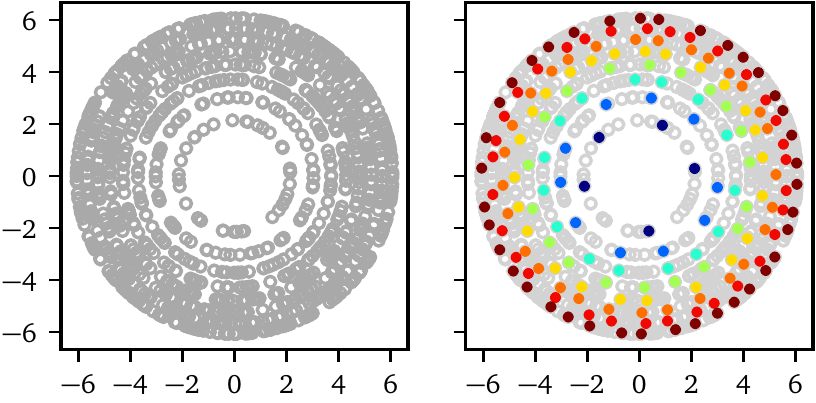}
  \captionof{figure}{Illustration of the proposed removal method performed individually for 8 \Nv-balls, each with different target numbers of points, $n$.  Gray circles show $7n$ points generated via scrambled \sobol\ sequence; the colored circles are the final $n$ points.}
  \label{fig:test2}
\end{minipage}
\end{figure}

In the present work, a~simple approach inspired by our recent work on Maximin and miniMax criteria
\cite{VorEli:Technometrics:20,EliVorSad:miniMax:ADES:20} is proposed. In order to obtain $n$ points that have no severe clusters and do not miss compact portions of directions, it suffices to generate a~larger pool of points, say $7n$, and keep selecting, one by one, points experiencing the maximum \emph{pressure} exerted on them. Imagine each point $i$ experiences pressures exerted by all other points. This pressure depends on the inverse distance between points $i$ and $j$, $j=1,\ldots,n$ and $i\neq j$. In particular, we consider the pressure on point $i$ exerted by point $j$ as $P_{i,j} = 1/d_{i,j}^{\Nv}$, where $d_{i,j}$ is the Euclidean distance between the two points.
A point $i$ receiving the maximum pressure, $P_i = \sum_{j=1, j\neq i}^{n} P_{i,j}$, is simply removed. The corresponding entries in the square distance/pressure matrix are deleted (masked), new pressures are computed via the summation of each row (or column), and the removal continues until the desired set of $n$ points remains.
Fig.~\ref{fig:test2} shows the original pool of points (empty gray circles) and the retained points after the removal procedure (colored solid circles) for eight ``onion layers''.

To prevent severe clusters or empty spaces being generated for the initial set of points, it is desirable to generate the directions using \ref{sec:rand:dir} in which the sampling probabilities $p$ are selected, e.g., via Quasi-Monte Carlo or randomized  Quasi-Monte Carlo sequences \cite{Halton:60,Sobol:67,Sobol:76,Niederreiter:87,Niederreiter:88,Nieder:RandNumGen_AND_QMC_1992,Faure:81,FangWangBentler:NT:SS:1994,Tezuka:95,Owen:ScramblSobolXing:98, Owen:Scramblings:99,LEcuyerLemieux:02}. These sequences provide points with very good uniformity in a~unit hypercube.

\section{Example network of predefined exploration points \label{sec:explor:table}}

\begin{figure}[!htb]
    \centering
    \includegraphics[width=8cm]{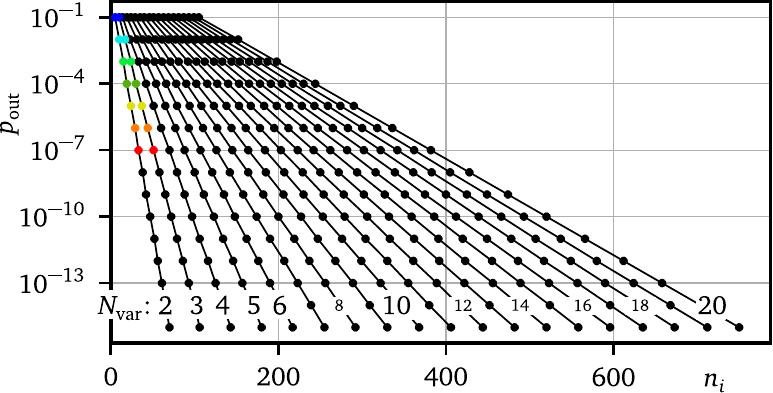}
    \caption{Proposed numbers of exploration points covering \Nv-balls according to Eq.~\eqref{Eq:ni}.}
 \label{fig:ExplorationSet:Numbers}
\end{figure}

In this section, we present Tab.~\ref{tab:ExplorationSet} and Fig.~\ref{fig:ExplorationSet:Numbers} with the point counts obtained via Eq.~\eqref{Eq:ni}. How these points cover circles (2D) and balls (3D) is visualized in Fig.~\ref{fig:ExplorationSet}.

\begin{table*}[htbp]
\small
  \centering
  \caption{Overview of the levels with the associated probabilities outside the \Nv-balls with radii $\rho$. The recommended numbers of surface points $n_i$ are obtained using Eq.~\eqref{Eq:ni}. The colored region corresponds to the visualization in Fig.~\ref{fig:ExplorationSet}}
    \begin{tabular}{rl|rr|rr|rr|rr|rr|rr|rr}
          &&
          \multicolumn{2}{c|}{\Nv=2} & \multicolumn{2}{c|}{\Nv=3} & \multicolumn{2}{c|}{\Nv=4} & \multicolumn{2}{c|}{\Nv=5} & \multicolumn{2}{c|}{\Nv=6} & \multicolumn{2}{c|}{\Nv=10} & \multicolumn{2}{c}{\Nv=20}
        \\
        \cmidrule{3-16}
    $i$ & $p_{\mathrm{out},i}$ & $n_i$& $\rho_i$& $n_i$& $\rho_i$& $n_i$& $\rho_i$& $n_i$& $\rho_i$& $n_i$& $\rho_i$& $n_i$& $\rho_i$& $n_i$& $\rho_i$\\
    \midrule
    {\color{l1} 1}    & {\color{l1} $10^{-1}$ }&{\color{l1} 5}    & {\color{l1} 2.15}  & {\color{l1} 10}   & {\color{l1} 2.50}  & 14    & 2.79  & 19    & 3.04  & 24    & 3.26  & 46    & 4.00  & 105   & 5.33 \\
   {\color{l2} 2}     & {\color{l2} $10^{-2}$} & {\color{l2} 10}    &  {\color{l2} 3.03}  &  {\color{l2} 17}    &  {\color{l2} 3.37}  & 23    & 3.64  & 31    & 3.88  & 38    & 4.10  & 69    & 4.82  & 152   & 6.13 \\
   {\color{l3} 3}     & {\color{l3} $10^{-3}$} & {\color{l3} 15}    & {\color{l3} 3.72}  & {\color{l3} 24}    & {\color{l3} 4.03} & 33    & 4.30  & 42    & 4.53  & 52    & 4.74  & 92    & 5.44  & 198   & 6.73 \\
   {\color{l4} 4}     & {\color{l4} $10^{-4}$} & {\color{l4} 19}    & {\color{l4} 4.29}  & {\color{l4} 30}    & {\color{l4} 4.59}  & 42    & 4.85  & 54    & 5.07  & 66    & 5.28  & 115   & 5.96  & 244   & 7.24 \\
   {\color{l5} 5}     & {\color{l5} $10^{-5}$} & {\color{l5} 24}    & {\color{l5} 4.80}  & {\color{l5} 37}    & {\color{l5} 5.09} & 51    & 5.34  & 65    & 5.55  & 79    & 5.75  & 138   & 6.43  & 290   & 7.68 \\
   {\color{l6} 6}     & {\color{l6} $10^{-6}$} &  {\color{l6} 29}    &  {\color{l6} 5.26}  &  {\color{l6} 44}    &  {\color{l6} 5.54}  & 60    & 5.78  & 77    & 5.99  & 93    & 6.19  & 161   & 6.85  & 336   & 8.09 \\
   {\color{l7} 7}     & {\color{l7} $10^{-7}$} & {\color{l7} 33}    & {\color{l7} 5.68}  & {\color{l7} 51}   & {\color{l7} 5.95}  & 70    & 6.18  & 88    & 6.39  & 107   & 6.58  & 184   & 7.23  & 382   & 8.46 \\
    8     & $10^{-8}$ & 38    & 6.07  & 58    & 6.33  & 79    & 6.56  & 100   & 6.77  & 121   & 6.95  & 207   & 7.59  & 428   & 8.81 \\
    9     & $10^{-9}$ & 42    & 6.44  & 65    & 6.70  & 88    & 6.92  & 111   & 7.12  & 135   & 7.30  & 230   & 7.93  & 474   & 9.14 \\
    10    & $10^{-10}$ & 47    & 6.79  & 72    & 7.04  & 97    & 7.26  & 123   & 7.45  & 148   & 7.63  & 253   & 8.26  & 520   & 9.45 \\
    11    & $10^{-11}$ & 52    & 7.12  & 79    & 7.36  & 106   & 7.58  & 134   & 7.77  & 162   & 7.95  & 276   & 8.56  & 566   & 9.74 \\
    12    & $10^{-12}$ & 56    & 7.43  & 86    & 7.68  & 116   & 7.89  & 146   & 8.08  & 176   & 8.25  & 299   & 8.86  & 612   & 10.03 \\
    13    & $10^{-13}$ & 61    & 7.74  & 93    & 7.97  & 125   & 8.18  & 157   & 8.37  & 190   & 8.54  & 322   & 9.14  & 658   & 10.30 \\
    14    & $10^{-14}$ & 65    & 8.03  & 100   & 8.26  & 134   & 8.47  & 169   & 8.65  & 204   & 8.82  & 345   & 9.41  & 704   & 10.56 \\
    15    & $10^{-15}$ & 70    & 8.31  & 106   & 8.54  & 143   & 8.74  & 180   & 8.92  & 217   & 9.09  & 368   & 9.68  & 750   & 10.82 \\
    \bottomrule
    \end{tabular}%
  \label{tab:ExplorationSet}%
\end{table*}%

\end{document}